\documentclass[twocolumn]{aastex61}
\pdfoutput=1 
\usepackage{amsmath,amstext}
\usepackage[T1]{fontenc}
\usepackage[figure,figure*]{hypcap}
\newcommand{\kms}{km~s$^{-1}$}
\newcommand{\doh}{$\Delta$O/H}

\def\D16{N2S2 (D16)~}
\def\O3N2{O3N2 (M13)~}
\def\N2{N2 (PP04)~}

\shorttitle{Mapping metallicity variations}
\shortauthors{Kreckel et al.}

\begin{document}

\title{Mapping metallicity variations across nearby galaxy disks}

\author{K.~Kreckel}
\affiliation{Max Planck Institut f\"{u}r Astronomie, K\"{o}nigstuhl 17, D-69117 Heidelberg, Germany}
\affiliation{Astronomisches Rechen-Institut, Zentrum f\"{u}r Astronomie der Universit\"{a}t Heidelberg, M\"{o}nchhofstra\ss e 12-14, 69120 Heidelberg, Germany}
\affiliation{kathryn.kreckel@uni-heidelberg.de}

\author{I.-T.~Ho}
\affiliation{Max Planck Institut f\"{u}r Astronomie, K\"{o}nigstuhl 17, D-69117 Heidelberg, Germany}
\author{G.~A.~Blanc}
\affiliation{The Observatories of the Carnegie Institution for Science, 813 Santa Barbara Street, Pasadena, CA 91101, USA}
\affiliation{Departamento de Astronom\'ia, Universidad de Chile, Casilla 36-D, Santiago, Chile}
\author{B.~Groves}
\affiliation{International Centre for Radio Astronomy Research, The University of Western Australia, 35 Stirling Hwy, 6009 Crawley, WA, Australia}
\affiliation{Research School of Astronomy and Astrophysics, Australian National University, Weston Creek 2611, Australia}
\author{F.~Santoro}
\affiliation{Max Planck Institut f\"{u}r Astronomie, K\"{o}nigstuhl 17, D-69117 Heidelberg, Germany}

\author{E.~Schinnerer}
\affiliation{Max Planck Institut f\"{u}r Astronomie, K\"{o}nigstuhl 17, D-69117 Heidelberg, Germany}

\author{F.~Bigiel}
\affiliation{Argelander-Institut f\"{u}r Astronomie, Universit\"{a}t Bonn, Auf
dem H\"{u}gel 71, 53121 Bonn, Germany}
\author{M.~Chevance}
\affiliation{Astronomisches Rechen-Institut, Zentrum f\"{u}r Astronomie der Universit\"{a}t Heidelberg, M\"{o}nchhofstra\ss e 12-14, 69120 Heidelberg, Germany}
\author{E.~Congiu}
\affiliation{The Observatories of the Carnegie Institution for Science, 813 Santa Barbara Street, Pasadena, CA 91101, USA}
\affiliation{INAF--Osservatorio Astronomico di Brera, via E. Bianchi 46, 23807 Merate (LC), Italy}
\author{E.~Emsellem}
\affiliation{European Southern Observatory, Karl-Schwarzschild-Str. 2, 85748 Garching, Germany}
\affiliation{Universit\'{e} Lyon 1, ENS de Lyon, CNRS, Centre de Recherche Astrophysique de Lyon UMR5574, 69230 Saint-Genis-Laval, France}
\author{C.~Faesi}
\affiliation{Max Planck Institut f\"{u}r Astronomie, K\"{o}nigstuhl 17, D-69117 Heidelberg, Germany}
\author{S.~C.~O.~Glover}
\affiliation{Institut f\"{u}r theoretische Astrophysik, Zentrum f\"{u}r Astronomie der Universit\"{a}t Heidelberg, Albert-Ueberle Str. 2, D-69120 Heidelberg, Germany}
\author{K.~Grasha}
\affiliation{Research School of Astronomy and Astrophysics, Australian National University, Weston Creek 2611, Australia}
\author{J.~M.~D.~Kruijssen}
\affiliation{Astronomisches Rechen-Institut, Zentrum f\"{u}r Astronomie der Universit\"{a}t Heidelberg, M\"{o}nchhofstra\ss e 12-14, 69120 Heidelberg, Germany}
\author{P.~Lang}
\affiliation{Max Planck Institut f\"{u}r Astronomie, K\"{o}nigstuhl 17, D-69117 Heidelberg, Germany}
\author{A.~K.~Leroy}
\affiliation{Department of Astronomy, The Ohio State University, 140 West 18th Ave, Columbus, OH 43210, USA}
\author{S.~E.~Meidt}
\affiliation{Sterrenkundig Observatorium, Universiteit Gent, Krijgslaan 281 S9, B-9000 Gent, Belgium}
\author{R.~McElroy}
\affiliation{Max Planck Institut f\"{u}r Astronomie, K\"{o}nigstuhl 17, D-69117 Heidelberg, Germany}
\author{J.~Pety}
\affiliation{IRAM, 300 rue de la Piscine, 38406 Saint-Martin-d'H\`eres, France} 
\affiliation{LERMA, Observatoire de Paris, PSL Research University, CNRS, Sorbonne Universites, Univ. Paris 06, 75005 Paris, France}
\author{E.~Rosolowsky}
\affiliation{4-183 CCIS, University of Alberta, Edmonton, Alberta, Canada}
\author{T.~Saito}
\affiliation{Max Planck Institut f\"{u}r Astronomie, K\"{o}nigstuhl 17, D-69117 Heidelberg, Germany}
\author{K.~Sandstrom}
\affiliation{Center for Astrophysics and Space Sciences, Department of Physics, University of California, San Diego, 9500 Gilman Drive, La Jolla, CA 92093, USA}
\author{P.~Sanchez-Blazquez}
\affiliation{Departamento de F\'{i}sica Te\'{o}rica, Universidad Aut\'{o}noma de Madrid, Cantoblanco, E-28049 Madrid, Spain}
\author{A.~Schruba}
\affiliation{Max-Planck-Institut f\"{u}r extraterrestrische Physik, Giessenbachstrasse 1, D-85748 Garching, Germany}

\begin{abstract}

The distribution of metals within a galaxy traces the baryon cycle and the buildup of galactic disks, but the detailed gas phase metallicity distribution remains poorly sampled. We have determined the gas phase oxygen abundances for 7,138 HII regions across the disks of eight nearby galaxies using
VLT/MUSE optical integral field spectroscopy as part of the PHANGS-MUSE survey. After removing the first order radial gradients present in each galaxy, we look at the statistics of the metallicity offset (\doh) and explore azimuthal variations.  Across each galaxy, we find low ($\sigma$=0.03-0.05 dex) scatter at any given radius, indicative of efficient mixing.  We compare physical parameters for those HII regions that are 1$\sigma$ outliers towards both enhanced and reduced abundances.  Regions with enhanced abundances have high ionization parameter, higher H$\alpha$ luminosity, lower H$\alpha$ velocity dispersion, younger star clusters and associated molecular gas clouds show higher molecular gas densities.  This indicates recent star formation has locally enriched the material.  Regions with reduced abundances show increased H$\alpha$ velocity dispersions, suggestive of mixing introducing more pristine material.  We observe subtle azimuthal variations in half of the sample, but can not always cleanly associate this with the spiral pattern.  Regions with enhanced and reduced abundances are found distributed throughout the disk, and in half of our galaxies we can identify subsections of spiral arms with clearly associated metallicity gradients.  This suggests spiral arms play a role in organizing and mixing the ISM.
\end{abstract}


\section{Introduction}

The production and build up of heavy elements from stellar nucleosynthesis over cosmic time has resulted in pronounced radial trends in chemical abundances within galaxies.  This is commonly constrained by measuring the current gas-phase oxygen abundance (metallicity) within HII regions \citep{Kennicutt1996}.  The predominantly negative radial trends in metallicity have been well established across large samples of galaxies \citep{Zaritsky1994, Sanchez2014, Kaplan2016, Belfiore2017, Poetrodjojo2018}, tracing the inside-out growth of galaxy disks \citep{Boissier1999}. 

In addition to tracking feedback and pollution from massive stars into the interstellar medium (ISM),  metals play a key role in the cooling of the ISM, and drive changes in fundamental local ISM properties (e.g., gas to dust ratio, HI-to-H$_2$ transition, CO-to-H$_2$ conversion factor).  Understanding the degree of chemical homogeneity in star-forming galaxies is key to make use of stellar metallicities for Galactic archaeology, including chemical tagging studies and assessments of burstiness and accretion events. 

This purely radial picture may be an oversimplification because spiral arms, bars, and stellar feedback drive mixing and gas flows. This has been theorized to happen through  bar-driven radial mixing \citep{DiMatteo2013}, spiral arm driven large-scale systematic streaming motions \citep{Grand2016, Sanchez-Menguiano2016}, kiloparsec-scale mixing-induced dilution due to the spiral density waves passage \citep{Ho2017}, thermal and gravitational instabilities \citep{Yang2012, Petit2015}, and interstellar turbulence \citep{deAvillez2002, Krumholz2018}. Many of these processes should affect azimuthal inhomogeneity in the gas-phase abundances, which could then impact the local physical conditions that regulate star formation.

With typically only 10-100 HII regions with measured abundances per galaxy \citep[e.g.][]{Pilyugin2014}, detailed exploration of azimuthal trends has been unfeasible for large samples of galaxies, and the detailed gas phase metallicity distribution remains poorly understood. With the introduction of new wide-field optical integral field unit (IFU) spectrographs, it has become possible to fully sample the gas-phase metallicity across galaxy disks.  Large optical IFU projects that survey hundreds to thousands of galaxies (CALIFA - the Calar Alto Legacy Integral Field Area Survey, \citealt{Sanchez2012}; MaNGA -  Mapping Nearby Galaxies at Apache Point Observatory , \citealt{Bundy2015}; SAMI- Sydney-AAO Multi-object Integral-field spectrograph, \citealt{Bryant2015}) trade sample size for spatial resolution, achieving $\sim$kpc spatial resolution.  While having the obvious advantage of large statistical samples, this can introduce undesirable systematic effects when inferring metallicity, as multiple ionization sources can contribute within a single resolution element \citep{Ho2018, Poetrodjojo2019}. 

Metallicity studies with optical IFU observations that isolate individual HII regions ($<$100 pc scales) have so far been limited to individual objects.  These studies have produced conflicting results, with some showing no evidence for significant large-scale azimuthal variations in relation to the spiral pattern \citep{ Kreckel2016}, while other galaxies reveal clear enrichment patterns along spiral structures  \citep{Sanchez-Menguiano2016, Vogt2017, Ho2017, Ho2018}. Our Milky Way galaxy itself reveals tantalizing evidence for azimuthal metallicity variations \citep{Balser2011, Balser2015}. 

The Physics at High Angular resolution in Nearby GalaxieS (PHANGS) collaboration\footnote{\url{http://www.phangs.org}} aims to understand the interplay of the small-scale physics of gas and star formation with galactic structure and galaxy evolution.  To this end, we are collecting observations of the molecular gas, ionized gas and stellar populations within nearby (D$<$17 Mpc) moderately inclined star-forming disk galaxies.
To constrain the local and large-scale metallicities of these galaxies, we have begun a large observing program employing the Very Large Telescope/Multi Unit Spectroscopic Explorer (VLT/MUSE) to mosaic the central disk of 18 galaxies with optical IFU observations (PI: Schinnerer).  Combined with our pilot survey that targeted NGC 628 \citep{Kreckel2016, Kreckel2018}, these 19 galaxies make up the PHANGS-MUSE survey. 

We present here initial results from the PHANGS-MUSE survey using observations of the first eight completed galaxies. With $\sim$50 pc resolution across the central star-forming disks of these galaxies, our observations are well matched to the typical spatial scales of HII regions (10-200pc; \citealt{Azimlu+2011}). Our MUSE observations can spatially isolate HII regions and, using multiple line diagnostics, distinguish photoionized regions from other ionizing sources (diffuse ionized gas, supernova remnants, planetary nebulae, shocks, outflows, and extended ionization from an active galactic nuclei). This combination of high spatial resolution with multiple line diagnostics mapped across the galaxy disks allows us to take a statistical approach to characterize metallicity variations  across galaxy disks and relate these to the local physical conditions of the ISM. 

Section \ref{sec:data} presents the data we use and describes the construction of our HII region catalogs.  In Section \ref{sec:strongline} we detail how we use emission lines to infer the physical conditions of HII regions, particularly the metallicity and ionization parameter.   Section \ref{sec:results} presents our results on the trends in metallicity with local physical conditions, the subtle evidence we observe for azimuthal variations, and a brief discussion of the impact of bars on the metal distribution. Section \ref{sec:discussion} discusses the implications of our results and we conclude in Section \ref{sec:conclusion}.

\section{Data}
\label{sec:data}
\subsection {Observations}
We use the MUSE spectrograph \citep{Bacon2010} at the VLT to mosaic the central star-forming disks of eight nearby spiral galaxies (Table \ref{tab:sample}) as part of the PHANGS-MUSE large observing program (PI: Schinnerer).  
This powerful optical IFU provides a 1\arcmin $\times$ 1\arcmin~ field of view with 0\farcs2 pixels and a typical spectral resolution of $\sim$2.5\AA~ over the nominal wavelength range, covering 4800-9300\AA.  
Our sample of eight galaxies includes observations of NGC 628, which have been described previously \citep{Kreckel2016, Kreckel2018}. 
For our large program observations, each position was observed in four rotations, with two sky pointings, and a total on-source integration time of 43 minutes.  For NGC 628 we used slightly different integration times (43-50 minutes).  The center pointing for NGC 2835 was observed as part of the MUSE Atlas of Disks survey (MAD; \citealt{Erroz-Ferrer2019}), with 60 minute exposure time.  NGC 4254 and NGC 4535 were observed using the wide-field adaptive optics (AO) mode.  We present here results using the first eight completed galaxies.  Table \ref{tab:sample} summarises the main properties of the galaxy sample, the angular resolution achieved in the observations, the number of HII regions in each galaxy and the physical size of the survey area covered in the disk.

Angular resolution varies between galaxies from 0\farcs5 (for the AO mode) to 1\farcs0,  and can vary slightly between fields (typically less than 0\farcs2) and as a function of wavelength (0\farcs1-0\farcs2 across the whole wavelength range). At the distances of our sample, this resolution corresponds to physical scales of 40--70~pc. As we integrate the spectrum over each entire HII region, the slight variations in seeing should not impact our results.

\begin{deluxetable*}{cccrrrrcccrr}
\centering
\tablecaption{Key parameters of our galaxy sample \label{tab:sample}}
\tablehead{
\colhead{Name} \vspace{-0.3cm}&
\colhead{D} &
\colhead{Type} &
\colhead{Log$_{10}$(M$_*$)} & 
\colhead{Log$_{10}$(SFR)} & 
\colhead{PA\tablenotemark{a}} &
\colhead{i\tablenotemark{b}} &
\colhead{$\rm R_{25}$\tablenotemark{c}} &
\colhead{Angular} &
\colhead{Physical} &
\colhead{Number of} & 
\colhead{Survey} 
\\
\colhead{} \vspace{-0.2cm} &
\colhead{} &
\colhead{} &
\colhead{} & 
\colhead{} & 
\colhead{} &
\colhead{} &
\colhead{} &
\colhead{resolution} &
\colhead{resolution} &
\colhead{HII regions} & 
\colhead{Area}
\\
\colhead{} &
\colhead{(Mpc)} &
\colhead{} &
\colhead{(M$_\odot$)} & 
\colhead{(M$_\odot$ yr$^{-1}$)} & 
\colhead{(degrees)} &
\colhead{(degrees)} &
\colhead{(arcmin)} &
\colhead{(arcsec)} &
\colhead{(pc)} &
\colhead{} &
\colhead{(kpc$^2$)}
}
\startdata
NGC\,0628 & $9.77\pm0.82$ & Sc & 10.2 & 0.26 & 20.9 & 8.7  &  5.0 & 1.0 & 47 & 1277 & 96 \\
NGC\,1087 & $14.4\pm4.8$ & SBc & 9.8 & 0.05 & 177.3 & 41.3 & 1.5 & 1.0 & 70 & 679 & 105\\
NGC\,1672 & $11.9\pm4.0$ & SBb (S) & 10.2 & 0.48 & 135.7 & 37.5 & 3.1 & 0.8 & 46 & 880 & 95\\
NGC\,2835 & $10.1\pm3.4$ & SBc & 9.6 & -0.08 & 1.6 & 47.8 & 3.2 & 0.9 & 44 & 699 & 60 \\
NGC\,3627 & $10.6\pm0.9$ & SBb (S3) & 10.6 & 0.55 & 174.1 & 55.0 &  5.1 & 1.0 & 51 & 692 & 76\\ 
NGC\,4254 & $16.8\pm5.6$ & Sc & 10.5 & 0.74 & 67.7  & 37.8 & 2.5 & 0.7 & 57 & 1824 & 191\\ 
NGC\,4535 & $15.8\pm2.3$ & SBc & 10.4 & 0.35 & 179.8  &    40.7 & 4.1 & 0.5 & 38 &  864 & 126\\ 
IC\,5332 &  $9.95\pm3.3$  & SBc & 9.6 & -0.31 & 74.7  &    24.0 & 3.0 & 0.8 & 39 &  223 & 42 \\
\enddata
\tablenotetext{a}{Position angle}
\tablenotetext{b}{inclination}
\tablenotetext{c}{the galactic radius at 25 mag arcsec$^{-2}$}
\tablecomments{The galaxy parameters are adopted from the z0mgs (z=0 multiwavelength galaxy survey; Leroy, Sandstrom et al. ApJS accepted) database (\url{https://github.com/akleroy/z0mgs}), where the original references can be found. R25 taken from HYPERLEDA. }
\end{deluxetable*}

Observations are reduced using a pipeline built on esorex and developed by the PHANGS team.\footnote{\url{https://github.com/emsellem/pymusepipe}} Between 5 and 12 MUSE pointings are observed per galaxy and each is reduced individually to produce a data cube.  As observations were taken under mostly photometric conditions, no additional absolute flux calibration corrections are applied except for on NGC 628 (as described in \citealt{Kreckel2017, Kreckel2018}). As all metallicity diagnostics are inferred from line ratios, our main results depend only on the relative flux calibration and are insensitive to the absolute flux calibration.   

The relative astrometric solution across each MUSE field-of-view is excellent ($\sim$0.04\arcsec). A single additional absolute astrometric correction for each pointing is obtained by comparing the white light image with reference images.  For the majority of our galaxies we used SDSS r-band images as our reference. 
For NGC1672 we used an R-band image acquired as part of the PHANGS H$\alpha$ imaging survey (Razza et al. in prep.) with astrometry determined from GAIA \citep{GAIA_DR2}, while for NGC2835 and IC5332 we used respectively an R-band image coming from the Danish 1.54m telescope \citep{Larsen1999} with an astrometric solution determined by \url{astrometry.net}.  For IC5332 we used an R-band image from the CTIO LVL optical survey \citep{Cook2014}. 
To determine the astrometric solution, both the MUSE image and the reference image were analyzed with the IDL procedure `find' (an adaptation of DAOPHOT) to identify point-like sources in the image.  As our MUSE fields are entirely contained within the disk of the galaxy, these often do not correspond to stars but to unresolved star-forming knots within the disk. We then cross-matched these point sources, typically finding 5-20 matches per MUSE field, and applied the average offset.  Typical shifts are 1-2\arcsec, and systematic uncertainties are estimated to be $\sim$0.2\arcsec, but are dependent on the absolute accuracy of the astrometric solution in the reference image.

In order to morphologically identify the HII regions by their H$\alpha$ line emission, we first produce maps covering the full field-of-view.  All data cubes are processed using the IDL toolkit LZIFU \citep{Ho2016} to  fit stellar spectral templates and extract emission line fluxes. The resulting emission line maps from all pointings are combined to produce full coverage emission line maps for each galaxy.   Three color images are presented for all eight galaxies in an atlas (Appendix \ref{appendix:atlas}, Figures \ref{fig:prettyfig} to \ref{fig:prettyfig_cont2}). We reach a typical 3$\sigma$ surface brightness sensitivity for H$\alpha$ of $1.5 \times 10^{-17}$ erg s$^{-1}$ cm$^{-2}$ arcsec$^{-2}$.

\subsection{HII region catalogs}
\label{sec:hiireg}
H$\alpha$ emission line maps are analyzed with HIIphot \citep{Thilker2000}, which has been designed to morphologically isolate HII regions from diffuse ionized gas, and provide a 2D mask that identifies separate regions within the map.  
After identifying `seeds' for the HII region peaks, their boundaries are set by identifying where the slope of the H$\alpha$ surface brightness distribution meets a user-defined termination criteria.  We have adjusted the settings of HIIphot to produce realistic HII region sizes and shapes across the eight galaxies in this sample, changing only the background surface brightness sensitivity level between galaxies.  We then use the resulting mask to extract integrated spectra from the  data cubes.  As a check, we repeated our analysis using spectra extracted from a circular $1\arcsec$ diameter aperture centered at the  position of each identified region. The derived line ratios and metallicities agree with those calculated from the HII region catalogs within 0.01 dex for 84\% of the sample. Thus, the exact boundaries set and choice of tuning parameters for HIIphot have only a small effect. 

These integrated region spectra are then fit with GANDALF (Gas AND Absorption Line Fitting; \citealt{Sarzi2006}), which simultaneously fits the stellar continuum (using pPXF - Penalized Pixel-Fitting; \citealt{Cappellari2004, Cappellari2017}) and the emission line fluxes (H$\beta$, [OIII] $\lambda$4959,5007, [OI]$\lambda$6300, H$\alpha$, [NII]$\lambda$6548,6583, [SII]$\lambda$6716,6731, [SIII]$\lambda$9069).  Emission lines are fit without tying together the line centers or kinematics, in order to account for the varying spectral resolution.  We fit the stellar continuum with stellar population models taken from the \cite{Tremonti2004} library of \cite{Bruzual2003} templates across a range of ages (5 Myr to 11 Gyr) and metallicities (Z = 0.004, 0.02 and 0.05).  Our results do not change significantly if we use another stellar population model (e.g.\ MIUSCAT stellar population synthesis models; \citealt{Vazdekis2012}). All line fluxes are corrected for extinction using the Balmer decrement to infer the reddening, assuming case B recombination and T$_{\rm e}$=10$^4$K, and apply the \cite{Fitzpatrick1999} extinction law assuming a value of R$_V$=3.1.  
This is relatively insensitive to the assumed electron temperature, with variations by a factor of two resulting in only $\sim$0.1 magnitude difference in the V-band extinction.  Using a different extinction law (e.g. \citealt{Cardelli1989}) does not significantly change our results.

We then employ a number of criteria to produce our HII region catalog:  
\begin{enumerate}
    \item Any region within 5\arcsec~ of the edge of the full mosaic is removed, to ensure the entire region is contained within the field of view. This excludes $\sim$7\% of the regions. 
    \item A S/N $>$5 is required for all strong emission lines used to infer the metallicity (H$\beta$, [OIII], H$\alpha$, [NII], [SII]), which excludes a further $\sim$5\% of the regions. 
    \item Two BPT \citep{Baldwin1981} emission line ratio diagnostics are employed to ensure the excitation is consistent with photoionization by massive stars (see Appendix \ref{appendix:bpt}).  In the [OIII]/H$\beta$ vs. [NII]/H$\alpha$ diagnostic, we require regions to be consistent with the  stringent empirical line \citep{Kauffmann2003}.  In the [OIII]/H$\beta$ vs. [SII]/H$\alpha$ diagnostic, we require regions to be consistent with the \cite{Kewley2001} line.  These BPT diagnostics remove another 7\% of the regions.
    \item While the above removes the majority of planetary nebulae and supernova remnants, we further remove regions (15 additional regions, $\sim$0.1\% of the total catalog) with large ($>$100 \kms) velocity dispersion that are likely compact supernova remnants.
\end{enumerate}
These cuts together remove a total of $\sim$10-20\% of regions from each galaxy, with the remainder comprising our HII region catalogs (Table \ref{tab:catalog1}). 
We list the total number of HII regions identified for each galaxy (ranging from 200-2000) in Table \ref{tab:sample}.

\begin{figure*}
\centering
\includegraphics[width=3.5in]{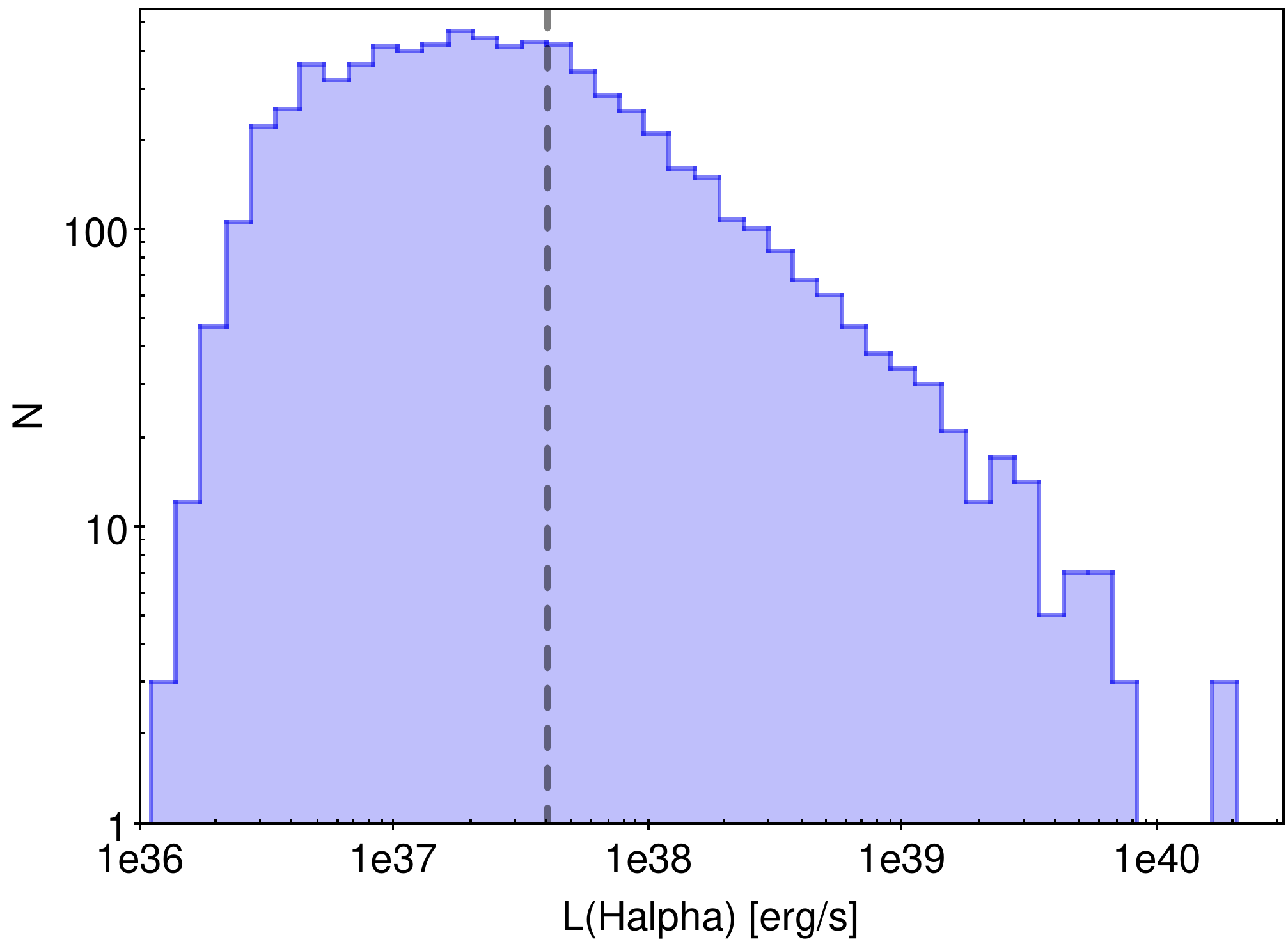}
\includegraphics[width=3.5in]{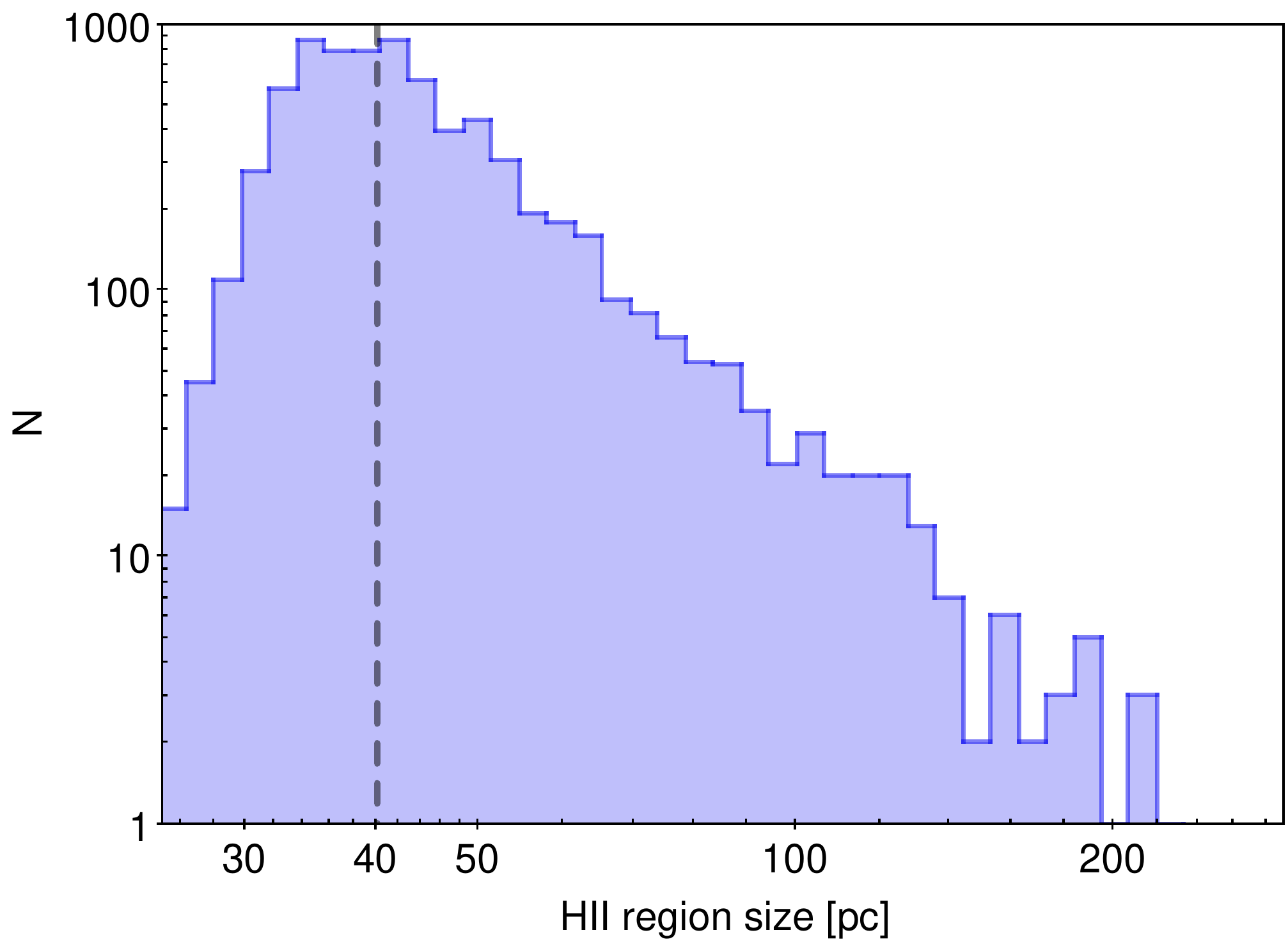}
\caption{Distribution of extinction-corrected H$\alpha$ luminosities (left) and HII region sizes (right) for the full HII region sample across all galaxies. Here, we have computed the sizes as the circular radius which would contain the same area as covered by the HII region masks. Dashed lines show the approximate completeness limit corresponding to the most distant galaxy in our sample.  Both distributions exhibit the expected power law behavior.  
\label{fig:lfn}}
\end{figure*}

By eye, HIIphot does an excellent job of identifying HII regions (Figures \ref{fig:prettyfig} to \ref{fig:prettyfig_cont2}), isolating the individual H$\alpha$ peaks and segmenting clustered complexes into discrete HII regions. 
However, it is also apparent that in rare circumstances the termination criteria is not correctly met at the HII region boundary, which results in a small number of regions that grow to unphysical sizes of up to $>$200 pc ($<$1\% of regions).  This is particularly clear for some of the inter-arm HII regions (e.g.\ on the eastern side of NGC 628). Taking the area of the irregular mask defining each HII region,  we infer a size corresponding to a circular radius that would encompass the same area.  Looking at this HII region size distribution (Figure \ref{fig:lfn}), we see the expected power-law behavior \cite[e.g.][]{Azimlu+2011} up to 120pc sizes, suggesting these unphysically large regions are not introducing a significant bias, and we retain these regions in our sample.   

The extinction-corrected H$\alpha$ luminosity function for our full sample is presented in Figure \ref{fig:lfn}, and shows the expected power law slope with a break at $\sim 10^{39}$ erg s$^{-1}$ \citep{Bradley2006}. Based on our 5$\sigma$ H$\alpha$ sensitivity for 1\arcsec\ diameter regions, we estimate typical completeness limits of L(H$\alpha$) $\simeq$ 4 $\times$ 10$^{37}$ erg s$^{-1}$, though they  vary given the range in galaxy distances.  This is roughly comparable to the ionizing flux produced by a single O5V star \citep{Schaerer1997}. All galaxies are fairly face-on or only moderately inclined (i$<$55$^{\circ}$), and the typical V-band extinctions we infer are quite moderate (1.0$\pm$0.3 mag), so we do not expect that we are missing a significant fraction of bright HII regions. 

All HII region positions, line fluxes, and associated uncertainties are tabulated in Table \ref{tab:catalog1}.  The full table is available online, with only a small sample shown here. Positions reflect the geometric center of each identified HII region, without any intensity weighting. The integrated H$\alpha$ luminosity and all emission line fluxes have been extinction corrected, as described above and are listed in the table in absolute flux units. 

\begin{deluxetable*}{ccccc}
\centering
\tablecaption{HII region catalog (online only, example data for 3 regions) \label{tab:catalog1}}
\tablehead{
\colhead{parameter} &
\colhead{unit} &
\colhead{} &
\colhead{} &
\colhead{} 
}
\startdata
ID & & ngc628\_0001  & ngc628\_0002  & ngc628\_0003  \\
ra\tablenotemark{a} & degrees &        24.175116 &        24.194861 &        24.170381 \\
dec\tablenotemark{a} & degrees &        15.784356 &        15.788300 &        15.793022 \\
R/R$_{25}$ & &  0.01 &  0.25 &  0.12 \\
E(B-V) & mag &  0.00$\pm$ 0.02 &  0.30$\pm$ 0.05 &  0.40$\pm$ 0.03 \\
L(H$\alpha$) & 1e36 erg s$^{-1}$ &     4.56$\pm$    0.03 &     5.08$\pm$    0.05 &     5.01$\pm$    0.04 \\
H$\beta$ & 1e4 erg s$^{-1}$ cm$^{-2}$ &     1.68$\pm$    0.16 &     1.55$\pm$    0.37 &     1.54$\pm$    0.22 \\
\textrm{[OIII]}$\lambda$5007 & 1e4 erg s$^{-1}$ cm$^{-2}$ &     0.69$\pm$    0.08 &     1.00$\pm$    0.25 &     0.32$\pm$    0.08 \\
H$\alpha$ & 1e4 erg s$^{-1}$ cm$^{-2}$ &     3.99$\pm$    0.21 &     4.45$\pm$    0.54 &     4.39$\pm$    0.33 \\
\textrm{[NII]}$\lambda$6583 & 1e4 erg s$^{-1}$ cm$^{-2}$ &     1.73$\pm$    0.10 &     1.40$\pm$    0.19 &     1.77$\pm$    0.15 \\
\textrm{[SII]}$\lambda$6716 & 1e4 erg s$^{-1}$ cm$^{-2}$ &     0.69$\pm$    0.05 &     1.04$\pm$    0.13 &     0.64$\pm$    0.05 \\
\textrm{[SII]}$\lambda$6731 & 1e3 erg s$^{-1}$ cm$^{-2}$ &     5.04$\pm$    0.38 &    6.90$\pm$    0.94 &     3.82$\pm$    0.37 \\
\textrm{[SIII]}$\lambda$9069 & 1e3 erg s$^{-1}$ cm$^{-2}$ &     6.93$\pm$    0.70 &     1.93$\pm$    0.53 &     1.68$\pm$    0.30 \\
12+log(O/H) & &  8.57$\pm$ 0.05 &  8.46$\pm$ 0.13 &  8.61$\pm$ 0.08 \\
\enddata
\tablenotetext{a}{J2000}
\end{deluxetable*}

\subsection {Ancillary data}

We compare our MUSE results to spectroscopic imaging of the CO~(2-1) line from the PHANGS-ALMA survey (PI E. Schinnerer). The sample selection, observations, and data reduction for PHANGS-ALMA are described in A. K. Leroy et al. (in preparation). Briefly, we observed each target with ALMA's main array in a compact configuration and with the 7-m arrray and single dish (``total power'') telescope. The combination of these observations yield CO line data cubes with resolution $\approx 1''{-}2''$ and sensitivity to emission on all spatial scales. The standard PHANGS-ALMA products include integrated intensity (``moment 0'') maps, maps of the peak intensity (``$T_{\rm peak}$'') at any velocity along each line of sight, intensity-weighted mean intensity (``moment 1'') maps, and several line width estimates. Here we use the velocity dispersion estimated from the rms scatter of emission about the intensity-weighted mean velocity (i.e., the ``moment 2'' line width). Use of an alternate parameterization of the molecular gas line width does not significantly change our results. 
The kinematic measurements at high resolution require reasonable S/N, and so have limited coverage in IC~5332 and NGC~2835, which both have relatively low stellar mass and correspondingly weak CO emission. Maps for six of our eight targets have appeared before  \citep{Sun2018,Kreckel2018,Utomo2018}, and we refer the reader to those papers for additional visualizations and details.

The surface brightness sensitivity of the CO maps becomes higher at coarser resolution. At the $\approx 7''$ resolution of ALMA's 7-m telescopes, we detect essentially all CO emission from all targets, including IC~5332 and NGC~2835. When computing the general distribution of molecular gas, we use a low resolution, high completeness version of the moment 0 maps created at this lower resolution of the 7-m data.

FUV emission was observed by the Galaxy Evolution Explorer (GALEX) satellite, with all maps taken from the GALEX archive using the longest exposure time available for each galaxy.  Most are taken from guest investigator programs with 1.7-21 ksec exposures.  NGC 1087 was observed as part of the Medium Imaging survey with a 3 ksec exposure, and NGC 1672 was observed as part of the Nearby Galaxy Survey \citep{GildePaz2007} with a 2.7 ksec exposure.  Only NGC 2835 has a significantly shorter (100 second) exposure time from the All-Sky Imaging Survey.  Images are used at the native 4.5\arcsec angular resolution.   FUV fluxes are extinction corrected based on the observed Balmer decrement  assuming a \cite{Fitzpatrick1999} extinction law, including a factor of two decrease in the reddening E(B-V) assumed for the stellar UV emission in relation to the nebular H$\alpha$ emission \citep{Kreckel2013}.

\section{HII region physical conditions}
We present a sample of eight galaxies with unprecedented statistics, identifing a total of 7138 HII regions, with measurement of 200-2000 HII region metallicities per galaxy.  We outline here our method for inferring the gas phase oxygen abundance (metallicity), and present the radial gradients observed within each galaxy. 
\label{sec:strongline}

\subsection{Strong line metallicities}
\label{sec:metals}
There is a long standing debate in the literature regarding how best to calculate the gas phase oxygen abundances using only strong emission lines, where the HII region electron density and temperature cannot be directly measured.  
Empirical calibrations are determined by comparing different strong line ratios with metallicities calculated from faint auroral line detections, that are used to estimate the electron temperature and therefore `directly' measure ionic abundances.  This $T_\textrm{e}$ method is often viewed as the `gold standard' for determining metallicity (\citealt{Skillman1989, Garnett2002, Stasinska2007, Bresolin2007, Berg2015, McLeod2019}; but see \citealt{Blanc2019}). 
Alternately, theoretical calibrations are determined by comparing different strong line ratios with photoionization model grids.  For the same datasets, systematic offsets of 0.2 dex are found between empirical and theoretical calibrations \citep{Kewley2008, Blanc2015}. While we do detect auroral line emission in $\sim$20-50 HII regions per galaxy (Ho et al. in prep.), this is an order of magnitude fewer than the number of regions for which we can identify and measure strong line fluxes. 

As our MUSE observations do not cover the [OII]$\lambda$3727 emission line, only a limited number of strong line prescriptions are available for determining the metallicity (see Appendix \ref{appendix:prescriptions}).  All calculations  in this paper adopt the empirical \cite{Pilyugin2016} S-calibration (Scal).   
It relies on the following three standard diagnostic line ratios: \\ 
$N_2$  = $I_{\rm [N\,II] \lambda 6548+ \lambda 6584} /I_{{\rm H}\beta }$,  \\
$S_2$  = $I_{\rm [S\,II] \lambda 6717+ \lambda 6731} /I_{{\rm H}\beta }$,  \\
$R_3$  = $I_{{\rm [O\,III]} \lambda 4959+ \lambda 5007} /I_{{\rm H}\beta }$.  \\
Here we have measured only the stronger line in both the [OIII] and [NII] doublets and assume a fixed ratio of 3:1.  The Scal prescription is defined separately over the upper and lower branches in log$N_{2}$, with almost all HII regions in our sample falling on the upper branch
(log$N_{2} \ge -0.6$), as
\begin{eqnarray}
\footnotesize
       \begin{array}{lll}
     {\rm 12+log(O/H)}  & = &   8.424 + 0.030 \, \log (R_{3}/S_{2}) + 0.751 \, \log N_{2}   \\  
                          & + &  (-0.349 + 0.182 \, \log (R_{3}/S_{2}) + 0.508 \log N_{2})   \\ 
                          & \times & \log S_{2}   \\ 
     \end{array}
\label{equation:ohsu}
\end{eqnarray}

and the lower branch
(log$N_{2} \le -0.6$) is calculated as
\begin{eqnarray}
\footnotesize
       \begin{array}{lll}
     {\rm 12+log(O/H)}  & = &   8.072 + 0.789 \, \log (R_{3}/S_{2}) + 0.726 \, \log N_{2}   \\  
                          & + &  (1.069 - 0.170 \, \log (R_{3}/S_{2}) + 0.022 \log N_{2})    \\ 
                          & \times & \log S_{2}   \\ 
     \end{array}
\label{equation:ohsl}
\end{eqnarray}

We prefer this metallicity calibration as it shows small intrinsic scatter  versus the auroral line metallicities for the same regions. This is quoted to be $<$0.1 dex across the full metallicity range \citep{Pilyugin2016}. A more recent investigation comparing metallicity calibrations  reports systematic $\sim$0.17 dex scatter for this prescription, as good or better than all other tested calibrations, and similar to the best results achieved from a machine learning approach \citep{Ho+2019}.   Furthermore, Scal relies on a combination of three emission line ratios, an important improvement over past calibrations, which allows it to correct for dependencies with the ionization parameter \citep{Pilyugin2016}.   
Finally, we find that this prescription systematically shows the smallest scatter (0.03-0.05 dex) around the radial metallicity gradient compared to alternate metallicity prescriptions (see Appendix \ref{appendix:prescriptions}).  
While this does not necessarily mean this method is the most accurate, we do not expect any method to imply a scatter lower than the true scatter.  Some further discussion on this point is presented in Section \ref{sec:dis:azimuth}. 
The Scal metallicity for each HII region is listed in Table \ref{tab:catalog1}.  Two-dimensional maps of the identified HII regions, each colored by the measured metallicity, are shown in Figures \ref{fig:prettyfig} to \ref{fig:prettyfig_cont2}. 

While a detailed exploration of the differences between  empirical and theoretical metallicity calibrations is beyond the scope of this paper, we find very good qualitative agreement between the metallicity trends and correlations presented here with an analysis performed using the \cite{Dopita2016} photoionization model based N2S2 prescription. The reported trends and corrleations also do not change qualitatively when using the Bayesian inference code IZI \citep{Blanc2015} to derive the joint and marginalized posterior probability density functions for metallicity and ionization parameter given a set of observed line fluxes and an input photoionization model. This further increases our confidence in the robustness of the presented results. We do note that our results change qualitatively when using the \cite{Marino2013} O3N2 and N2 calibrations.  However, because these use smaller dimensionality in the line ratios, we believe both of these alternative calibrations retain the degeneracy between metallicity and ionization parameter, enough to cloud the subtle trends presented here.  Select results using additional calibrations are shown in Appendix \ref{appendix:prescriptions}, along with a more detailed discussion of how results presented in this paper would change if the analysis was performed with an alternate calibration. 

\subsection{Radial metallicity gradients}
\label{sec:radgrad}
The rightmost panels in our atlas (Appendix \ref{appendix:atlas}, Figures \ref{fig:prettyfig} to \ref{fig:prettyfig_cont2}) map the Scal metallicity of each HII region in each galaxy. These figures show clearly both the different mean metallicity between galaxies (linked to the mass-metallicity relation; \citealt{Tremonti2004}), and also the known trend of metallicity decreasing with galactic radius. On top of this however, there appear to be higher-order trends with suggestive links to galactic features like bars and spiral arms.

\begin{figure*}
\centering
\includegraphics[width=7in]{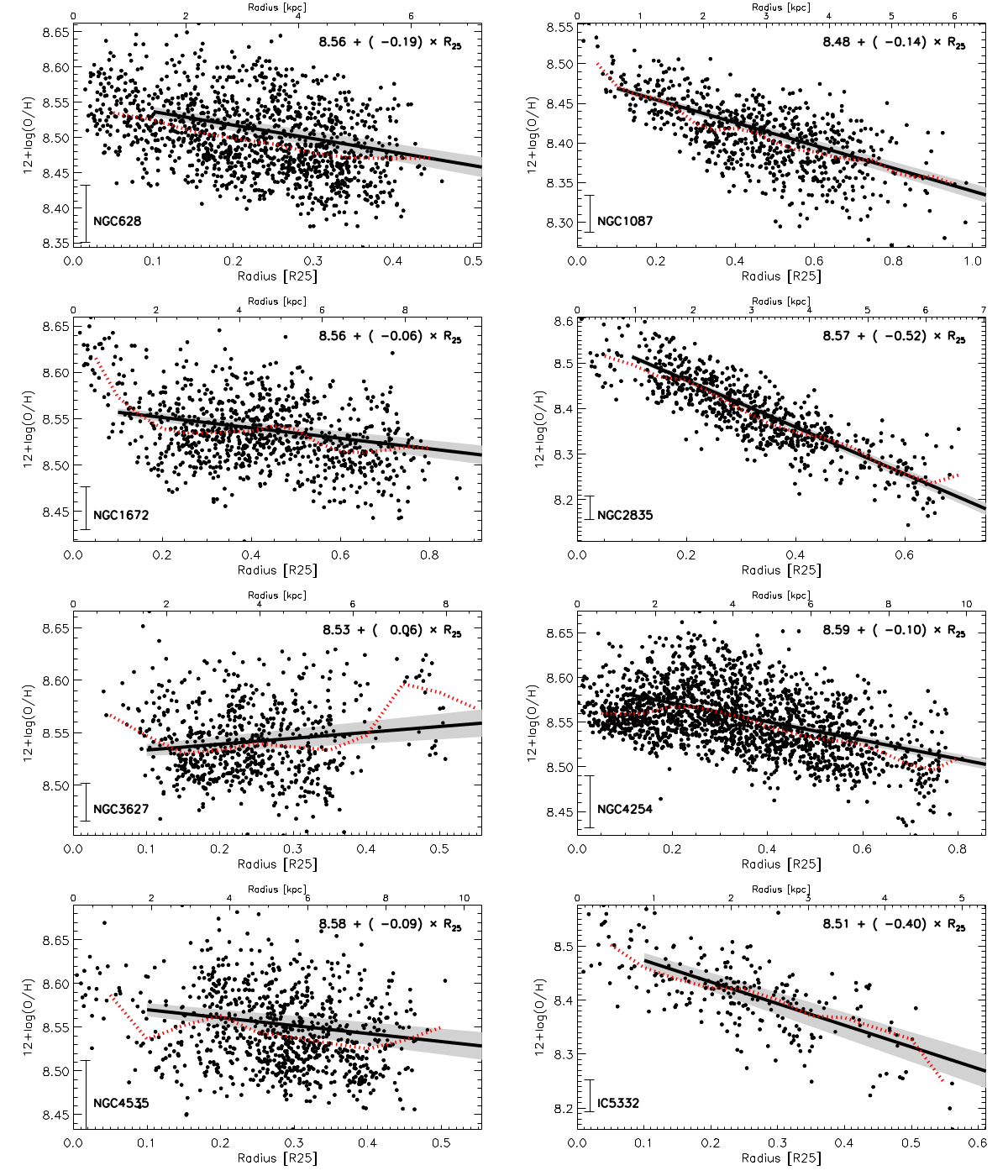}
\caption{Radial metallicity gradients for all eight galaxies in our sample.  All metallicities are calculated using the \cite{Pilyugin2016} Scal strong line prescription. Uncertainties in the metallicity from propagating line flux errors (lower left corners) are similar (0.02-0.3 dex) to the systematic uncertainties in the calibration (0.02 dex).  Overplotted are linear fits (excluding the central 0.1 R$_{25}$) to the radial gradient (solid black), with the  uncertainty in the fit shown as a grey bar, and the median calculated for 0.05 R$_{25}$ wide radial bins (dotted red).  The two radial trends show agreement overall.  NGC~1672, NGC~3627 and NGC~4535 exhibit flat or slightly increasing gradients, but note that all three galaxies contain a pronounced stellar bar that extends across much of the field of view (see Figures \ref{fig:prettyfig}-\ref{fig:prettyfig_cont2}). The remaining galaxies exhibit declining gradients.  All galaxies show small (0.03-0.05 dex) scatter about the radial gradient. 
\label{fig:gradients}}
\end{figure*}

We perform a first order linear fit to the radial gradient (Figure \ref{fig:gradients}). We include in the fit the determined errors for all line flux measurements, including the uncertainty in the flux measurements due to the uncertainty in our measurement of E(B-V). We exclude the inner 0.1R$_{25}$ when performing our fit, as it could be biased by  nuclear ionization sources.  We also calculate the running median in radial bins, and find very good agreement with our simple linear fit.  Coefficients for radial gradients in each galaxy are shown in Figure \ref{fig:gradients} and tabulated in Appendix \ref{appendix:prescriptions} (Table \ref{tab:gradients}).  All galaxies show small (0.03-0.05 dex) systematic scatter after accounting for the radial trends.

\subsection {Ionization Parameter}

The structure of an HII region is set by the ionizing radiation provided by the central star(s) acting on the surrounding ISM. For the simplistic case of a spherically symmetric HII region that has achieved equilibrium, this balance can be parameterized in terms of the dimensionless ionization parameter, U = Q(H$^{0}$)/(4$\pi$R$^2$$n$$c$), where Q(H$^0$) is the number of hydrogen-ionizing photons (E $>$ 13.6 eV) emitted per second, R is the empty (wind-blown) radius, $n$ is the hydrogen density, and $c$ is the speed of light. Direct interpretation of this parameter is challenging, as there are degeneracies between the ionizing flux produced by a single very young star as opposed to a cluster of slightly older stars. Moreover,  detailed studies of resolved HII regions reveal asymmetric and broken shells surrounding the ionizing sources \citep{Pellegrini2011, Pellegrini2012}. Nonetheless, this parameter is still useful as it encodes information about the local physical conditions, with higher values typically indicating a stronger ionizing source or a lower ionized gas density.

Photoionization modeling has been used to constrain the ionization parameter through different diagnostic line ratios \citep{Kewley2002, Dors2011}.  [OIII]$\lambda$5007/([OII]$\lambda$3726+[OII]$\lambda$3729) has frequently been used to constrain the ionization parameter, but has a strong dependence on metallicity.  ([SIII]$\lambda$9069+[SIII]$\lambda$9532)/([SII]$\lambda$6716+[SII]$\lambda$6731) also shows a positive correlation with ionization parameter, and is a much more direct diagnostic of the ionization parameter, with only a weak dependence on metallicity and no sensitivity to the ISM pressure between between 4 cm$^{-3}$ K $<$ log(P/k) $<$ 7 cm$^{-3}$ K.  This makes [SIII]/[SII] the best ionization parameter diagnostic available in the optical, although the long wavelength [SIII]$\lambda$9069,9532 lines have not often been observed. With MUSE we observe the [SIII]$\lambda$9069 line flux, and for our calculations we assume the fixed theoretical line ratio of [SIII]$\lambda$9532/[SIII]$\lambda$9069 = 2.5  \citep{Vilchez1996}, which is well determined from the transition probabilities.
Figure \ref{fig:gradients_q} presents the radial gradients of [SIII]/[SII], tracing the radial ionization parameter gradient.  All gradients are very close to being flat within the inner part of the galaxy disks, as also found in \cite{Poetrodjojo2018}.

We choose in this paper to present results related to the ionization parameter directly in terms of the observed line ratios, and remind the reader that a positive  linear correlation is expected between log([SIII]/[SII]) and U.   Various prescriptions (e.g. \citealt{Kewley2002, Dors2011}) are available in the literature to convert a measured [SIII]/[SII] line ratio to a value for U (or q = U/c), however they differ by an order of magnitude.

\subsection{Diffuse ionized gas}
The diffuse ionized gas (DIG) is seen in disk galaxies as an extended, relatively smooth background in emission line maps \citep{Zurita2000}.  The DIG contributes $\sim$40-60\% of the total H$\alpha$ flux observed within any disk galaxy \citep{Ferguson1996, Zurita2000, Thilker2002, Oey2007, Chevance2019}, and even at the location of individual HII regions has been shown to contribute at significant levels (20--80\%, \citealt{Kreckel2016}) that vary by local environment (e.g. arm/interarm).  DIG is detected predominantly in the H$\alpha$, H$\beta$, [NII] and [SII] line maps, as these lines are strongly emitted in the low density, high temperature diffuse gas \citep{Haffner2009}.  One major advantage of the high physical resolution of the maps employed in this work is that we can spatially distinguish HII regions from the surrounding diffuse emission, which when blended over larger apertures has been shown to artificially flatten metallicity gradients \citep{Poetrodjojo2019}.  
The emission line fluxes and H$\alpha$ luminosities reported here do not include any correction for the diffuse ionized gas background contributing to the integrated line flux, as seen in projection against any given HII region. A major goal of the PHANGS collaboration is to develop and compare spectroscopic, morphological, and hybrid estimators of the DIG. At this time, we lack such rigorous methods to identify the DIG, and we view imposing such a correction as premature.  As such, we have chosen to exclude such a correction from our reported line fluxes and subsequent analysis as it would strongly depend on our choice of HII region boundary.  

We estimate the effect this un-subtracted diffuse background could have on our inferred metallicities by using our emission line maps to measure the median line flux per pixel over an annular aperture surrounding each HII region in our masks. The aperture size is scaled corresponding to the observed HII region size, and is typically 0.6-2\arcsec\ wide. Here we have masked any neighboring HII regions to ensure minimal contamination of our measurement. Our DIG flux fractions are consistent with those obtained through other methods \citep{Hygate2019,Chevance2019}. As in \cite{Kreckel2016}, we observe order of magnitude variations in DIG surface brightness across each disk at the locations of HII regions, with the most H$\alpha$ luminous HII regions showing lower fractional DIG contribution ($\sim$10-20\%) compared to fainter HII regions, with contributions as high as $\sim$80\%.  For this calculation, we identify a subset of regions where the H$\alpha$ DIG background is 50\% or less.  For all emission lines we then subtract the estimated DIG level from our HII region line fluxes. We then use these DIG-subtracted line fluxes to calculate the metallicity as described in Section \ref{sec:metals}.  We find that subtracting the DIG emission would result in a slight ($\sim$0.02 dex) systematic increase in our inferred metallicities, and an increased systematic scatter (0.05-0.1 dex) across the sample.  This increased scatter washes out some of the weaker azimuthal variations we see, but does not qualitatively change the strongest trends we observe in NGC 1672  and NGC 1087. 

This diffuse background more strongly impacts our inferred ionization parameters via our measurement of the [SIII]/[SII] line ratios, as [SIII] is predominantly confined to the HII regions while [SII] is emitted strongly in the diffuse gas. Correcting for the DIG contribution to the [SII] line would result in a $\sim$70\% increase in the [SIII]/[SII] line ratios but does not qualitatively change the trends described in the following sections.

\section{Results}
\label{sec:results}

\subsection{Trends with local physical conditions}

\begin{figure*}
\centering
\includegraphics[width=3.5in]{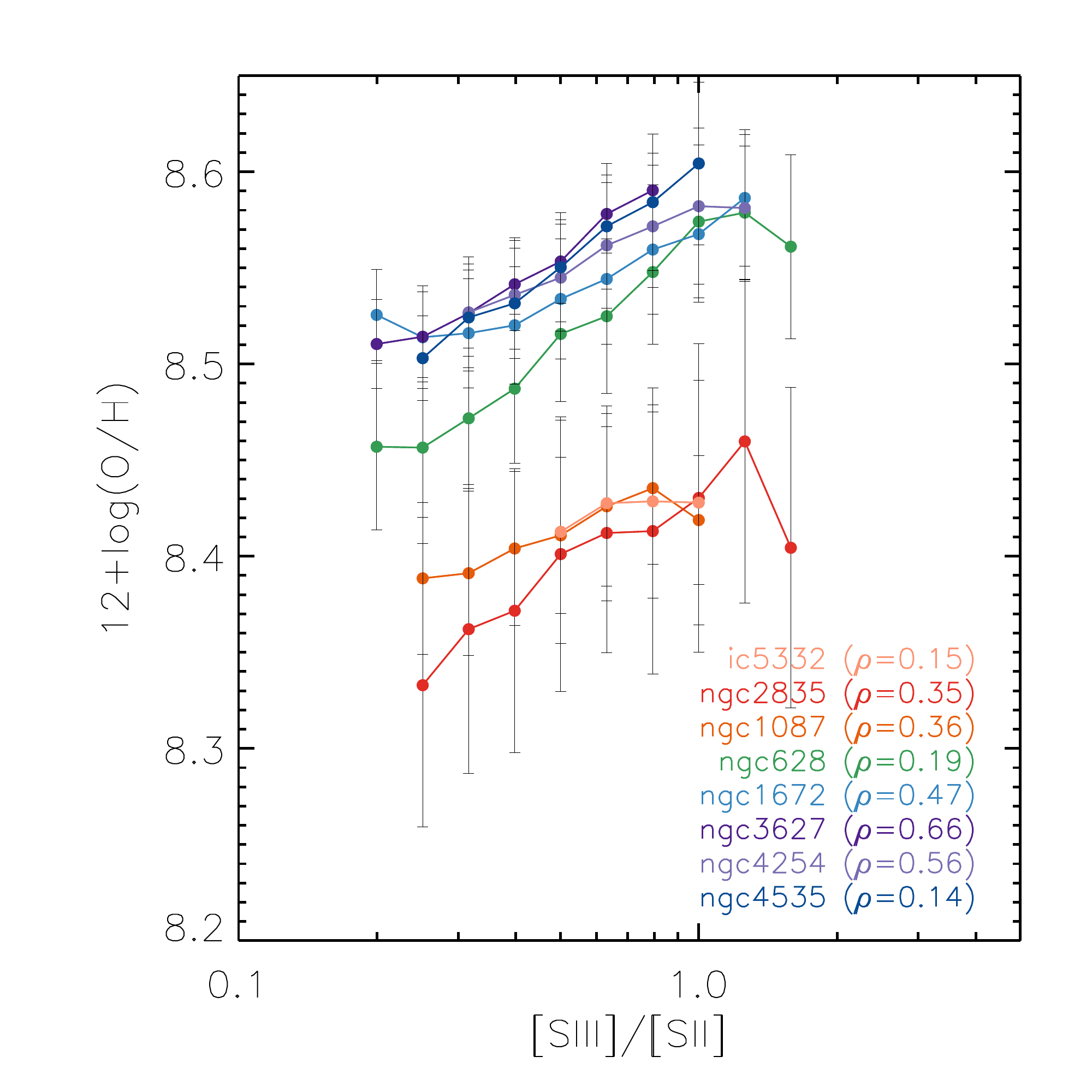}
\caption{
The metallicity of each HII region as a function of the  ionization parameter (as traced by [SIII]/[SII]).  Galaxies are color-coded by their total stellar mass from low (red) to high (blue) masses.  The Spearman's rank correlation coefficient ($\rho$) is listed for each, and all have high statistical significance (p$<$0.001).  All galaxies exhibit a positive correlation, with the offset between them set by the stellar mass - metallicity relation.
\label{fig:q_z}}
\end{figure*}

Within each of our galaxies we observe a positive correlation between the metallicity and ionization parameter of the HII regions (Figure \ref{fig:q_z}), with an offset between different galaxies that is set by the difference in stellar mass of the systems. As the ionization parameter shows no radial gradient, this is not simply a result of correlated observables. Similarly, while many emission line diagnostics can be influenced by both the metallicity and the ionization parameter, we have selected a metallicity prescription that has been defined by carefully fitting a three dimensional parameter space of diagnostic line ratios in order to minimize the degeneracies.

This positive correlation conflicts with theoretical predictions, which state that as stellar atmospheres of massive O stars become cooler with increasing metallicity there will be enhanced line and wind blanketing, resulting in a negative correlation \citep{Massey2005}.  Most previous studies have found either negative correlations \citep{Bresolin1999, Nagao2006} or no correlation \citep{Kennicutt1996, Garnett1997, Dors2011, Poetrodjojo2018}. However, they combined observations from different galaxies.  Our analysis shows a clear trend within individual galaxies, that would be obscured when taken together if one studied only a few regions from each of several galaxies.  
A positive correlation has previously been observed in a sample of luminous infrared galaxies \citep{Dopita2014}.  These authors discuss this result mainly as a positive correlation between star formation rate surface density and ionization parameter, which they attribute to a difference in geometry.
In this case, dense star-forming regions could contain inclusions of molecular clouds undergoing violent radiation pressure dominated photo-ablation close to the exciting stars, resulting in a positive correlation between star formation surface density and ionization parameter. See Section \ref{sec:dis:localprops} for further discussion on this.

To account for the metallicity offsets between galaxies introduced by the  mass-metallicity relation, we subtract off the fitted radial metallicity gradients, and consider the residuals from the linear trend (\doh, measured in logarithmic units).  
Apparent from our visualization of the metallicities across the galaxy disks (Figures \ref{fig:prettyfig} to \ref{fig:prettyfig_cont2}) is the tendency for the more luminous (and more spatially extended) HII regions to be more enriched.  The left panel of Figure \ref{fig:offsets} traces this trend directly, demonstrating that \doh~ increases systematically as a function of HII region luminosity for all galaxies in our sample, with a Spearman's rank correlation coefficient ($\rho$) of 0.34 and high statistical significance (p$<$0.001).  In total, we see a 0.03 dex increase in metallicity for the most luminous (L$_{\rm H\alpha}$ $>$ 10$^{38.5}$ erg/s) regions.  

Moreover, we observe a much tighter correlation between \doh~ and ionization parameter ([SIII]/[SII]).  
We observe a stronger correlation ($\rho$=0.63) for the sample as a whole, as well as improved correlation compared to the trends for individual galaxies between 12+log(O/H) and ionization parameter (Figure \ref{fig:q_z}).  This suggests the local physical conditions are very tightly linked to the localized enrichment of the ISM.

\subsection{Physical conditions of regions with enhanced and reduced abundances}
\label{sec:trends}
We split the sample of HII regions into those with (a) typical metallicities (-1$\sigma$ $<$ \doh $<$ 1$\sigma$), (b) enhanced metallicity ($1\sigma$ $<$ \doh), and (c) reduced metallicity (\doh $<$ -1$\sigma$). In Figure \ref{fig:histos} we show how the distributions of six physical parameters differ between the three samples.  For all parameters explored, we find that the distribution of HII regions with enriched/reduced metallicity is significantly different from the average sample.  They show a Kolmogorov-Smirnov (KS) test probability value (p-value) less than 1e-6, allowing us to reject the null hypothesis that the two samples are drawn from the same distribution.

As shown in Figure \ref{fig:offsets}, the clearest correlation we see is with ionization parameter ([SIII]/[SII]).  
As this traces the relative density of ionizing photons to the surrounding ISM gas density, we expect to find additional dependencies on the stellar cluster age and on the local gas properties.  In general, we expect HII regions with younger, more massive star clusters and a denser surrounding medium to be characterized by higher ionization parameter values.

The H$\alpha$-to-FUV flux ratio has been shown to correlate with star cluster age \citep{Sanchez-Gil2011}, with younger clusters having larger flux ratios. We use GALEX imaging to measure the FUV flux at the position of each HII region.  In measuring the H$\alpha$ flux we have convolved the H$\alpha$ images to matched 4.5\arcsec resolution.  We find that the HII regions with enhanced metallicities tend towards higher extinction corrected H$\alpha$-to-FUV flux ratios (younger star cluster ages). A similar trend is seen if we instead use the equivalent width of H$\alpha$ as a tracer of star cluster age. 

To probe the local ISM gas conditions, we use PHANGS-ALMA maps of the CO emission (Leroy et al. in prep.) at 1\arcsec~ spatial resolution to measure the local molecular gas surface density.  While the angular resolution is well matched to our MUSE observations, this molecular material is necessarily not co-located with the ionized gas of the HII region, and on our working $\sim$50 pc scales is likely associated with molecular clouds and associations neighboring the HII regions (see also \citealt{Kreckel2018}), potentially representing the remnant of the cloud from which the HII region formed \citep[see][]{Chevance2019,Kruijssen2019}.  We find that the HII regions with enhanced abundances are at locations where we measure higher CO integrated intensity, indicating a higher molecular gas surface density compared to the HII regions with reduced abundances.

A change in the local ISM conditions is also seen when looking at the dust column along the line of sight (traced by the color excess, E(B-V)), which is generally thought to be well mixed with the gas.  Here we find that the HII regions with enhanced metallicities see higher dust columns along the line of sight, consistent with the previously described association with higher CO integrated intensity, and implying higher gas surface densities. We note that this increased extinction is not enough to significantly bias our ability to detect enriched but low luminosity HII regions.  Increased electron densities are also directly seen corresponding to the HII regions with enhanced metallicity by comparing the density sensitive line ratio [SII]$\lambda$6716/[SII]$\lambda$6731, though in most regions the ratio suggests electron densities at or below the low-density limit ($<$100 cm$^{-3}$).

The velocity dispersion in the H$\alpha$ line indicates the dynamical state of the ionized gas. In contrast to the HII regions with enhanced abundances, the  HII regions with decreased abundances exhibit increased velocity dispersions (deconvolved from the instrumental resolution).  This could indicate a difference in the large scale mixing acting on the ISM, which would be needed to introduce less enriched material.  We also see a slight (2 \kms) average increase in line width in the molecular gas phase for the most enriched regions (not shown), suggestive of the ongoing impact of stellar feedback or association with more massive molecular clouds, but with less statistical significance (p-value of 0.002). This further suggests a (weak) local correlation between the H$\alpha$ luminosity and the molecular gas line width. 

To explore the spatial distribution of HII regions with abnormal abundances, we show in an atlas (Appendix \ref{appendix;azimuthal}, Figures \ref{fig:unwrap628} to \ref{fig:unwrap5332}) the azimuthal and spatial distribution of the metal offsets and the positions of these abnormal populations with respect to the $\sim 500$~pc scale bulk CO emission (as measured by ALMA) of each galaxy.  Here, our deliberate choice of low resolution observations is designed to achieve high completeness in the integrated intensity. 
We use this CO map to trace the gas response to the underlying gravitational potential and other large scale hydrodynamical effects.  

While the azimuthal trends of HII regions with reduced and enhanced abundances are discussed in more detail in the following section, we highlight here the difference between the spatial distribution of the two populations across the galaxy disks.  Both populations are found across the full disks, but the HII regions with reduced abundances are slightly more clustered in their distribution.  Future work will explore in more detail the statistical properties of the HII region metallicity distribution. Here we simply quantify this by calculating the average distance of the three nearest neighbors (n$_3$) of all HII regions with enhanced and reduced abundances.  Computing this statistic using a larger number of neighbors shows similar trends but (naturally) larger clustering sizes.  These trends are not dependent on the  threshold we choose to separate the regions with enhanced and reduced abundances. As the enriched regions are only slightly larger on average in size than the total population (49~pc in radius compared to 45~pc), this cannot account for the systematic differences in clustering scale that we observe.   

\begin{figure*}
\centering
\includegraphics[height=3.5in]{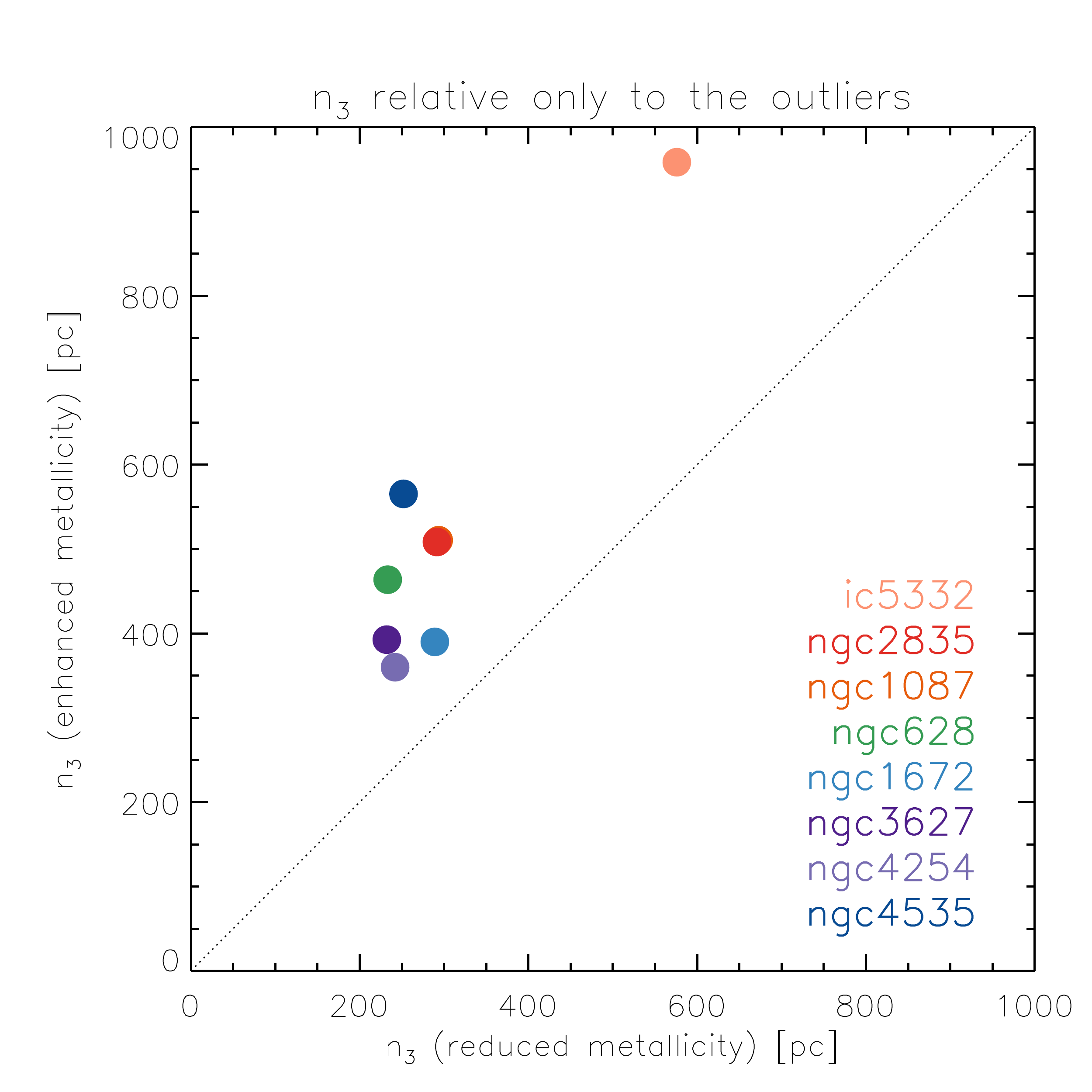}
\includegraphics[height=3.5in]{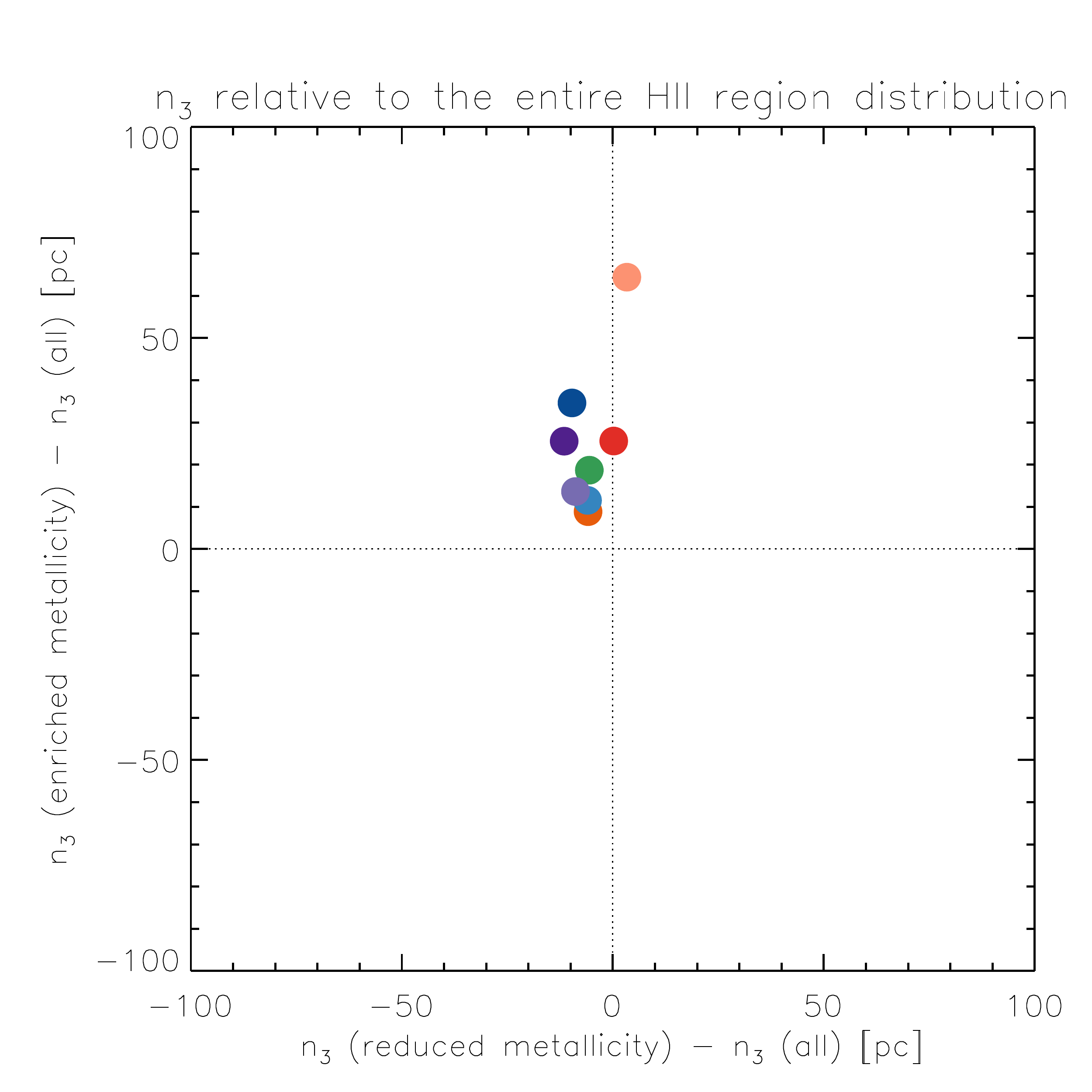}
\caption{Clustering of the HII regions with enhanced and reduced metallicities in relation to themselves (left) and in relation to the full HII region population (right), as parameterized by the average distance to the three nearest neighbors (n$_3$). The mean value for each galaxy is shown. 
We also normalize the right hand figure to the clustering of the entire HII region population across the whole galaxy.  Galaxies are color coded by their total stellar mass from low (red) to high (blue) masses. HII regions with enhanced abundances relative to the radial gradient are systematically less clustered.  IC 5332, our least massive and lowest star formation rate galaxy, shows less clustered star formation in general and a fairly flocculent morphology compared to the rest of the sample (Figure \ref{fig:prettyfig_cont2}), but still shows an increase in clustering for the HII regions with enhanced abundances. 
\label{fig:n3}}
\end{figure*}

Figure \ref{fig:n3} (left) shows a large scatter across this measure, but most galaxies show clustering on $\sim$300~pc scales of the reduced metallicity regions while enhanced metallicity regions are more widely distributed (n$_3$ $\sim$ 400-600~pc; see also Figure \ref{fig:zoomaz}). This suggests a connection between clustering and mixing, where isolated young regions with a history of star formation have the chance to become enriched \citep{Ho2017, Ho2018}. We further compare the clustering of the outliers in relation to the full HII region population (Figure \ref{fig:n3}, right), relative to the average, and again find that the HII regions with enhanced metallicities are less clustered.  

\subsection{Azimuthal Metallicity Variations}
\label{sec:aztrends}
We find that the most enriched HII regions are distributed across the disk,  and simple statistical approaches where we try comparing changes in \doh~ with large scale environment (arm/interarm/bar), overdensities in stellar mass surface density, or dynamical tracers of the spiral pattern are inconclusive. For this paper, we simply present a few visualizations of how changes in \doh~ are distributed spatially across the disk as an atlas in Appendix \ref{appendix;azimuthal} (Figures \ref{fig:unwrap628} to \ref{fig:unwrap5332}). These images demonstrate the wealth of information we now have, but also the complicated nature of disentangling the preliminary azimuthal trends we see from their complicated environmental dynamics that vary widely between galaxies due to differences in their disk structure.  

To match the metallicity maps presented in Appendix \ref{appendix:atlas}, we similarly take our HII region mask and paint on the measured \doh~ corresponding to each HII region.  NGC 628, NGC 1087, NGC 1672 and NGC 2835 all show regular azimuthal variations at select rings.  Such clear trends are not necessarily seen at all radii. NGC 3627 shows large scale variations with much less regular patterns, but the remaining three galaxies show no obvious azimuthal trends. Subsections of the four galaxies showing the clearest signatures of azimuthal variations are highlighted in Figure \ref{fig:zoomaz}.  The morphology would suggest that this implies some correlation with the spiral structure, but when we compare the azimuthal metallicity variations with the total extinction-corrected H$\alpha$ luminosity around the azimuthal ring (tracing the spiral structure), the connection is not so clear.  

Enriched regions on the spiral arm often form a relatively narrow ridge, implying that the total count of such regions is not high (in contrast to high numbers in a thicker ridge). This may partially explain why its not so easy to pick out trends when considering the population as a whole.  This is particularly apparent in NGC 1087 and NGC 1672, but also to a lesser extent in NGC 628, NGC 2835, NGC 3627 and NGC 4535. 
The narrowness of this feature is in good agreement with the highly filamentary alignment of star-forming giant molecular clouds seen along spiral arms at high ($\sim$50 pc) spatial resolution in nearby galaxies \citep{Schinnerer2017, Kreckel2018, Kruijssen2019}, and consistent with the association of the enriched regions with local conditions exhibiting high dust column and gas density (see Section \ref{sec:trends}). 
The regions with reduced abundances show a wider, more clustered distribution (see also Figure \ref{fig:n3}).  They are often seen on the trailing side of the spiral arm (particularly on the eastern arm of NGC 1672).  This is in very good agreement with the azimuthal variations identified by \citet{Ho2017, Ho2018}. 

In conclusion, we qualitatively identify half of our sample (NGC 628, NGC 1087, NGC 1672 and NGC 2835) as having azimuthal variations that are tenuously associated with the spiral pattern.  The link to gas density can also be interpreted as indirect evidence for the influence of spiral arms, which tend to yield locally enhanced gas densities.  
Establishing more conclusively whether a correlation with the spiral pattern exists in a given galaxy will require an informed view of the dynamical structure specific to that disk.

\section{Discussion}
\label{sec:discussion}

In this first look at a sample of galaxies with thousands of spatially isolated ($<$100~pc scale) HII regions, we can begin to explore systematic trends and patterns in the abundance distributions.

\subsection{Localized enrichment within galaxy disks}
\label{sec:dis:localprops}
\begin{figure*}
\centering
\includegraphics[width=7in]{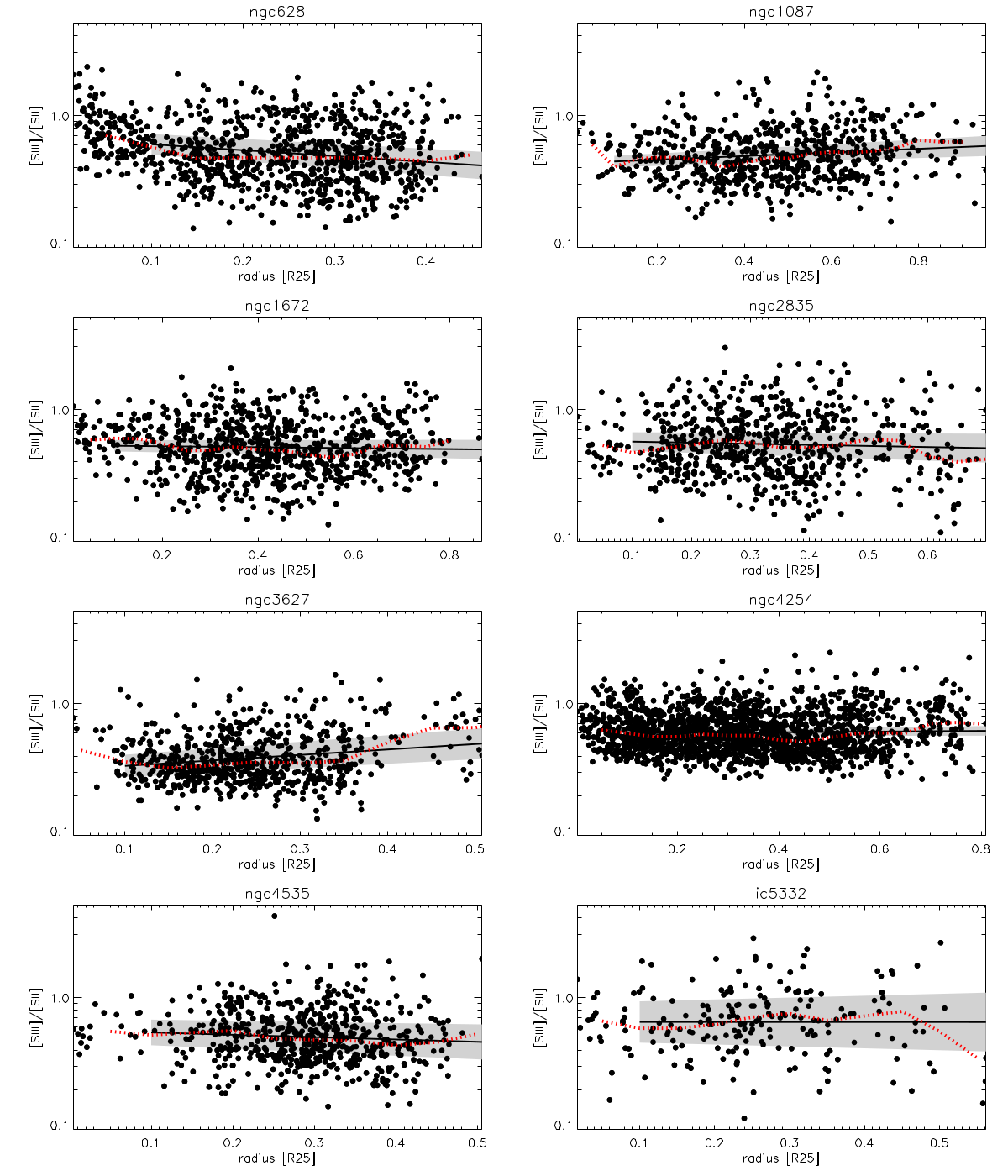}
\caption{Radial trends in ionization parameter (as traced by [SIII]/[SII])  for all eight galaxies in our sample.  Typical uncertainties in [SIII]/[SII] are 0.05 ($\sim$10\%).  
Overplotted are linear fits (excluding the central 0.1 R$_{25}$) to the radial gradient, with the 1$\sigma$ uncertainties shown as a grey bar, and a median calculated for radial bins (dotted red).  Most galaxies are consistent with a flat gradient.  
\label{fig:gradients_q}}
\end{figure*}

\begin{figure*}
\centering
\includegraphics[width=7in]{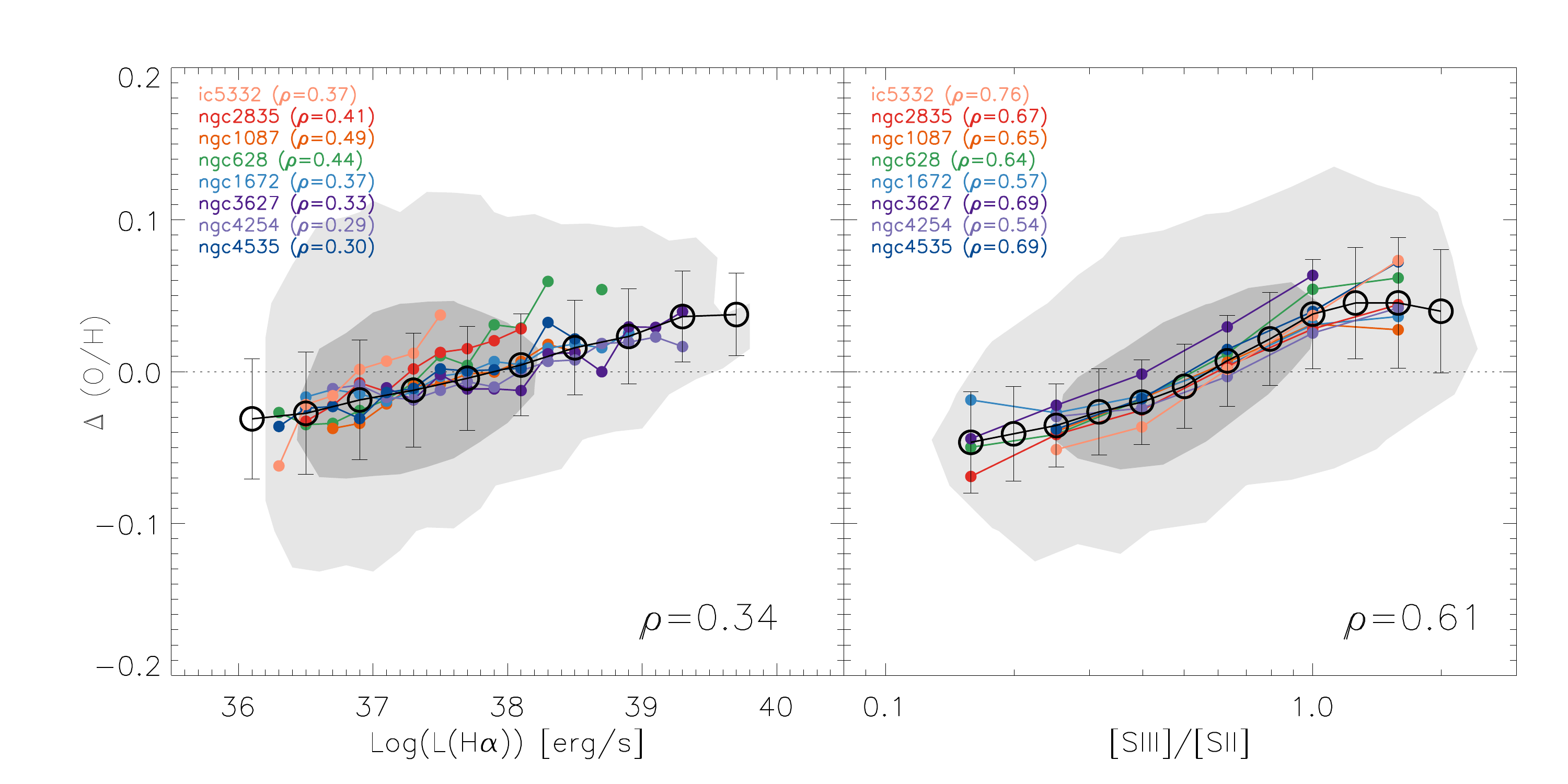}
\caption{Trends in the metallicity offset from a radial gradient (\doh) as a function of extinction-corrected H$\alpha$ luminosity (left) and ionization parameter (as traced by [SIII]/[SII], right).  Galaxies are color coded by their total stellar mass from low (red) to high (blue) masses. The median and 1$\sigma$ scatter is overplotted in black.  Background contours show the full distribution (98\% and 68\%).  While we observe correlations with L(H$\alpha$), we observe the strongest systematic correlation with the [SIII]/[SII] ratio, which is a tracer of the ionization parameter. The Spearman's rank correlation coefficient ($\rho$; all with high statistical significance, p$<$0.001) is calculated for each galaxy individually (top left) as well as for the sample as a whole (lower right).  
\label{fig:offsets}}
\end{figure*}

\begin{figure*}
\centering
\includegraphics[height=2in]{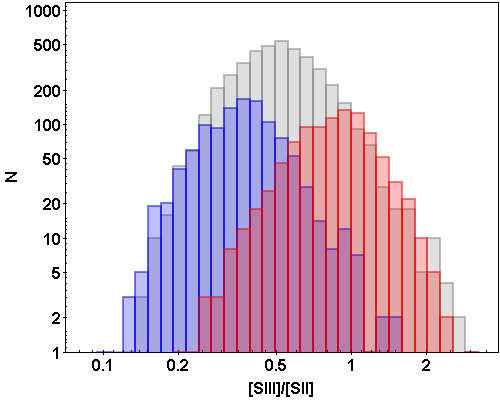}
\includegraphics[height=2in]{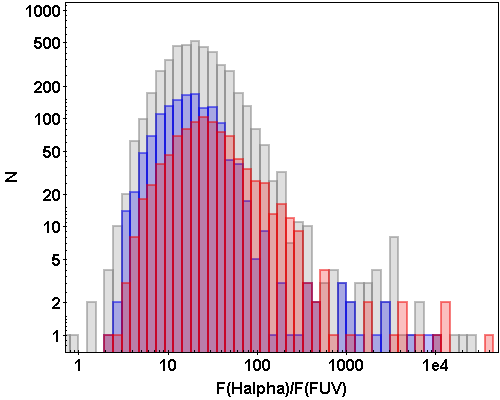}
\includegraphics[height=2in]{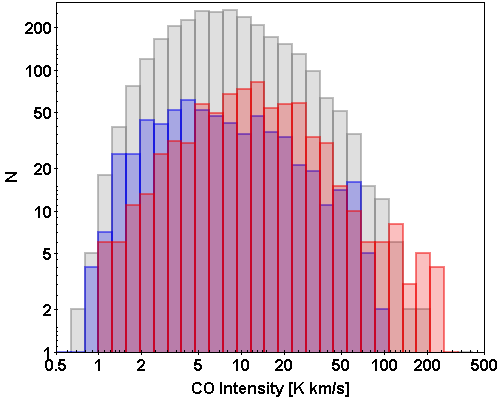}
\includegraphics[height=2in]{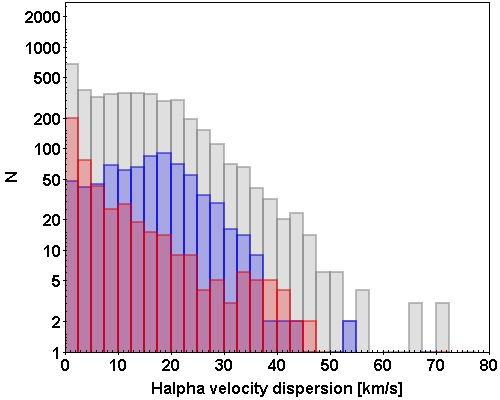}
\includegraphics[height=2in]{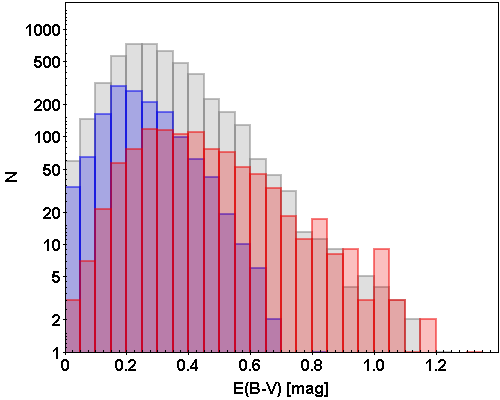}
\includegraphics[height=2in]{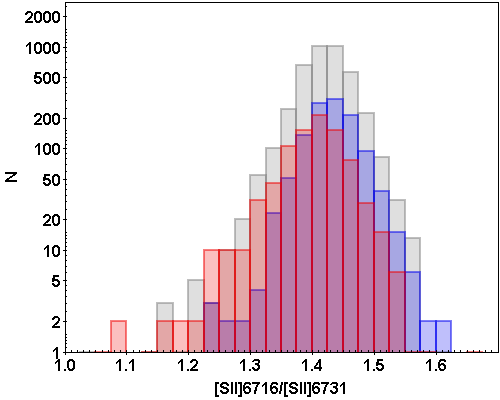}
\caption{Distribution of six physical parameters of HII regions with typical abundances (grey) as compared to HII regions where the abundances are enhanced (\doh $>$ 1$\sigma$, red), or reduced (\doh $<$ -1$\sigma$, blue).  All distributions are significantly different (KS test p-value$<$1e-6). 
Compared to the depleted regions, enriched regions have markedly high [SIII]/[SII] line ratios (indicating higher ionization parameter), as is also shown in Figure \ref{fig:offsets}.  They also have systematic offsets towards higher extinction corrected H$\alpha$-to-FUV flux ratio (indicating younger star cluster ages), higher CO intensity (indicating higher molecular gas densities), lower H$\alpha$ velocity dispersions (indicating less mixing), higher E(B-V) (indicating more dust along the line of sight), and lower [SII]$\lambda$6716/[SII]$\lambda$6731 line ratios (indicating higher electron densities).  
\label{fig:histos}}
\end{figure*}

We observe clear correlations between increased abundances relative to the radial trend (\doh) and the local physical conditions of the ISM (Figures \ref{fig:offsets} and \ref{fig:histos}). The most fundamental positive correlation arises from changes in the ionization parameter (traced here by [SIII]/[SII]), which tracks changes in the ionizing source, the ISM density and HII region geometry.  Given how flat the radial ionization parameter gradient is (Figure \ref{fig:gradients_q}), the high degree of correlation (Spearman's rank correlation coefficient ($\rho$) of 0.60) between \doh~ and [SIII]/[SII] is quite striking.  

Without removing the radial metallicity gradient, a trend between metallicity and ionization parameter is still apparent (Figure \ref{fig:q_z}).  Here, each galaxy appears offset in metallicity due to the well established relation between stellar mass and metallicity \citep{Tremonti2004}, and the correlation for any individual galaxy is weaker compared to the correlation within that galaxy when instead considering \doh. Given that previous work generally has compared HII region metallicities and ionization parameters by combining results from different galaxies with differing global metallicities \citep{Bresolin1999, Dors2011}, it is not surprising that the trends we now identify within individual galaxies would be washed out. 

A similar positive correlation between metallicity and ionization parameter was reported in \cite{Dopita2014} for their analysis of resolved regions within ten Luminous Infrared Galaxies (LIRGs). They observed similar trends between star formation rate and ionization parameter as well as between metallicity and ionization parameter, and argue that the trend arises as more active star-forming regions have a different distribution of molecular gas which favour higher ionization parameters.  However, we are observing these trends within normal star-forming disk galaxies and, as we see in Figure \ref{fig:offsets}, the observed trend is much weaker with the HII region H$\alpha$ luminosity, and the high metallicity outliers are not more strongly clustered (Figure \ref{fig:n3}).  Also, while we do see that more enriched regions preferentially fall along the ridge of spiral arms in some regions of some galaxies, they are also found distributed throughout the disk, with no strong systematic trends with galaxy environment (see also Section \ref{sec:dis:azimuth}).  
This suggests another explanation is required. 

As we show in Figure \ref{fig:histos}, other local conditions correlate with the local changes in metallicity.  Compared to the depleted regions, enriched regions have higher H$\alpha$-to-FUV flux ratios (indicating younger star cluster ages), higher E(B-V) (indicating more dust along the line of sight), and lower [SII]6716/[SII]6731 line ratios (indicating larger electron densities).  They also show higher CO intensity (indicating higher molecular gas surface densities) and slightly larger CO line widths (indicating kinematic disturbance in the local ISM). We note that the ISM conditions inferred from the CO are on matched angular scales (1\arcsec $\sim$ 50pc) but this material by nature does not co-exist with the ionized HII region, so making a direct link between the physics occurring in these two different ISM phases is not straightforward. The trends, however, are very interesting, as they might represent two different evolutionary phases of the same underlying star-forming region even if they are somewhat spatially offset \citep[e.g.][]{kruijssen18}.

As the HII region metallicity generally reflects the ISM abundances locally before it was illuminated by the newest generation of stars, these trends are suggestive of a star formation history that has locally enriched the ISM.  Recent work has suggested that star-forming complexes can host extended continuous star formation histories spanning $\sim$10 Myr  \citep{Ramachandran2018, Rahner2018}.  We see small systematic scatter in \doh, and HII regions with enhanced abundances typically show offsets of $\sim$0.05 dex. We perform a simple calculation to explore how much enrichment could be expected from a single generation of stars.  

For a cloud of gas mass $M_{\rm g}=5.8\times$10$^5$~M$_\sun$ (the typical GMC mass in NGC 628; \citealt{Kreckel2018}) and an integrated star formation efficiency per star formation event of 
$\epsilon$=0.05 (measured within the PHANGS sample; \citealt{Chevance2019}), this would produce $M_{\rm *,produced}=M_{\rm g}$ $\times$ $\epsilon$ = 2.9$\times$10$^4$~M$_\sun$ of new stars. Then, given an oxygen yield of y$_o$ = 0.00313 (based on observations; \citealt{Kudritzki2015}), when the massive stars die they return  $M_{\rm O, produced}=M_{\rm *,produced}$ $\times$ y$_o$ = 91~M$_\sun$ of oxygen to the surrounding ISM.  Assuming perfect mixing within the cloud, this gives an upper limit on how much extra oxygen (by mass) can be added in one generation.   Converting this to the change in oxygen abundance, this corresponds to an increase in log(O/H) of $\sim$0.02 dex with each generation of star formation. 
However, the oxygen yield is quite uncertain, and taking a theoretically motivated, higher oxygen yield of y$_o$=0.04 \citep{Vincenzo2016} would result in an increase in log(O/H) of $\sim$0.18 dex.   This demonstrates that the level of enrichment we report in this sample of eight galaxies is roughly consistent with what could be produced by a single generation of star formation.

We believe this is unlikely to be due to self-enrichment from a single burst of star formation, but could be consistent with an extended 5-10 Myr continuous star formation history. Typical supernova timescales (4 Myr; e.g. \citealt{Leitherer2014}) and subsequent mixing timescales over $<$100 pc scales (2 Myr; \citealt{deAvillez2002}) are roughly equivalent to the expected HII region lifetime (6-8 Myr; e.g. \citealt{Kennicutt1998, Chevance2019}). Better constraints on the HII region ages are needed to distinguish these scenarios.   Simulations indicate that larger scale ($\sim$kpc) mixing is relatively inefficient (taking 100-350 Myr; \citealt{deAvillez2002, Roy1995}), making it plausible that subsequent generations of stars could form before mixing has finished.

\subsection{Subtle evidence for azimuthal variations}
\label{sec:dis:azimuth}
In all galaxies we observe a very small systematic scatter (0.03-0.05 dex) after subtracting the radial gradients. This scatter is generally smaller than the amplitude of the radial variation, which typically ranges from 0.1 to 0.3 dex over the radial range we observe.  It has been previously shown that strong line methods systematically produce smaller scatter in their radial gradients than direct temperature methods \citep{Arellano-Cordova2016}, but this very small systematic scatter with respect to the radial gradient is similar to what has been found using direct temperature methods to measure the metallicity in the Milky Way \citep{Esteban2018}.  It is, however, much smaller  than what has been found with direct temperature methods for three nearby galaxies by the CHAOS project ($\sim$0.1 dex;  \citealt{Croxall2015, Berg2015, Croxall2016}) and in M33 (0.11 dex; \citealt{Rosolowsky2008}).  

We find azimuthal variations in half of the galaxies in our sample, but these trends are not always obviously associated with the spiral pattern.  In all cases the azimuthal trends are more pronounced on only one spiral arm. 
In general, we find that  HII regions with enhanced or reduced metallicity are located across the full disk. Because of this, we have failed to distinguish any systematic correlations between changes in \doh~ and environmental parameters (arm/interarm masks, stellar mass surface density offset). This suggests that while the spiral pattern plays a role in organizing the ISM, it  alone does not establish the azimuthal variations we observe, with changes also being driven by the strong correlations with the local physical conditions described in Section \ref{sec:dis:localprops}.

\begin{figure*}
\centering
\includegraphics[width=7in]{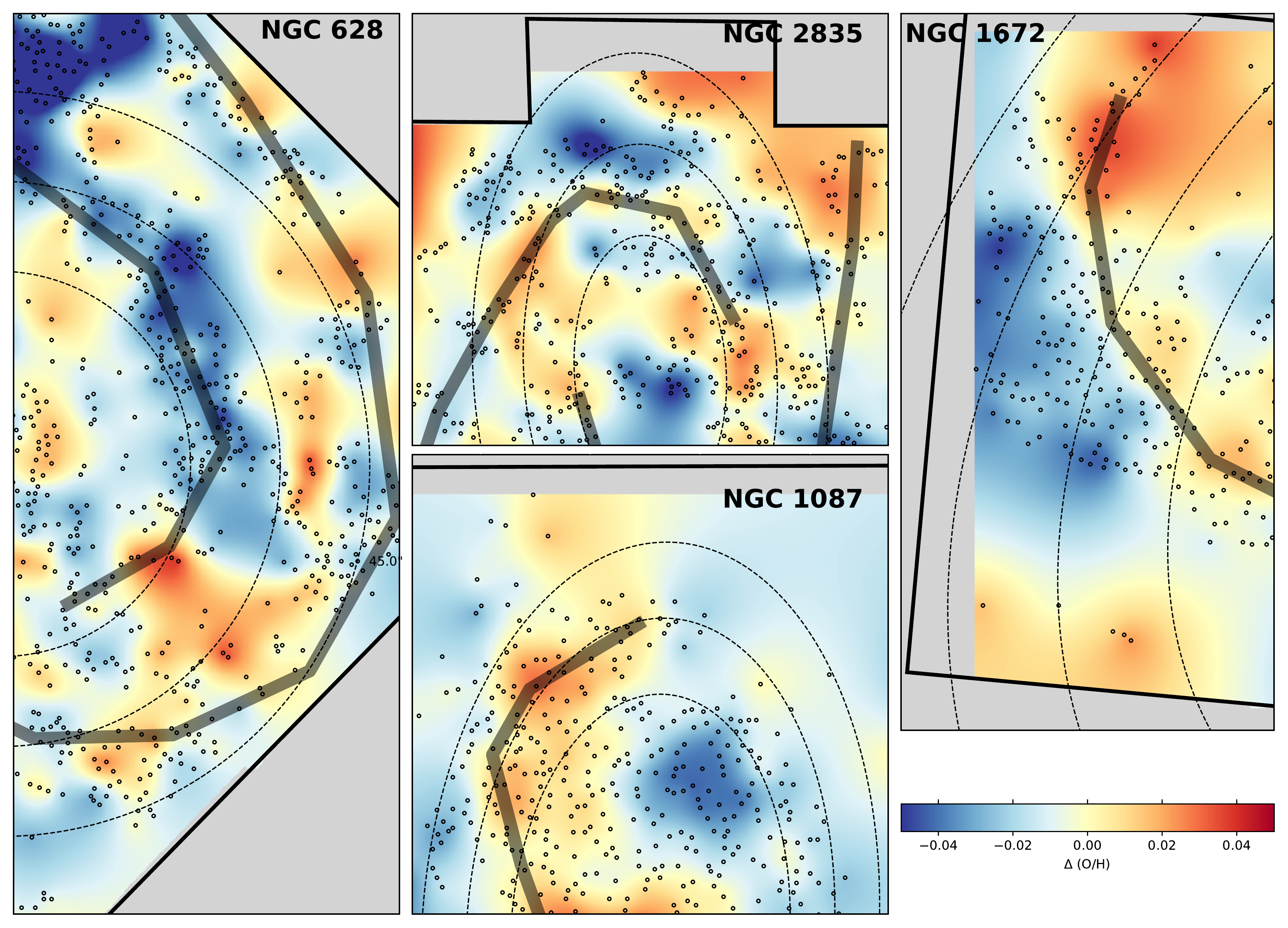}
\caption{Zooming in on subsections of galaxies with the most pronounced azimuthal variations.  Dashed ellipses in each galaxy mark fixed 1kpc wide rings.   The individual measurements of \doh\ for each HII region (circles) are interpolated onto a two dimensional surface (shown in colorscale) using kriging and an exponential model.  The brightest star-forming regions along the spiral arm have been traced by eye to provide a rough outline of the spiral pattern (overlaid in grey). 
Variations are seen in relation to the spiral arm in some cases (NGC 1672, NGC 1087) but more generally along fixed azimuth in other case (NGC 628, NGC 2835).  These variations are not perfectly symmetric, and are found in half of our sample. 
\label{fig:zoomaz}}
\end{figure*}

NGC 1672 presents a particularly interesting case (Figure \ref{fig:zoomaz}), as the eastern arm shows a very clear ridge along the spiral arm with enhanced abundances.  This appears very similar to the situation observed in the nearby galaxy NGC 1365 \citep{Ho2017}, with its widely separated spiral pattern, for which those authors proposed a carousel model where enrichment can proceed slowly in small pockets as gas passes through the interarm region, with enrichment peaking at the spiral arm.  Passage through the spiral arm then triggers large-scale mixing, introducing more pristine gas from the surroundings and hence systematically decreasing the abundances.  NGC 6754 \citep{Sanchez-Menguiano2016} presents another case of azimuthal abundances attributed to the spiral pattern, however in this case the drop in abundances is attributed to radial streaming motions that introduce more pristine gas to decrease the abundances, with trends differing on the leading and trailing sides of the arm.

The azimuthal variations reported here present a more complicated picture.  Half of our eight galaxies show azimuthal variations, though often the case is much clearer across one half of the galaxy. NGC 1087 and NGC 1672 clearly show enhanced abundances along the spiral arm, but one is more flocculent while the other has a strong bar and a more pronounced spiral pattern.  They also show different distributions for the enhanced regions, with NGC 1672 exhibiting a narrow ($<$100~pc) ridge while NGC 1087 has a broader ($\sim$300~pc) enhanced region along the arm.   NGC 628 and NGC 2835 show some regular azimuthal variations, but they do not correspond to the peaks in star formation.   One significant difference between this study and previous work is our focus on the inner portions of the disk (R$<$0.5~R$_{25}$), whereas some recent theoretical work has suggested that azimuthal trends should be more pronounced in the outer disks near the corotation radius, where the relative velocity with respect to the pattern speed of the disk is close to zero \citep{Spitoni2019}. 

\cite{Sanchez-Menguiano2019} carried out a detailed investigation of the local relation between star formation rate and metallicity, where for both quantities a radial relation has been removed.  They find systematic trends, that transition from a negative correlation at lower stellar masses to a positive correlation at higher stellar masses. While it is the case that the three most massive galaxies in our sample do not show azimuthal variations, the stellar masses for all galaxies in our sample do not range over more than an order of magnitude, so it is difficult to draw strong conclusions.  Further progress could be made with more detailed dynamical models of each individual galaxy, but is beyond the scope of this paper.

\subsection{Barred galaxies with flat metallicity gradients}
Stellar bars drive gas inflows  \citep{Kormendy2004, Jogee2005}, transporting metal-poor gas from the outer disk towards the galactic center.   However, observations have presented conflicting evidence regarding whether or not barred spiral galaxies systematically exhibit flatter metallicity gradients \citep{Vila-Costas1992, Martin1994, Zaritsky1994, Dutil1999, Henry1999, Sanchez2014, Kaplan2016, Balser2011}.

While an investigation of the connection between the metallicity gradients reported for this sample of eight galaxies and their morphological structure is beyond the scope of this work, flattened gradients are quite pronounced for three galaxies in our sample (NGC 1672, NGC 3627 and NGC 4535), particularly within the inner 0.1-0.3 R$_{25}$ (Figure \ref{fig:gradients}).  All three of these show very strong stellar bars, which dominate the morphology over this inner radial range.   NGC 3627 is also strongly interacting, and NGC 4535 is a member of the Virgo Cluster.  Three other galaxies in our sample are also classified as barred (NGC 1087, NGC 2835 and IC 5332; Table \ref{tab:sample}), exhibiting weaker stellar bars, and show no obvious flattening of their metallicity gradients.  We draw no conclusions in this work, but find the connection between bar strength and flattened radial gradients intriguing.

\subsection{Physical interpretation of line ratios}
\begin{figure*}
\centering
\includegraphics[width=3.5in]{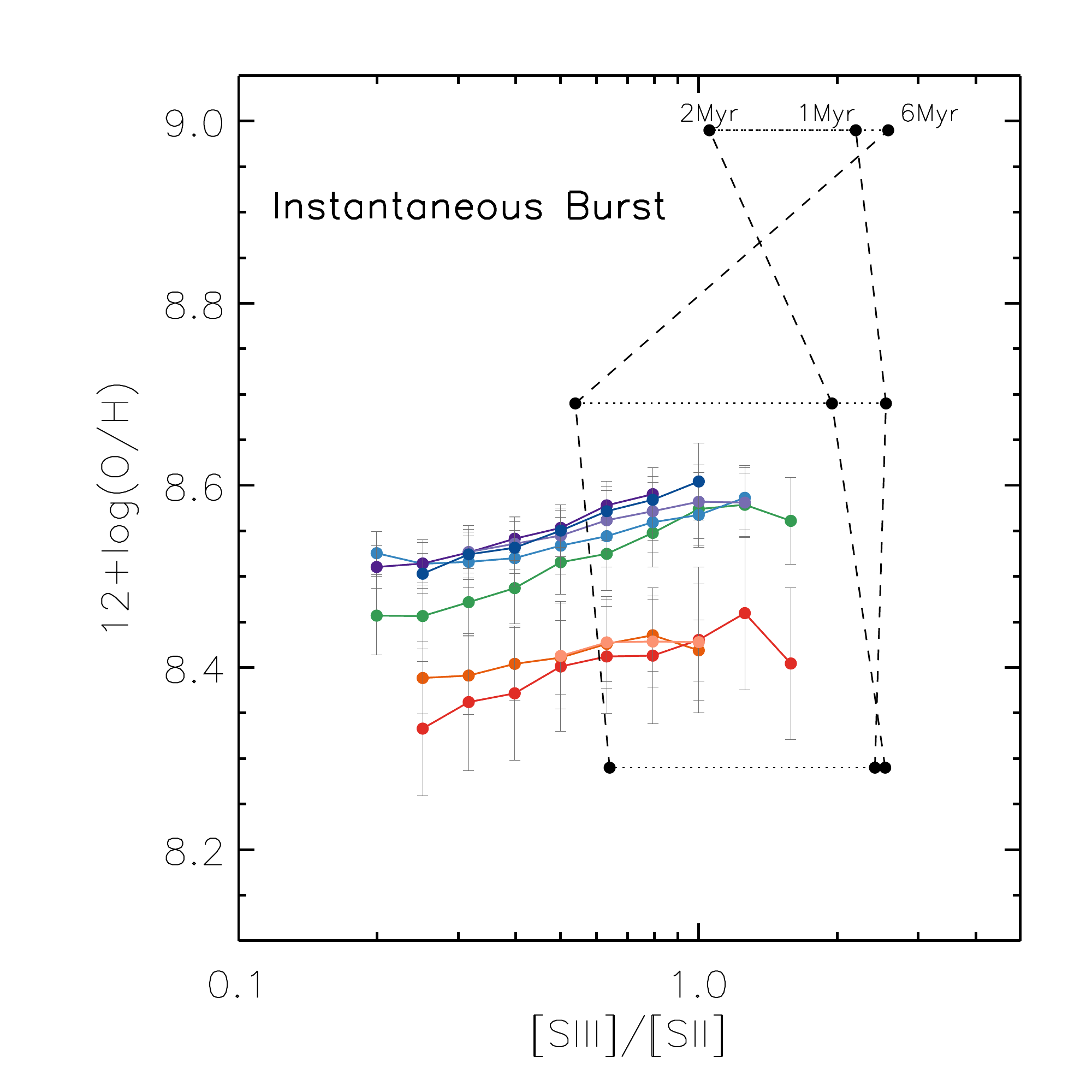}
\includegraphics[width=3.5in]{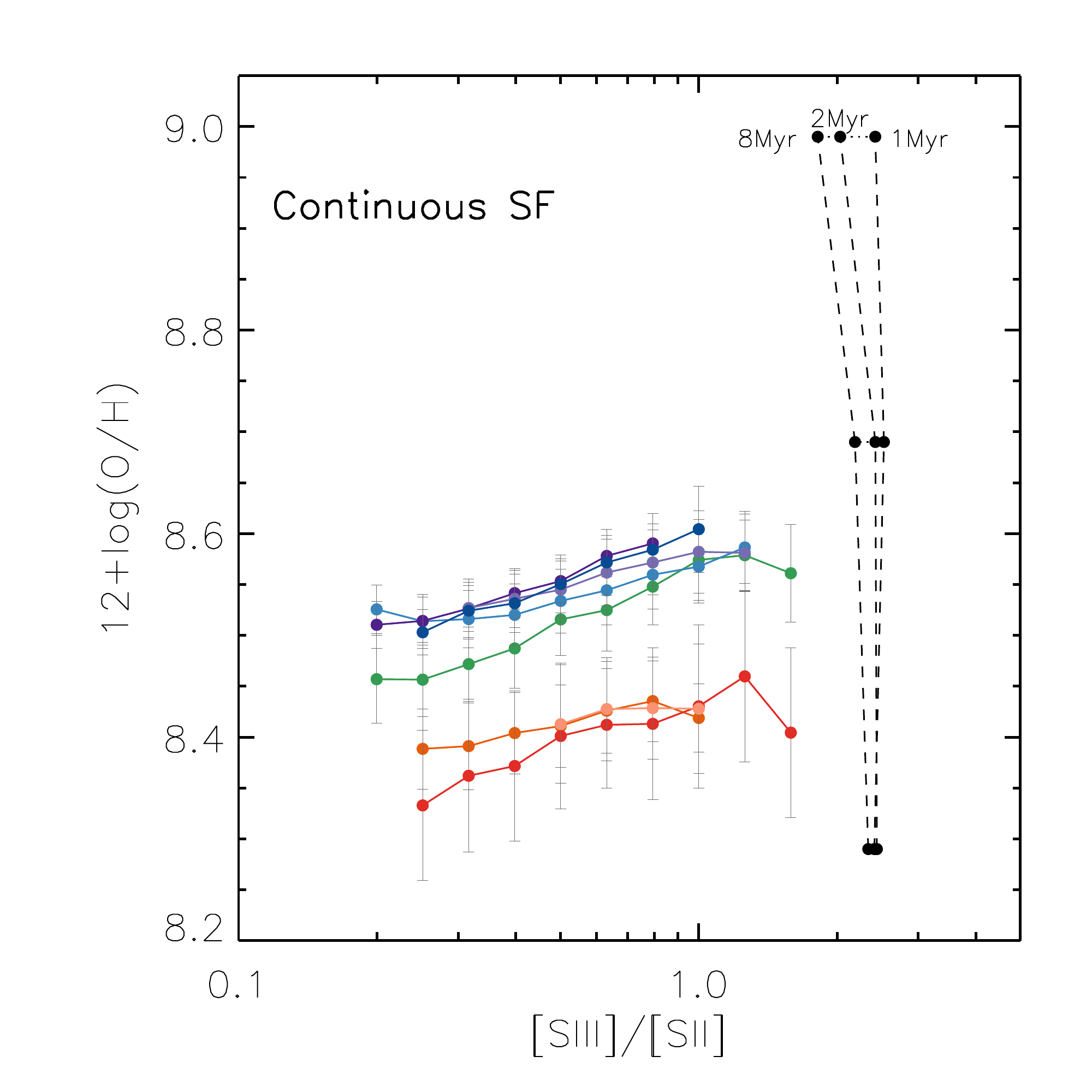}
\caption{
Comparison showing how changes in the age and metallicity of the stellar population are expected to affect the observed [SIII]/[SII] line ratio for both an instantaneous burst (left) or continuous star formation history (right) at fixed ionization parameter.  Here we show a grid from the \cite{Levesque2010} photionization models (n=100 cm$^{-3}$, q=2$\times$10$^7$) over a range of ages (dashed lines) and metallicities (horizongal dotted lines;  at fixed 0.4, 1 and 2 Z$_\odot$, assuming a solar value of 12+log(O/H)=8.69; \citealt{Asplund2009}) to demonstrate how the observed line ratios are expected to vary. This is overplotted on our observed relation (see also Figure \ref{fig:q_z}).  In the instantaneous burst models, age variations at fixed metallicity could theoretically account for some of the observed variation in [SIII]/[SII].  In the continuous star formation models, very little change in [SIII]/[SII] is expected at fixed metallicity (the dotted line is barely visible). Note that while the selected q value provides the best match to observations given the available models, it still does not well cover the observed range of [SIII]/[SII].  
\label{fig:q_z_photionmodel}}
\end{figure*}

We interpret the trends observed as changes in metallicity, however in a practical sense what we are tracing is changes in the line ratios integrated across HII regions.  The physical interpretations of these line ratio diagnostics are not straightforward, with many showing degeneracies between changes in metallicity and ionization parameter, and are expected to then correlate with variations in star/cluster age, ISM density and ISM pressure.  Figure \ref{fig:q_z_photionmodel} explores how \cite{Levesque2010} photoionization models at fixed electron density ($n=100$~cm$^{-3}$) and ionization parameter ($q=2\times$10$^7$) do produce variations in [SIII]/[SII] as a function of both metallicity and stellar population age.  Here we have taken the output line ratios from the ``high'' mass-loss tracks at three fixed metallicities (0.4, 1, and 2Z$_\odot$).  Assuming an instantaneous burst of star formation, this age dependency can result in significant (factor of 3-5) changes in the predicted [SIII]/[SII] line ratio.  Taking models with a continuous star formation history results in almost no change in [SIII]/[SII] with age.  

More detailed photoionization and stellar population modeling, along with follow-up UV observations to constrain the stellar population age, are necessary to disentangle these effects.  \cite{Shabani2018} found no dependence of star cluster age with distance to the spiral arm in two of the three nearby galaxies in their case study, including NGC 628, though their youngest age bin encompasses entirely this crucial 1-10 Myr regime.  We note that the strong correlation between \doh~ and [SIII]/[SII] persists even if we attempt to clean our sample of some of these population extremes.  Omitting regions with young ages (high H$\alpha$-to-FUV), high electron density (low [SII]6716/[SII]6731), high dust column (high E(B-V)) or close to the galaxy center does not fundamentally change the observed trend.  More work is needed to explore fully the interconnection of these parameters and the impact each could have on the metallicity prescriptions.

An important final caveat to our discussion is that the trends we report are heavily dependent on the choice of metallicity prescription used. Appendix \ref{appendix:prescriptions} explores this in more detail, but we emphasize here that our results are robustly reproduced if we instead use the \cite{Dopita2016} photo ionization model-based N2S2 prescription. Given that this second prescription takes a very different approach compared to the empirical \cite{Pilyugin2016} prescription employed in this work, and should exhibit different biases, this builds further confidence in our results. 

\section{Conclusion}
\label{sec:conclusion}

In the first results from the PHANGS-MUSE survey at $\sim$50~pc resolution we are able to cleanly separate individual HII regions within our optical IFU data cubes and measure line fluxes from a spectrum integrated over individual HII regions.  By applying the \cite{Pilyugin2016} strong line prescription, we measure gas phase oxygen abundances for 7,138 HII regions across the disks of eight nearby galaxies.  All galaxies are well fit by a simple linear metallicity gradient with small (0.03-0.05 dex) scatter. 

Subtracting that radial trend, we compare physical parameters for those HII regions that have enhanced and reduced abundances. Regions with enhanced abundances have high ionization parameter, higher H$\alpha$ luminosity, younger star clusters and associated molecular gas clouds show high CO line intesnity.  This indicates recent star formation, possibly the same that powers the HII regions, has locally enriched the material. Regions with reduced abundances show increased H$\alpha$ velocity dispersions, suggestive of mixing introducing more pristine material.  We also identify a positive correlation between gas phase metallicity and ionization parameter, in conflict with theoretical predictions.   Taken together, the trends we identify suggest the local physical conditions are tightly linked to the localized enrichment of the ISM. 

We observe subtle evidence for systematic azimuthal variations in half of the sample, though often in only one half of the galaxy, and we cannot cleanly associate this with the spiral pattern in all cases.  However, in NGC 1672 and NGC 1087 the strong correlations with spiral structure indicate that the spiral arms can play an important role in organizing and mixing the ISM.  This presents a complicated picture of the interplay between galaxy dynamics and enrichment patterns, which we aim to more fully address in the future with the full PHANGS-MUSE sample.

\acknowledgments
We thank the referee for helpful comments that improved this work.  
K.K.\ gratefully acknowledges funding from the German Research Foundation (DFG) in the form of an Emmy Noether Research Group (grant number KR4598/2-1, PI Kreckel). 
B.G.\ gratefully acknowledges the support of the Australian Research Council as the recipient of a Future Fellowship (FT140101202).  
FS, ES, CF, PL, RMcE, and TS acknowledges funding from the European Research Council (ERC) under the European UnionÕs Horizon 2020 research and innovation programme (grant agreement No. 694343).
M.C.\ and J.M.D.K.\ gratefully acknowledge funding from the Deutsche Forschungsgemeinschaft (DFG) through an Emmy Noether Research Group (grant number KR4801/1-1) and the DFG Sachbeihilfe (grant number KR4801/2-1). J.M.D.K.\ gratefully acknowledges funding from the European Research Council (ERC) under the European Union's Horizon 2020 research and innovation programme via the ERC Starting Grant MUSTANG (grant agreement number 714907). E.R.\ acknowledges the support of the Natural Sciences and Engineering Research Council of Canada (NSERC), funding reference number RGPIN-2017-03987. S.C.O.G.\ acknowledges support from the DFG via SFB 881 ``The Milky Way System'' (sub-projects B1, B2 and B8) and also via Germany's Excellence Strategy EXC-2181/1 - 390900948 (the Heidelberg STRUCTURES Excellence Cluster).J.P. and A.H. acknowledge funding from the Programme National ``Physique et Chimie du Milieu Interstellaire'' (PCMI) of CNRS/INSU with INC/INP, co-funded by CEA and CNES. FB acknowledge funding from the European Union's Horizon 2020 research and innovation programme (grant agreement No 726384). 

Based on observations collected at the European Organisation for Astronomical Research in the Southern Hemisphere under ESO programme IDs 094.C-0623(A),  098.C-0484(A), 1100.B-0651(A) and 1100.B-0651(B). 

This paper makes use of the following ALMA data:
ADS/JAO.ALMA\#2012.1.00650.S, 
ADS/JAO.ALMA\#2015.1.00925.S, 
ADS/JAO.ALMA\#2015.1.00956.S, 
ADS/JAO.ALMA\#2017.1.00392.S, 

ALMA is a partnership of ESO (representing its member states), NSF (USA) and NINS (Japan), together with NRC (Canada), MOST and ASIAA (Taiwan), and KASI (Republic of Korea), in cooperation with the Republic of Chile. The Joint ALMA Observatory is operated by ESO, AUI/NRAO and NAOJ. The National Radio Astronomy Observatory is a facility of the National Science Foundation operated under cooperative agreement by Associated Universities, Inc.

Based on observations made with the NASA Galaxy Evolution Explorer. GALEX is operated for NASA by the California Institute of Technology under NASA contract NAS5-98034.  

Data analysis made use of Topcat \citep{Taylor2005}. This research has made use of NASA's Astrophysics Data System. This research has made use of the NASA/IPAC Extragalactic Database (NED),
which is operated by the Jet Propulsion Laboratory, California Institute of Technology,
under contract with the National Aeronautics and Space Administration.

\bibliographystyle{yahapj}

\clearpage

\clearpage

\appendix
\section{Image Atlas}
\label{appendix:atlas}
We present an atlas of images for each of the eight galaxies in our initial PHANGS-MUSE sample. All presented images are constructed from our MUSE data cubes, including the gri-band three color images.  For all galaxies, corresponding images are at matched image stretch and scaling.  Each figure shows the gri-band three color image (bottom left), overlaid emission line maps (red: H$\alpha$, green: [OIII], blue: [SII]; bottom right), H$\alpha$ emission (top left) and HII regions with their measured metallicities (top right). 

\clearpage

\begin{figure*}
\centering
\includegraphics[width=7in]{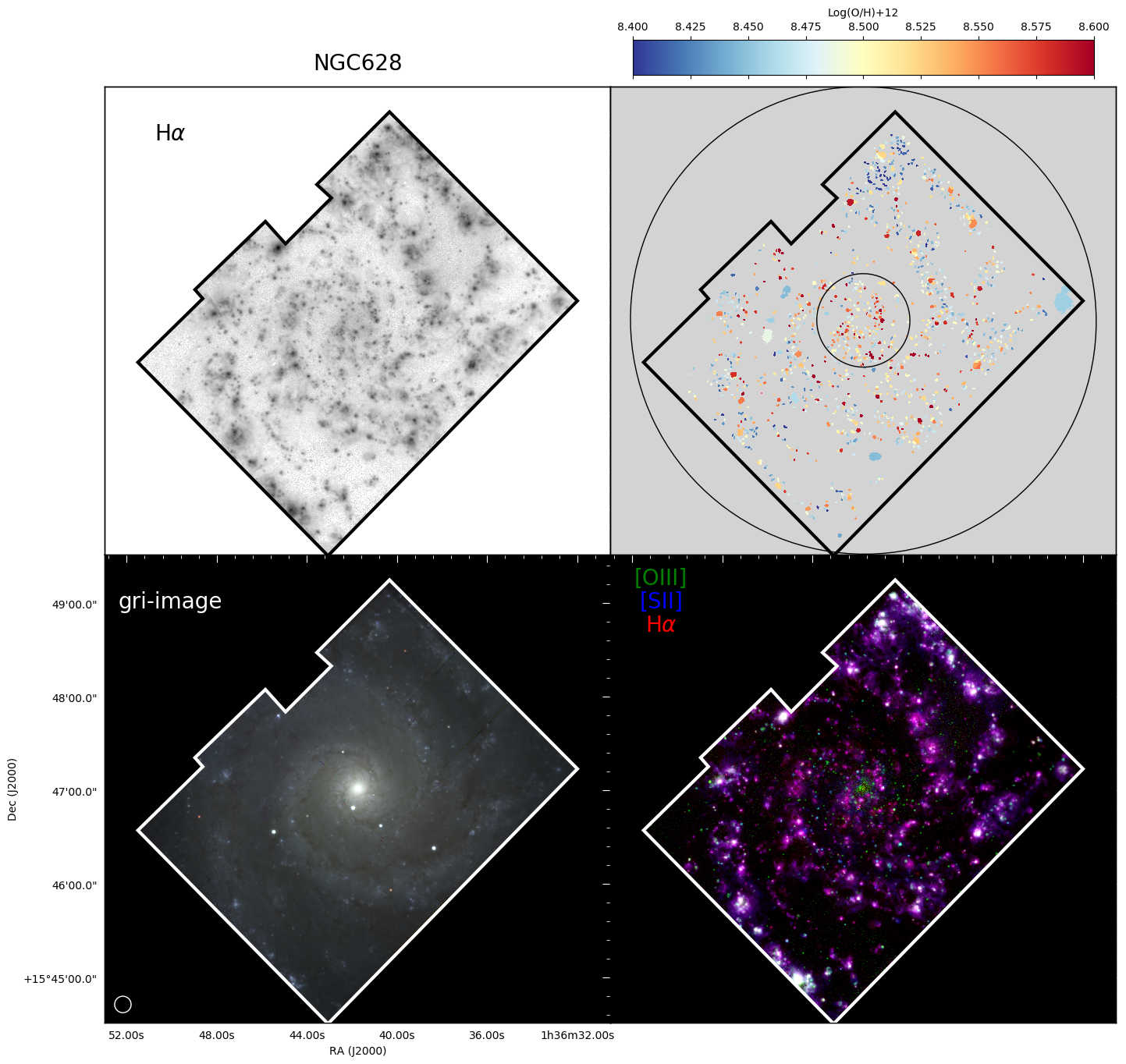}
\caption{Mapping metallicity variations within the disks of NGC 628.  Simulating a gri-band three color image (bottom left) and overlaying emission line maps (red: H$\alpha$, green: [OIII], blue: [SII]; bottom right) reveals the diversity of environments (bulge, spiral arms, dust lanes) and physical conditions across the disk.  The H$\alpha$ maps (top left) reveal a rich collection of HII regions and diffuse ionized gas. Using HIIphot to identify HII regions, we paint the HII region mask with the measured metallicity (top right). Overlaid are ellipses at fixed 0.1R$_{25}$ and 0.5R$_{25}$. The circle in the lower left indicates a 500~pc diameter scale size, observations have a typical physical resolution of $\sim$50pc (see also Table \ref{tab:sample}).
\label{fig:prettyfig}}
\end{figure*}

\begin{figure*}
\centering
\includegraphics[width=7in]{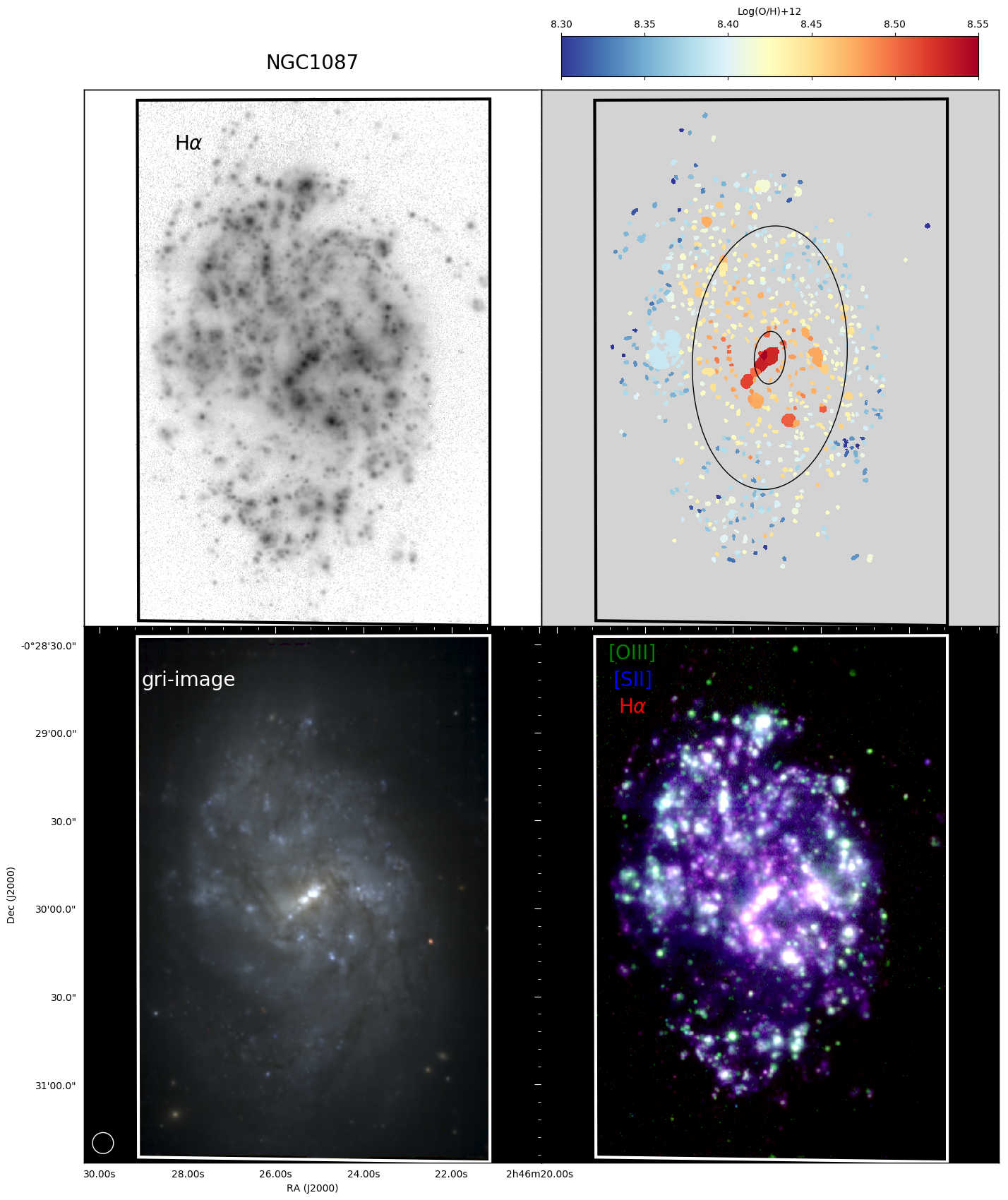}
\caption{As in Figure \ref{fig:prettyfig}, for NGC 1087.
\label{fig:prettyfig1087}}
\end{figure*}

\begin{figure*}
\centering
\includegraphics[width=7in]{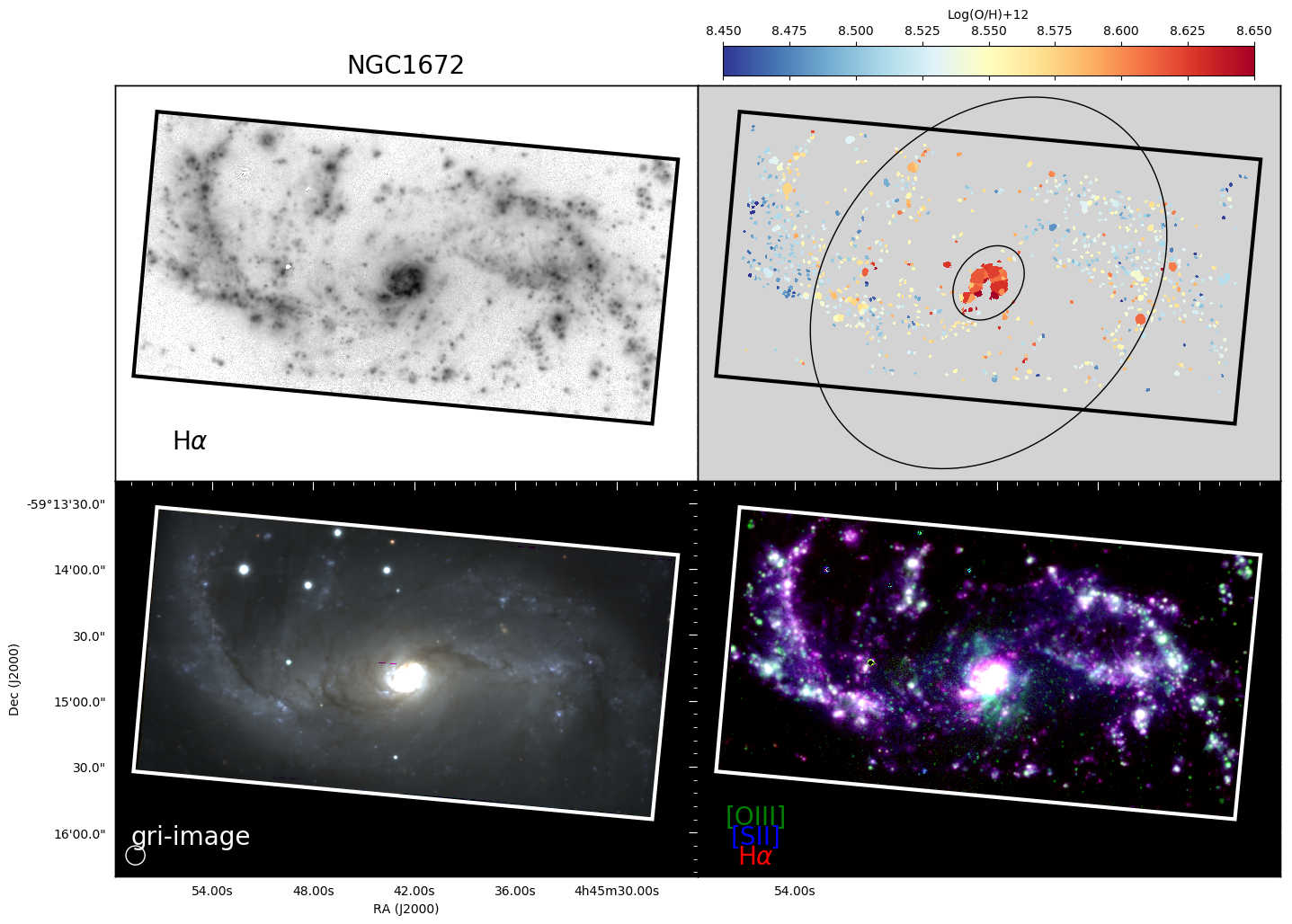}
\caption{As in Figure \ref{fig:prettyfig}, for NGC 1672.
\label{fig:prettyfig1672}}
\end{figure*}

\begin{figure*}
\centering
\includegraphics[width=7in]{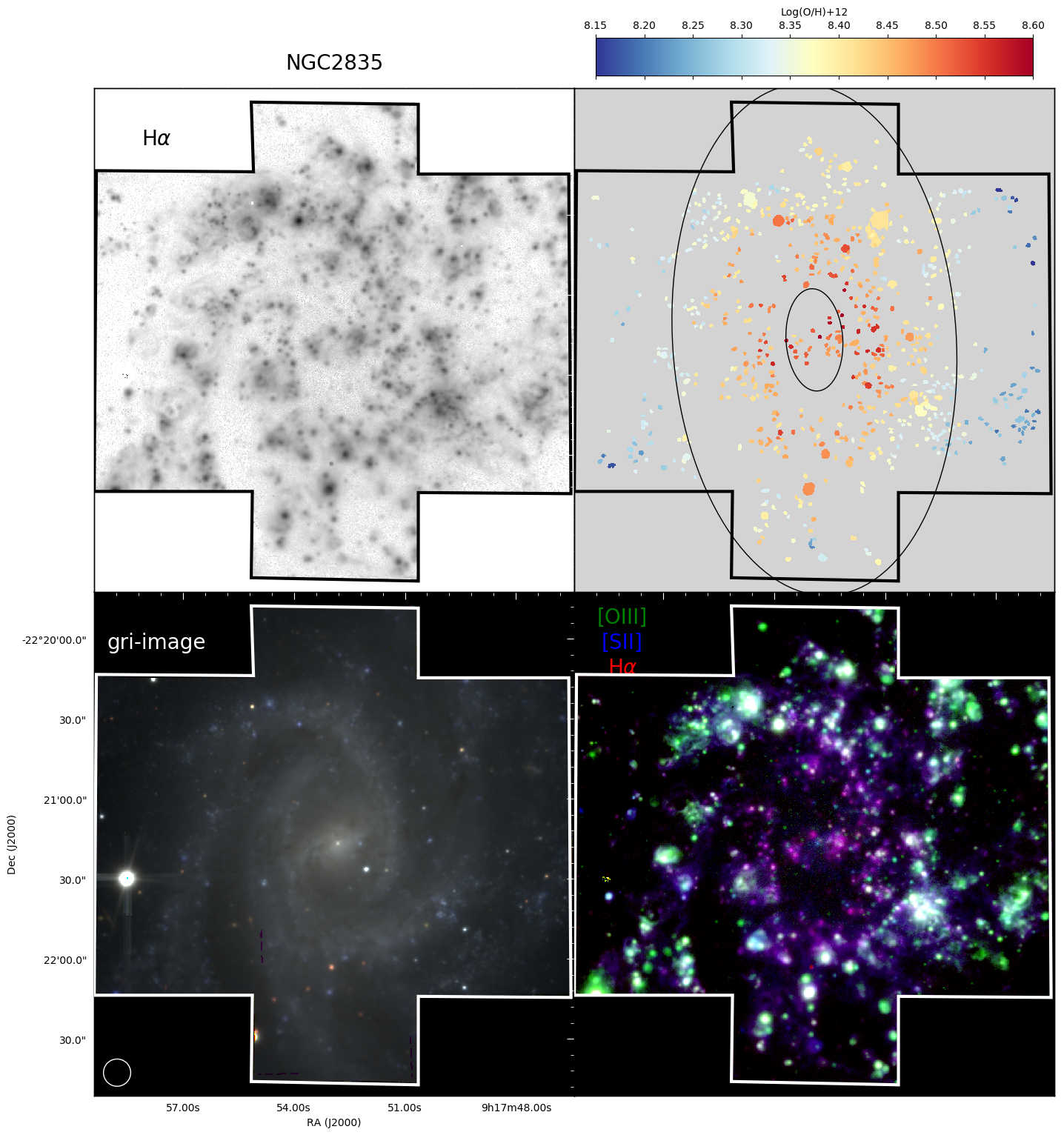}
\caption{As in Figure \ref{fig:prettyfig}, for NGC 2835.
\label{fig:prettyfig2835}}
\end{figure*}

\begin{figure*}
\centering
\includegraphics[width=6in]{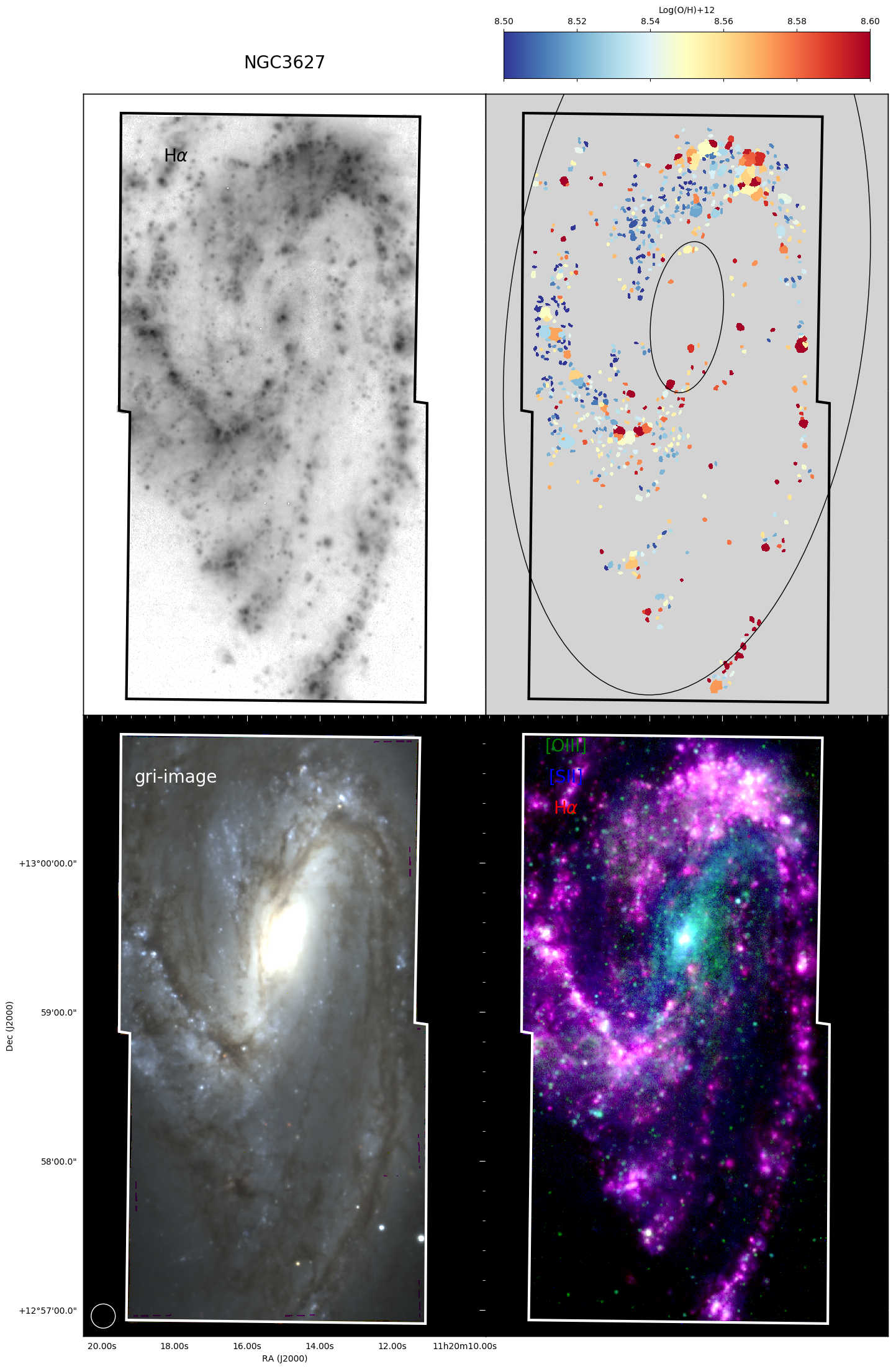}
\caption{As in Figure \ref{fig:prettyfig}, for NGC 3627.
\label{fig:prettyfig3627}}
\end{figure*}

\begin{figure*}
\centering
\includegraphics[width=7in]{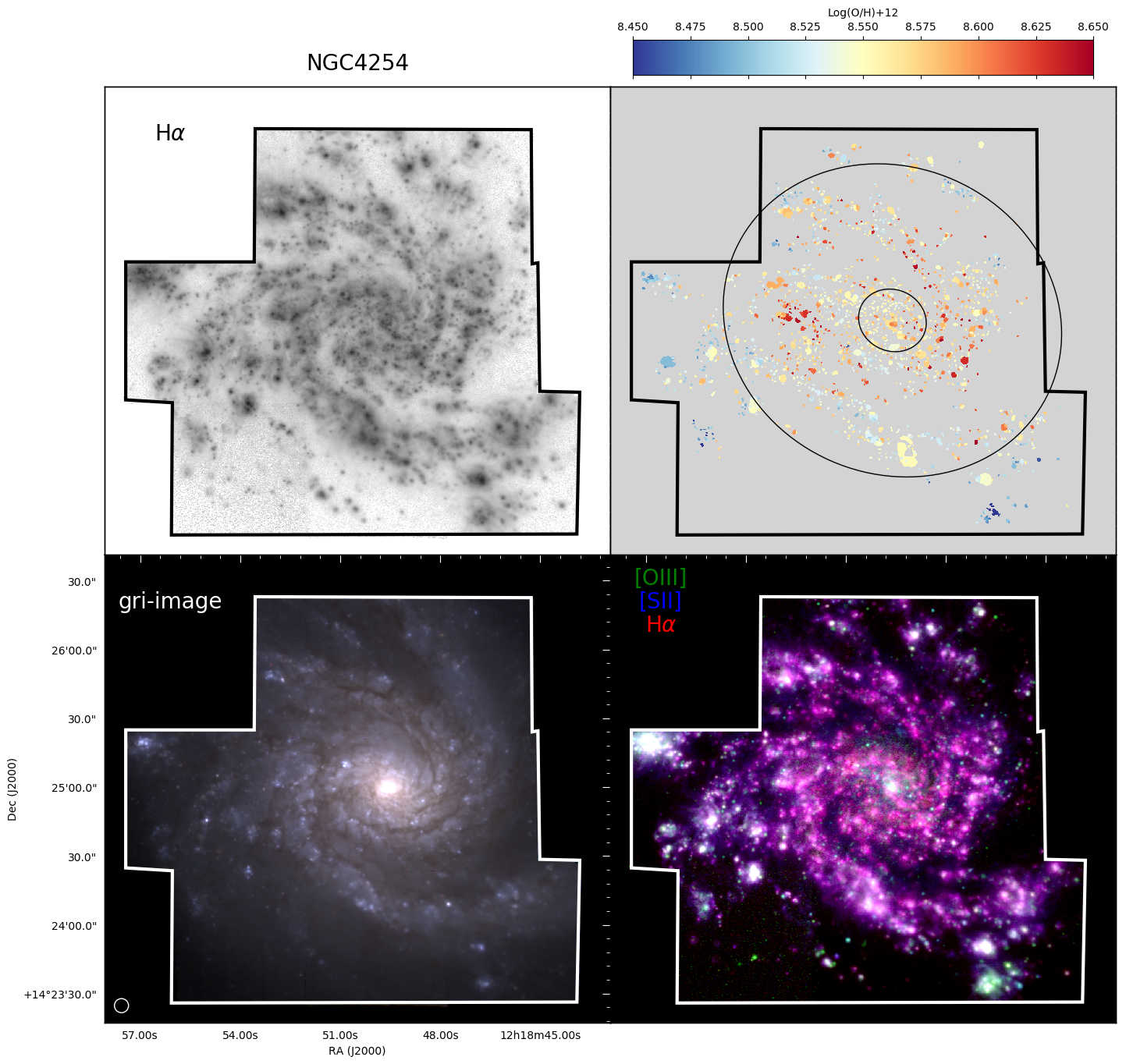}
\caption{As in Figure \ref{fig:prettyfig}, for NGC 4254.
\label{fig:prettyfig4254}}
\end{figure*}

\begin{figure*}
\centering
\includegraphics[width=6in]{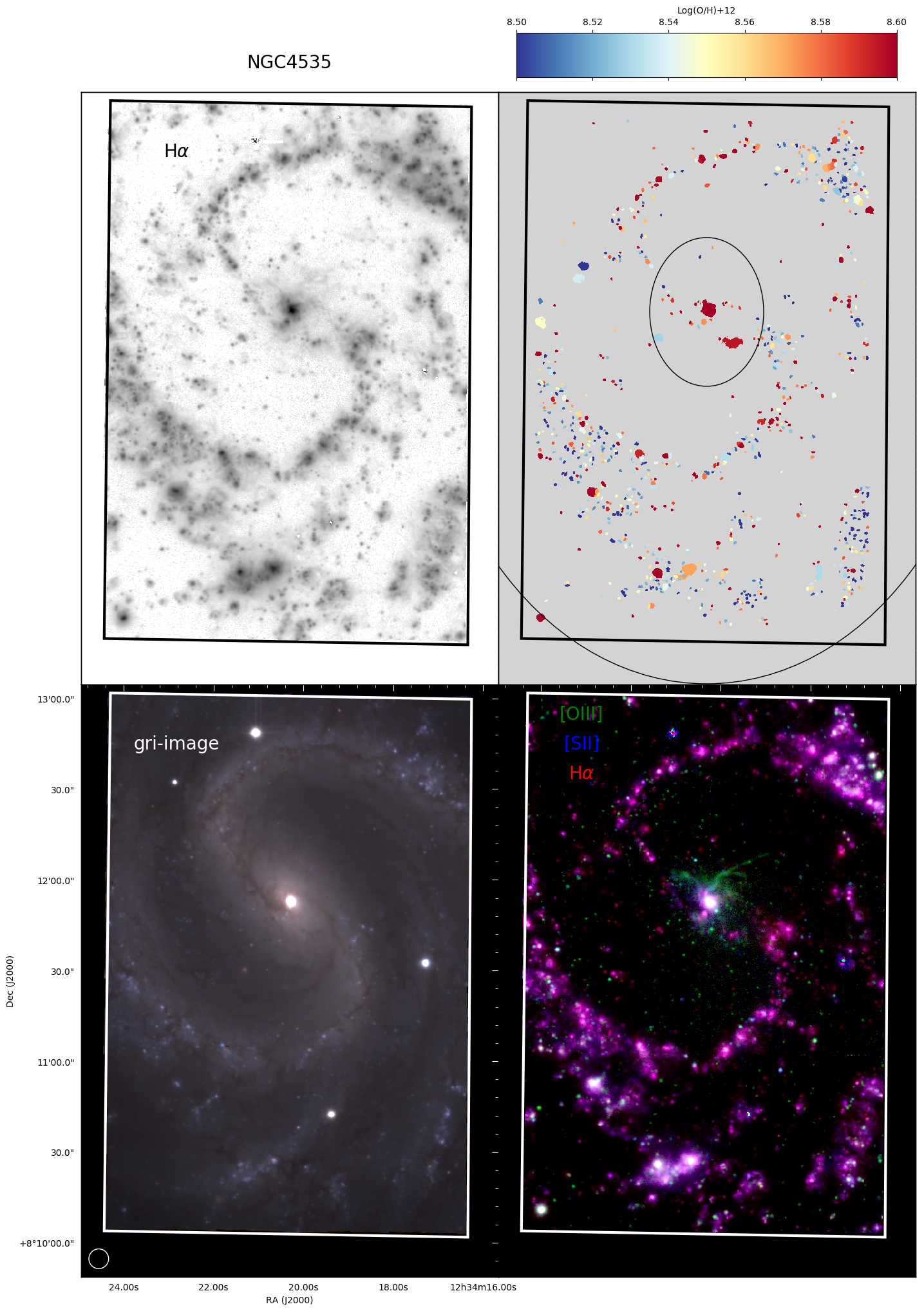}
\caption{As in Figure \ref{fig:prettyfig}, for NGC 4535.
\label{fig:prettyfig4535}}
\end{figure*}

\begin{figure*}
\centering
\includegraphics[width=7in]{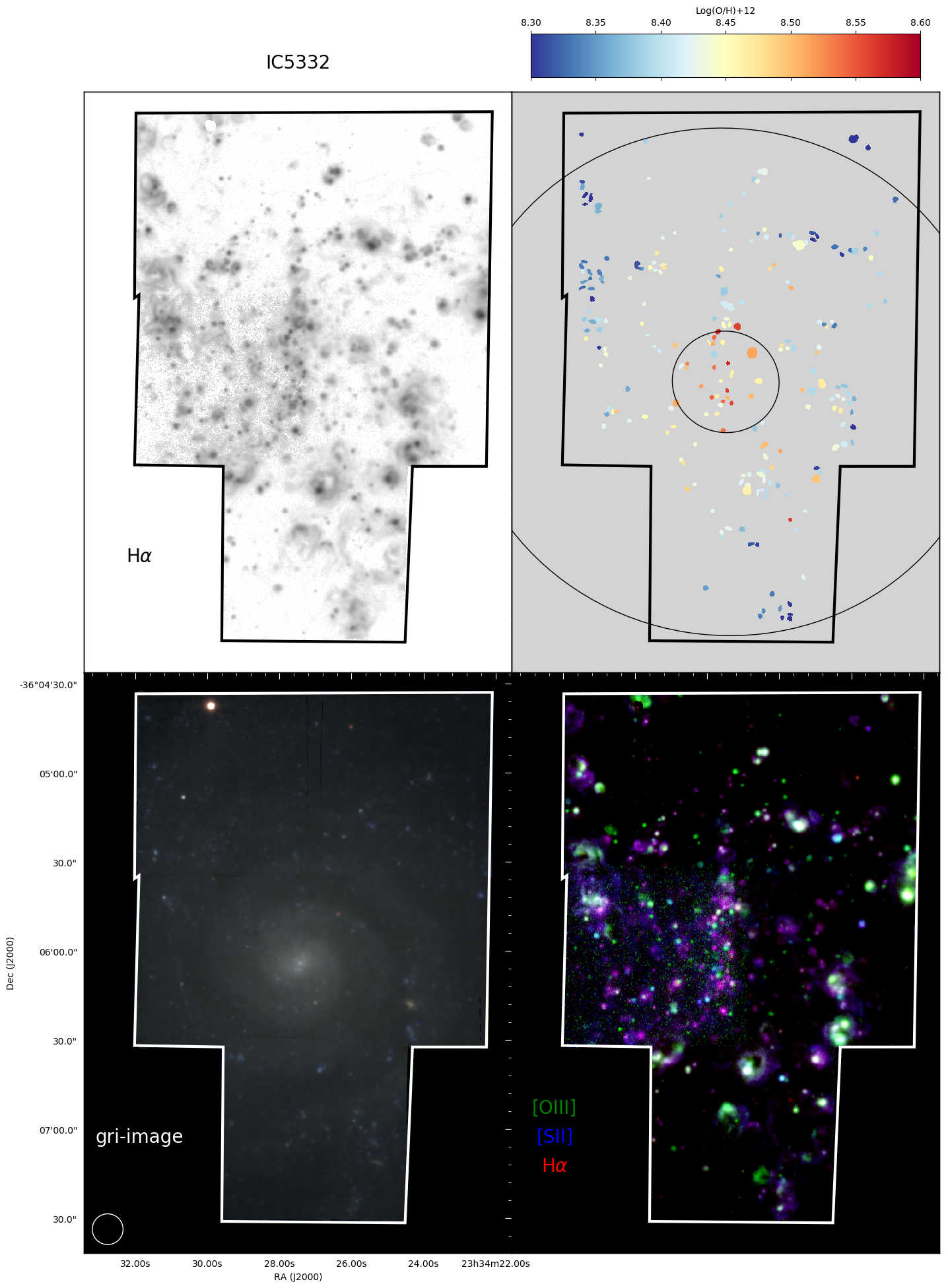}
\caption{As in Figure \ref{fig:prettyfig}, for IC 5332.
\label{fig:prettyfig_cont2}}
\end{figure*}

\clearpage

\section{Excitation Diagnostic Diagrams}
\label{appendix:bpt}
\begin{figure*}
\centering
\includegraphics[width=7in]{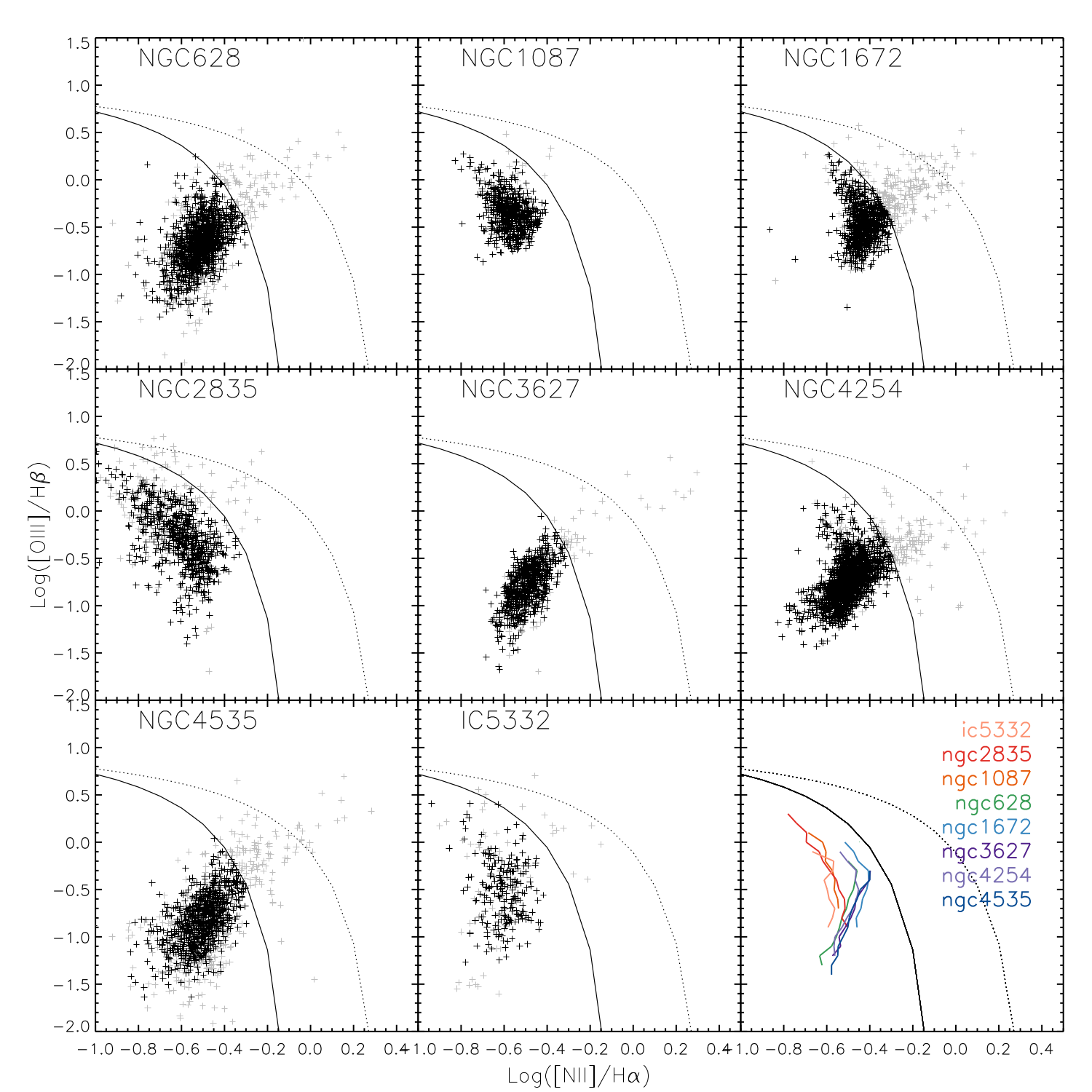}
\caption{Diagnostic line ratios [OIII]/H$\beta$ as a function of [NII]/H$\alpha$ \citep{Baldwin1981} for all eight galaxies in our sample.  HII regions (black) populate a distributed but contained region in this parameter space.  In grey are regions morphologically identified by HIIphot but that fail at least one of our selection criteria (see Section \ref{sec:hiireg}). Regions consistent with photoionization are expected to fall below the empirical \cite{Kauffmann2003} line (solid) and the theoretical \cite{Kewley2001} line (dotted). Median trends for each galaxy are plotted in the lower right panel, with galaxies color coded by their stellar mass.  
\label{fig:bpt1}}
\end{figure*}

\begin{figure*}
\centering
\includegraphics[width=7in]{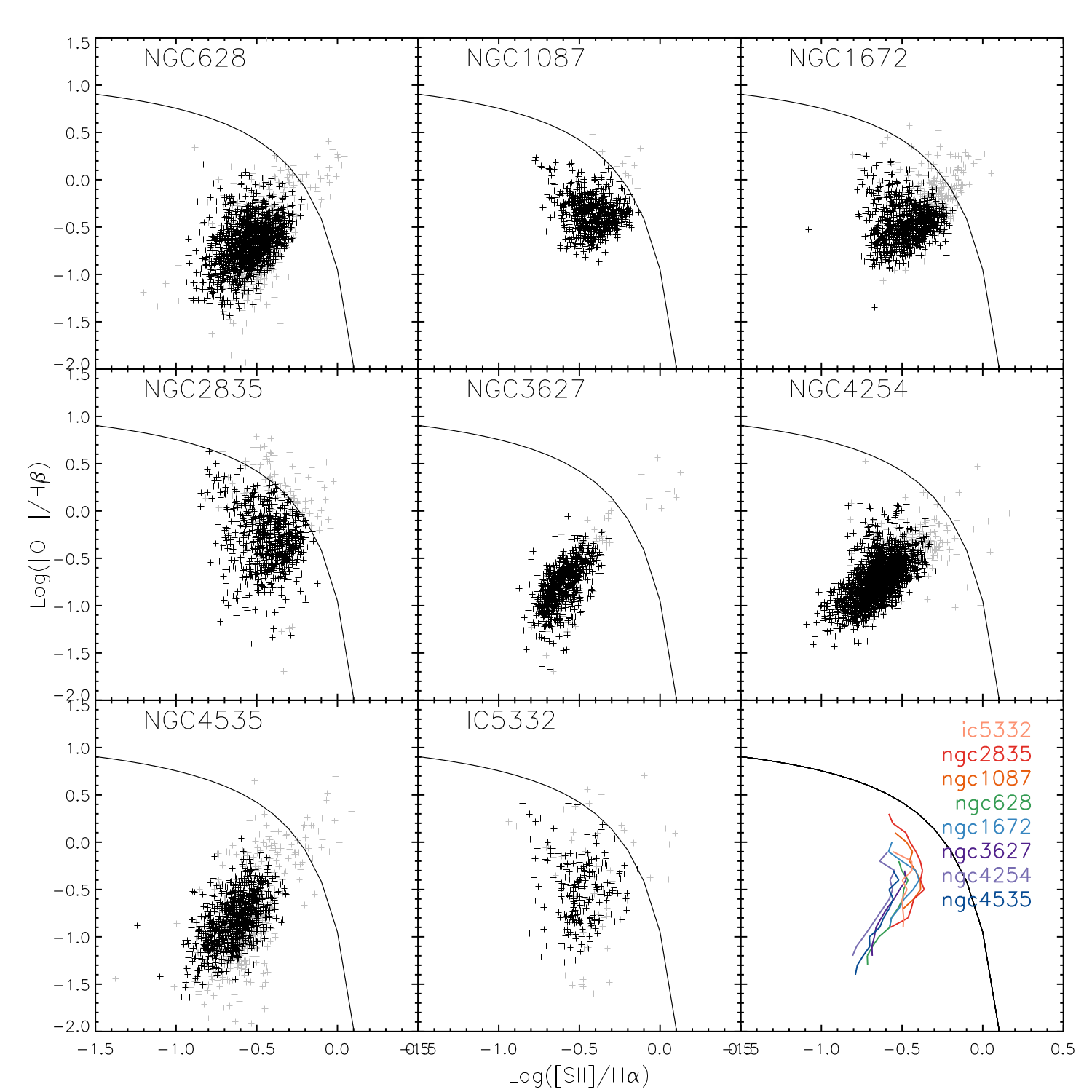}
\caption{Diagnostic line ratios [OIII]/H$\beta$ as a function of [SII]/H$\alpha$ \citep{Baldwin1981} for all eight galaxies in our sample.  HII regions (black) populate a distributed but contained region in this parameter space.  In grey are regions morphologically identified by HIIphot but that fail at least one of our selection criteria (see Section \ref{sec:hiireg}). Regions consistent with photoionization are expected to fall below the theoretical \cite{Kewley2001} line (solid). Median trends for each galaxy are plotted in the lower right panel, with galaxies color-coded by their stellar mass.  
\label{fig:bpt2}}
\end{figure*}

We show BPT \citep{Baldwin1981}  diagnostic diagrams  for all eight galaxies. 
In defining our HII region catalog (Section \ref{sec:hiireg}), we employ two BPT emission line ratio diagnostics to ensure the excitation is consistent with photoionization by massive stars.  In the [OIII]/H$\beta$ vs. [NII]/H$\alpha$ diagnostic (Figure \ref{fig:bpt1}), we require regions to be consistent with the  stringent empirical \cite{Kauffmann2003} line.  In the [OIII]/H$\beta$ vs [SII]/H$\alpha$ diagnostic (Figure \ref{fig:bpt2}), we require regions to be consistent with the \cite{Kewley2001} line.  

Each diagram shows all regions morphologically identified by HIIphot, and in grey we indicate HII regions which are excluded due to one of our rejection criteria listed in Section \ref{sec:hiireg}.  

\clearpage

\section{Metallicity Prescriptions}
\label{appendix:prescriptions}
\begin{figure*}
\centering
\includegraphics[width=7in]{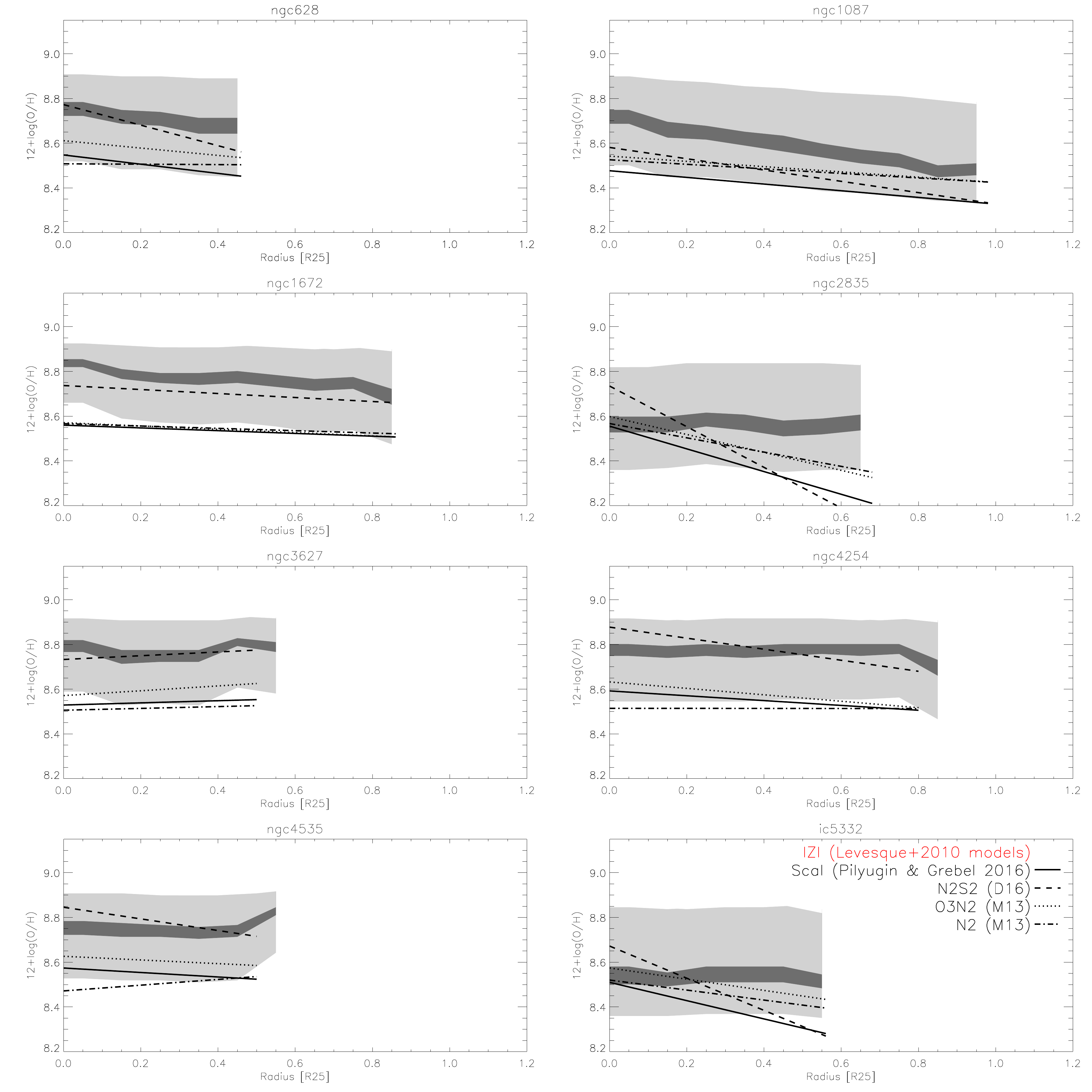}
\caption{
Fits to the radial metallicity gradients using four different prescriptions.  We show here the Scal prescription (solid lines; \citealt{Pilyugin2016}), which we have assumed for this work, the \cite{Dopita2016} N2S2 photoionization model based prescription (D16; dashed lines), and the commonly used empirical \cite{Marino2013} O3N2 (dotted lines) and N2 (dash-dot lines) prescriptions. Coefficients for each linear fit are given in Table~\ref{tab:gradients}. We overplot these on outputs from the Bayesian inference code IZI \citep{Blanc2015}, which derives the joint and marginalized posterior probability density functions for metallicity and ionization parameter.  Here we have radially binned and summed the marginalized PDFs to produce a probabilistic metallicity gradient for each galaxy.  Both the central 10\% (dark grey) and 68\% (light grey) percentile bands are shown. 
\label{fig:gradients_appendix}}
\end{figure*}

Strong line metallicity prescriptions produce well established differences in inferred metallicity and radial gradients in galaxies \citep{Kewley2008, Poetrodjojo2019}. 
In Figure \ref{fig:gradients_appendix} we present best-fitting radial metallicity gradients using four different prescriptions. 
This includes the Scal (solid lines; \cite{Pilyugin2016})  prescription, which we have adopted in this work, as well as the \cite{Dopita2016} N2S2 (D16; dashed lines) photoionization model based prescription, which gives qualitatively similar results.  For completeness and to facilitate consistent comparisons with the exisiting literature, we also include the commonly used empirical \cite{Marino2013} O3N2 (dotted lines) and N2 (dash-dot lines) prescriptions. 
As in Section \ref{sec:radgrad}, each radial gradient is fit using all HII regions with radius above $0.1 {\rm R}_{25}$. 

Finally, for further comparison with the results of photoionization models, we overplot each linear radial gradient on outputs from the Bayesian inference code IZI \citep{Blanc2015}.  This code uses a set of observed line fluxes combined with an input photoionization model to derive the joint and marginalized posterior probability density functions for metallicity and ionization parameter.  Here, we have compared all our observed strong emission lines ([OIII], H$\beta$, [NII], H$\alpha$, [SII], [SIII]) with the \cite{Levesque2010}  ``high'' mass-loss tracks assuming a constant star formation history over the past 6 Myr and a Salpeter IMF with a 100 M$_\odot$ upper cutoff.  
We then radially bin and sum the marginalized PDFs to produce a probabilistic metallicity gradient for each galaxy.
All gradients across all different methods show relatively good agreement in slope, and systematically $\sim$0.2 dex lower values in all empirical calibrations compared to the photoionization based methods.

In Table \ref{tab:gradients} we list the coefficients of all radial gradient fits for each galaxies, using each of the four different metallicity prescriptions. R$_0$ is the y-intercept, and the gradient is given both in units of R$_{25}$ and in units of kpc. We further list the scatter with respect to the linear fit ($\sigma$(O/H)).   
Here, $\sigma$(O/H) reflects the standard deviation with respect to the linear radial gradient fit, whereas
\doh\ is the residual for a given point from the fit.  
This scatter is systematically lower ($\sim$0.04 dex) in all empirical calibrations (Scal, O3N2, N2) compared to the N2S2 (D16) photoionization model based calibration ($\sim$0.09 dex).

\begin{figure*}
\centering
\includegraphics[width=7in]{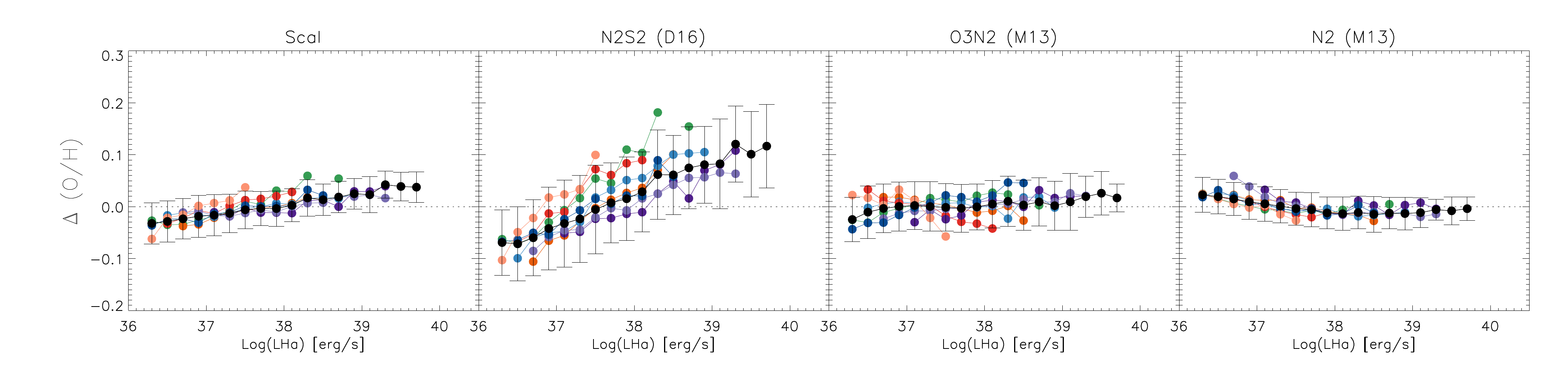}
\includegraphics[width=7in]{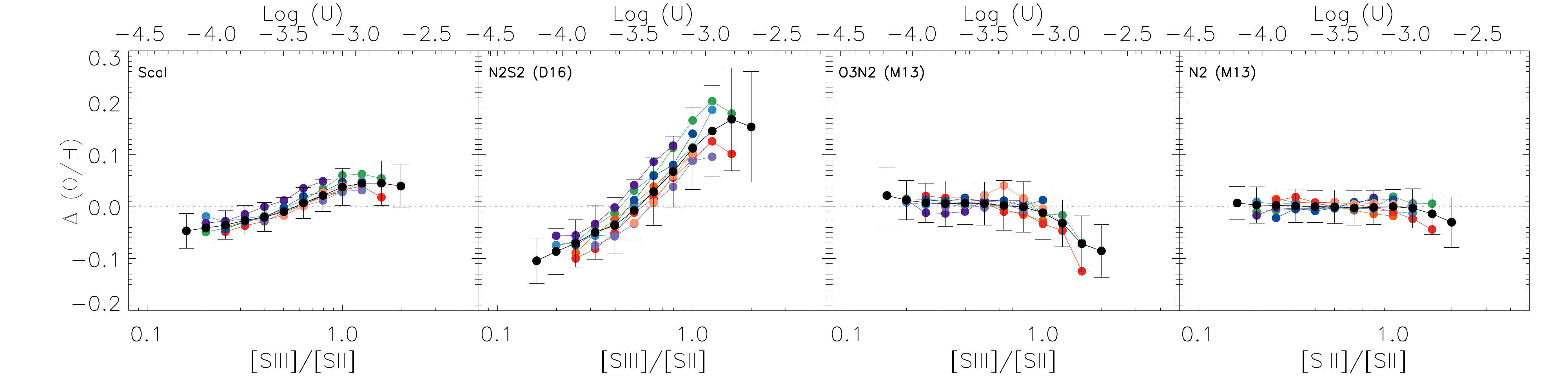}
\caption{Trends between \doh~ and L(H$\alpha$) (top) and [SIII]/[SII] (bottom; a proxy for the ionization parameter) for four different metallicity prescriptions. The Scal (adopted throughout this paper) and N2S2 (D16) prescriptions show positive correlations, with the ionization parameter exhibiting a more fundamental correlation. O3N2 (M13) and N2 (M13) show no trend with either parameter.  
Different galaxies are represented by different colors as in  Figure \ref{fig:bpt1}. 
\label{fig:offsets_appendix}}
\end{figure*}

While the radial gradients appear quite consistent across calibrations, and the scatter $\sigma$(O/H) is also similar between empirical calibrations, the different calibrations show very different trends between \doh~ and L(H$\alpha$) and [SIII]/[SII] (a proxy for the ionization parameter).  In Figure \ref{fig:offsets_appendix} we show this trend using each of the four prescriptions for each galaxy individually and for the entire sample.  Both the Scal and N2S2 (D16) calibrations exhibit a positive correlation with both L(H$\alpha$) and [SIII]/[SII] that is discussed in this paper.  On the other hand, O3N2 and N2 show a quite flat trend.  However, because these are both using smaller dimensionallity in the line ratios, we believe both of these presciptions retain some of the degeneracy between metallicity and ionization parameter, enough to cloud the subtle trends presented here.  

\begin{figure*}
\centering
\includegraphics[width=7in]{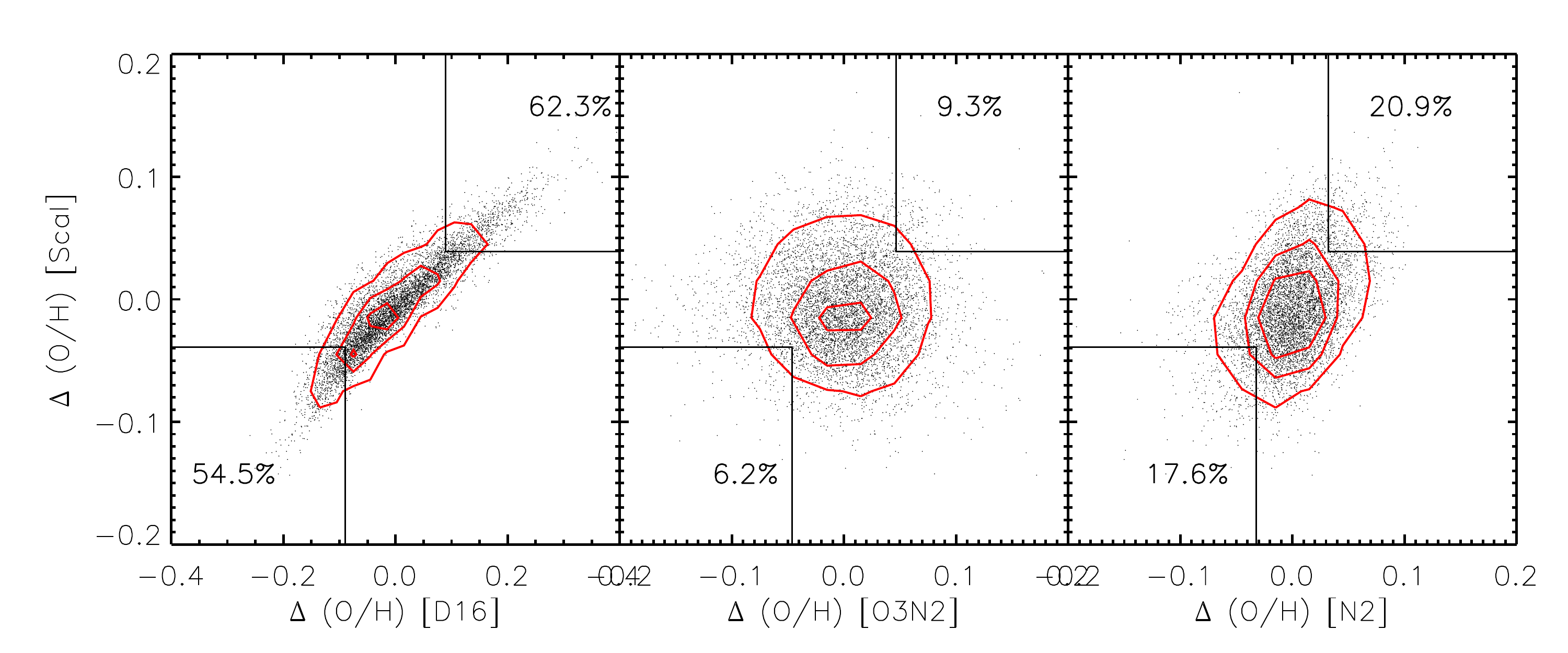}
\caption{For the same HII regions, we compute \doh~ using different metallicity prescriptions in order to compare the overlap of the outlier populations.  Each plot marks with boxes the 1$\sigma$ outliers in each calibration, with the fraction of outliers overlapping between samples given in each box.  Scal and N2S2 (D16, left) show good agreement, with 50-60\% of 1$\sigma$ outliers agreeing between the two prescriptions.  N2 (M13, right) shows poor agreement ($\sim$20\%), and O3N2 (M13, center) shows no correlation ($<$10\% agreement in outliers). 
\label{fig:scatter}}
\end{figure*}

A final demonstration of the strong discrepancies between calibrations is shown in Figure \ref{fig:scatter}.  Here, we calculate \doh~ for the same HII regions using different prescriptions, and compare the 1$\sigma$ outliers.  Between the Scal and N2S2 (D16) calibrations we find a clear correlation, with 50-60\% of outliers overlapping between the two calibrations.  N2 (M13) shows  poor agreement with  Scal, with only $\sim$20\% of the outliers agreeing between the samples.  O3N2 shows no correlation with Scal, having $<$10\% of the outliers overlapping.  While both O3N2 and N2 quantitatively reproduce the radial metallicity gradient slope, they appear to be insensitive to much of the small scale physics related to small metallicity variations recovered by both the Scal and N2S2 (D16).

\begin{deluxetable*}{@{\extracolsep{4pt}}l cccc cccc cccc cccc}
\rotate
\tabletypesize{\tiny}
\centering
\tablecaption{Radial gradient fits (in appendix)  \label{tab:gradients}}
\tablehead{
\colhead{} &
\multicolumn{4}{c}{Scal} &
\multicolumn{4}{c}{N2S2 (D16)} &
\multicolumn{4}{c}{O3N2 (M13)} &
\multicolumn{4}{c}{N2 (PP04)} \\
\cline{2-5} \cline{6-9}  \cline{10-13} \cline{14-17} 
\colhead{Name} &
\colhead{R$_0$} &
\colhead{Gradient} &
\colhead{Gradient} & 
\colhead{$\sigma$(O/H)} &
\colhead{R$_0$} &
\colhead{Gradient} &
\colhead{Gradient} & 
\colhead{$\sigma$(O/H)} &
\colhead{R$_0$} &
\colhead{Gradient} &
\colhead{Gradient} & 
\colhead{$\sigma$(O/H)} &
\colhead{R$_0$} &
\colhead{Gradient} &
\colhead{Gradient} & 
\colhead{$\sigma$(O/H)} \\
\colhead{} & 
\colhead{} & 
\colhead{[dex } & 
\colhead{[dex } & 
\colhead{[dex]} &
\colhead{} & 
\colhead{[dex } & 
\colhead{[dex } & 
\colhead{[dex]} &
\colhead{} & 
\colhead{[dex } & 
\colhead{[dex } & 
\colhead{[dex]} &
\colhead{} & 
\colhead{[dex } & 
\colhead{[dex } & 
\colhead{[dex]} \\
\colhead{} & 
\colhead{} & 
\colhead{ R$_{25}^{-1}$]} & 
\colhead{ kpc$^{-1}$]} & 
\colhead{} &
\colhead{} & 
\colhead{ R$_{25}^{-1}$]} & 
\colhead{ kpc$^{-1}$]} & 
\colhead{} &
\colhead{} & 
\colhead{ R$_{25}^{-1}$]} & 
\colhead{ kpc$^{-1}$]} & 
\colhead{} &
\colhead{} & 
\colhead{ R$_{25}^{-1}$]} & 
\colhead{ kpc$^{-1}$]} & 
\colhead{} 
}
\startdata
ngc628 & 8.556 & -0.191 & -0.0412 &  0.05 & 8.771 & -0.456 & -0.0913 &  0.11 & 8.611 & -0.164 & -0.0329 &  0.05 & 8.508 & -0.009 & -0.0018 &  0.04 \\
 &  $\pm$ 0.005 &  $\pm$  0.018 &  $\pm$  0.0037 &  &  $\pm$ 0.011 &  $\pm$  0.041 &  $\pm$  0.0082 &  &  $\pm$ 0.005 &  $\pm$  0.017 &  $\pm$  0.0035 &  &  $\pm$ 0.003 &  $\pm$  0.012 &  $\pm$  0.0025 &  \\
ngc1087 & 8.483 & -0.144 & -0.0994 &  0.03 & 8.580 & -0.251 & -0.1676 &  0.08 & 8.542 & -0.118 & -0.0784 &  0.04 & 8.526 & -0.102 & -0.0678 &  0.02 \\
 &  $\pm$ 0.003 &  $\pm$  0.006 &  $\pm$  0.0048 &  &  $\pm$ 0.009 &  $\pm$  0.017 &  $\pm$  0.0114 &  &  $\pm$ 0.005 &  $\pm$  0.009 &  $\pm$  0.0061 &  &  $\pm$ 0.003 &  $\pm$  0.005 &  $\pm$  0.0035 &  \\
ngc1672 & 8.563 & -0.056 & -0.0199 &  0.03 & 8.737 & -0.089 & -0.0287 &  0.10 & 8.571 & -0.074 & -0.0238 &  0.04 & 8.567 & -0.052 & -0.0169 &  0.03 \\
 &  $\pm$ 0.003 &  $\pm$  0.007 &  $\pm$  0.0022 &  &  $\pm$ 0.009 &  $\pm$  0.019 &  $\pm$  0.0062 &  &  $\pm$ 0.004 &  $\pm$  0.009 &  $\pm$  0.0030 &  &  $\pm$ 0.002 &  $\pm$  0.005 &  $\pm$  0.0017 &  \\
ngc2835 & 8.565 & -0.516 & -0.1586 &  0.04 & 8.734 & -0.907 & -0.2834 &  0.09 & 8.598 & -0.401 & -0.1253 &  0.06 & 8.567 & -0.319 & -0.0997 &  0.03 \\
 &  $\pm$ 0.004 &  $\pm$  0.012 &  $\pm$  0.0038 &  &  $\pm$ 0.010 &  $\pm$  0.027 &  $\pm$  0.0085 &  &  $\pm$ 0.006 &  $\pm$  0.018 &  $\pm$  0.0056 &  &  $\pm$ 0.003 &  $\pm$  0.009 &  $\pm$  0.0028 &  \\
ngc3627 & 8.528 &  0.056 &  0.0095 &  0.03 & 8.733 &  0.084 &  0.0164 &  0.07 & 8.571 &  0.108 &  0.0212 &  0.04 & 8.506 &  0.041 &  0.0080 &  0.03 \\
 &  $\pm$ 0.004 &  $\pm$  0.016 &  $\pm$  0.0031 &  &  $\pm$ 0.009 &  $\pm$  0.034 &  $\pm$  0.0066 &  &  $\pm$ 0.005 &  $\pm$  0.019 &  $\pm$  0.0037 &  &  $\pm$ 0.004 &  $\pm$  0.015 &  $\pm$  0.0029 &  \\
ngc4254 & 8.592 & -0.104 & -0.0425 &  0.03 & 8.873 & -0.241 & -0.0966 &  0.08 & 8.632 & -0.142 & -0.0568 &  0.04 & 8.513 &  0.000 &  0.0001 &  0.04 \\
 &  $\pm$ 0.002 &  $\pm$  0.004 &  $\pm$  0.0019 &  &  $\pm$ 0.004 &  $\pm$  0.011 &  $\pm$  0.0045 &  &  $\pm$ 0.002 &  $\pm$  0.006 &  $\pm$  0.0022 &  &  $\pm$ 0.002 &  $\pm$  0.005 &  $\pm$  0.0020 &  \\
ngc4535 & 8.579 & -0.091 & -0.0244 &  0.04 & 8.844 & -0.258 & -0.0628 &  0.09 & 8.626 & -0.082 & -0.0200 &  0.05 & 8.471 &  0.129 &  0.0314 &  0.04 \\
 &  $\pm$ 0.006 &  $\pm$  0.018 &  $\pm$  0.0045 &  &  $\pm$ 0.012 &  $\pm$  0.041 &  $\pm$  0.0100 &  &  $\pm$ 0.006 &  $\pm$  0.021 &  $\pm$  0.0051 &  &  $\pm$ 0.005 &  $\pm$  0.018 &  $\pm$  0.0043 &  \\
ic5332 & 8.514 & -0.401 & -0.1362 &  0.05 & 8.671 & -0.717 & -0.2391 &  0.11 & 8.574 & -0.251 & -0.0836 &  0.07 & 8.520 & -0.224 & -0.0746 &  0.03 \\
 &  $\pm$ 0.010 &  $\pm$  0.035 &  $\pm$  0.0119 &  &  $\pm$ 0.022 &  $\pm$  0.076 &  $\pm$  0.0255 &  &  $\pm$ 0.014 &  $\pm$  0.049 &  $\pm$  0.0163 &  &  $\pm$ 0.006 &  $\pm$  0.020 &  $\pm$  0.0065 &  \\
\enddata
\end{deluxetable*}

\clearpage

\section{Azimuthal Variations Atlas}
\label{appendix;azimuthal}

For each galaxy, we present two visualizations of the complete spatial distribution of \doh.  
Removing any size or luminosity information, we take the position of each HII region and place it onto a polar projection, with the azimuthal angle on the x-axis and the natural log of the radius on the y-axis (upper-left panels).
Here we color code each region by \doh.  In polar plots, spiral arms with fixed pitch angle appear as straight lines.  As an alternate visualization (upper-right panels), we also take the individual measurements of \doh\ for each HII region (circles) and interpolate them onto a two dimensional surface using kriging and an exponential model. We mark an approximate spiral arm location by tracing the brightest star-forming regions along the arm by eye (overlaid in grey), though we note here this is very approximate and not sufficient for detailed environmental comparisons.

We mark in each of these images a ring at fixed radius, matched between both images, and show below the polar plot how \doh~ varies azimuthally.  These rings are $0.1 {\rm R}_{25}$ wide, and selected to be completely contained within the mask but demonstrating the strongest trend for azimuthal variations.  To guide the eye, we overplot the median \doh\ measured in 30$^\circ$ wide bins, and color the positive and negative residuals with respect to \doh=0 (top left, middle panel).   We compare this with the total extinction-corrected luminosity around the same azimuthal ring in matched 30$^\circ$ wide bins (top left, bottom panel).  

As a final visualization, we also map the locations of the HII regions that are 1$\sigma$ outliers towards both enhanced and reduced abundances (bottom panels).  We show these populations in relation to the $\sim$500pc scale bulk CO emission, roughly tracing the underlying gravitation potential and the spiral pattern. 

Discussion of the trends identified within each galaxy and across the galaxy sample is contained in Sections \ref{sec:aztrends} and \ref{sec:dis:azimuth}. 

\clearpage

\begin{figure*}
\centering
\includegraphics[height=3.5in]{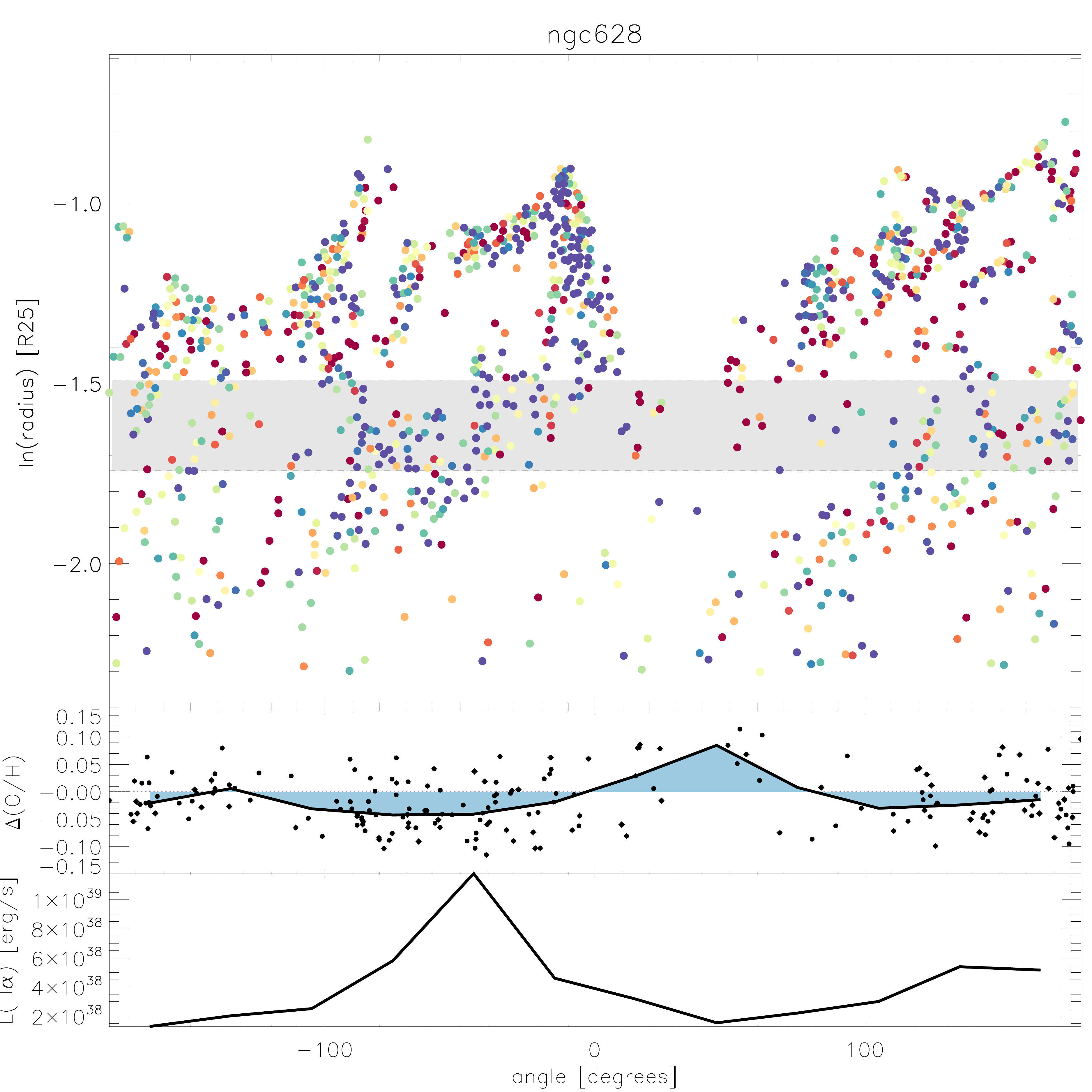}
\includegraphics[height=3.5in]{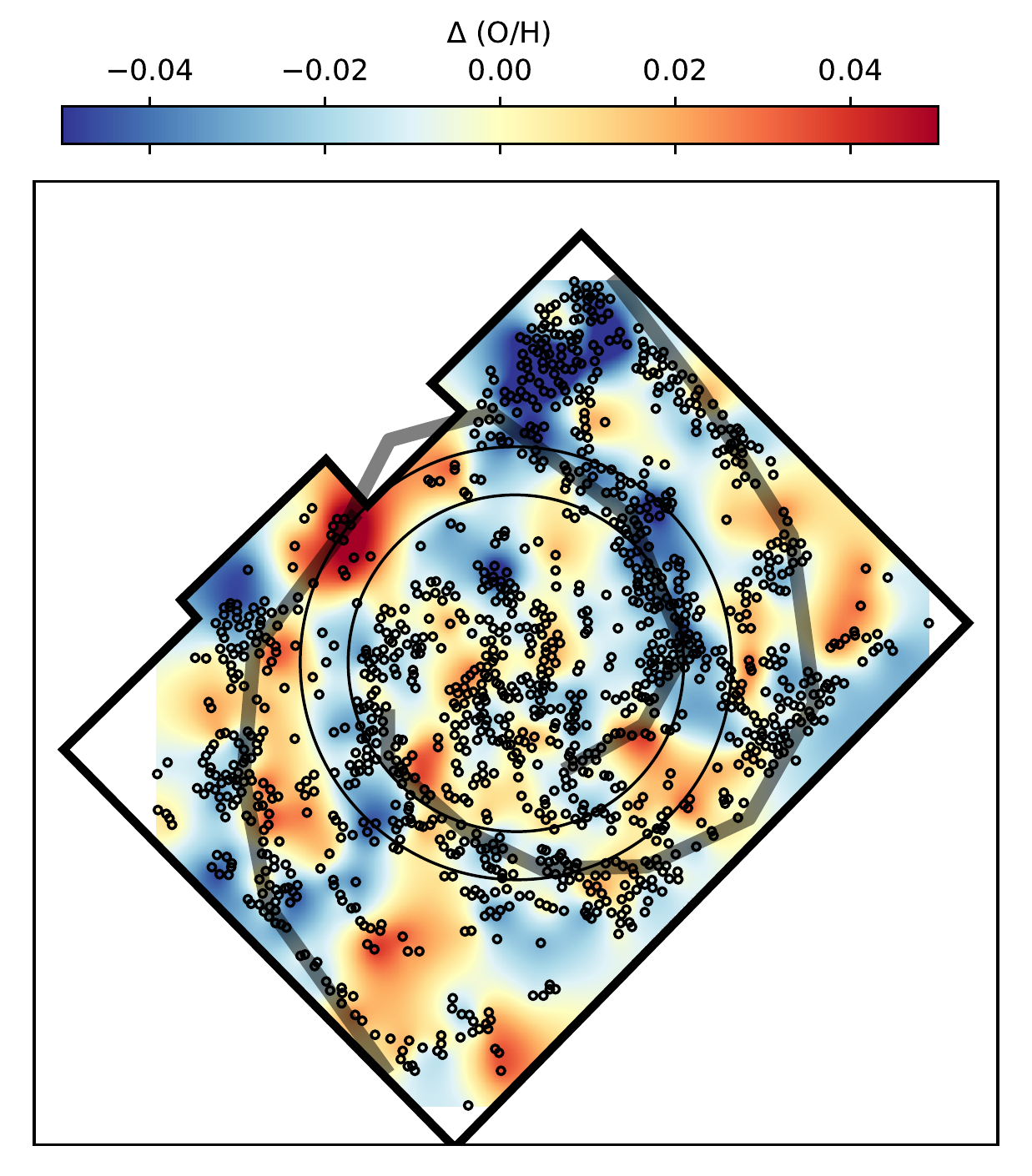}
\includegraphics[width=6in]{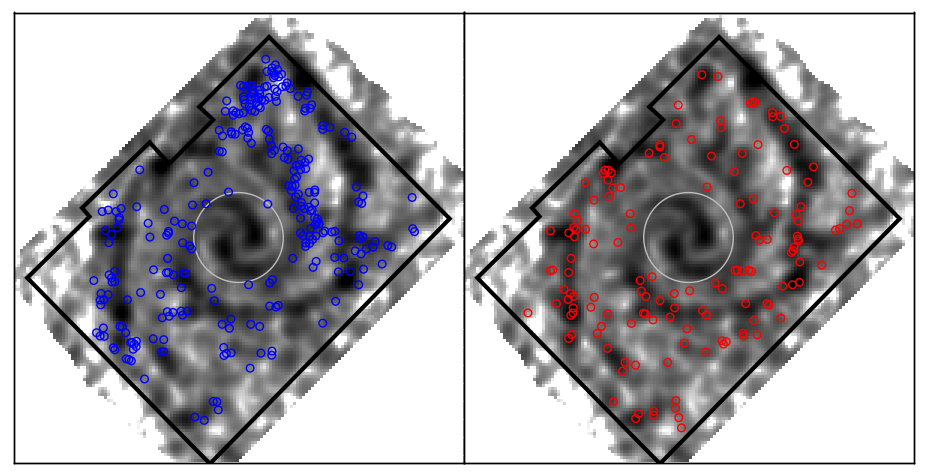}
\caption{Visualizing the azimuthal variations within NGC 628. Top left: We unwrap the spiral structure into a polar projection, plotting the azimuthal angle on the x-axis and the natural log of the radius on the y-axis.  We  indicate each HII region as a dot (removing any biases from size or luminosity) color-coded by \doh. 
We choose one ring at fixed deprojected radius that best exemplifies  azimuthal variations in \doh~ and how it relates to the azimuthal H$\alpha$ extinction-corrected luminosity. These rings are $0.1 {\rm R}_{25}$ wide, and selected to be completely contained within the mask but demonstrating the strongest trend for azimuthal variations.  To guide the eye, we overplot the median \doh\ measured in 30$^\circ$ wide bins, and color the positive and negative residuals with respect to \doh=0 (top left, middle panel).   We compare this with the total extinction-corrected luminosity around the same azimuthal ring in matched 30$^\circ$ wide bins (top left, bottom panel).
Top right: The individual measurements of \doh\ for each HII region (circles) are interpolated onto a two dimensional surface using kriging and an exponential model.  We mark an approximate spiral arm location by tracing the brightest star-forming regions along the arm by eye (overlaid in grey). 
The same ring at fixed deprojected radius is also drawn.  Bottom: We overlay the positions of all regions with enhanced (red) or reduced (blue) abundances onto an image of the bulk CO emission to highlight their position in relation to the spiral structure. 
NGC 628 shows tentative evidence for azimuthal variations, however in the ring at fixed deprojected radius it appears the enhancement is anti-correlated with the spiral pattern.  
Results for the remaining galaxies are shown in the following figures.  We tentatively identify azimuthal variations in half the galaxies shown, though a correlation with spiral structure is not always clear.
\label{fig:unwrap628}}
\end{figure*}

\begin{figure*}
\centering
\includegraphics[height=3.5in]{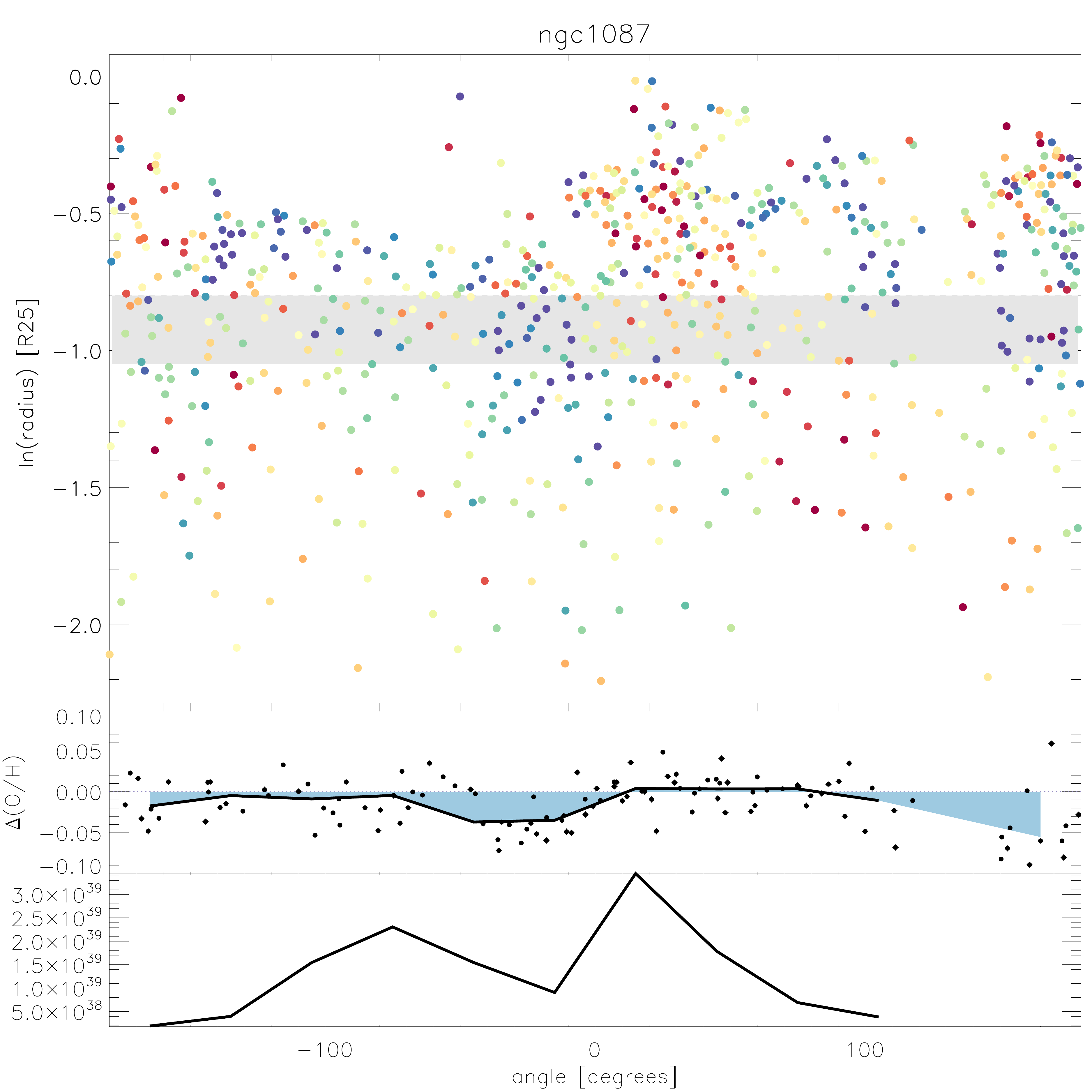}
\includegraphics[height=3.5in]{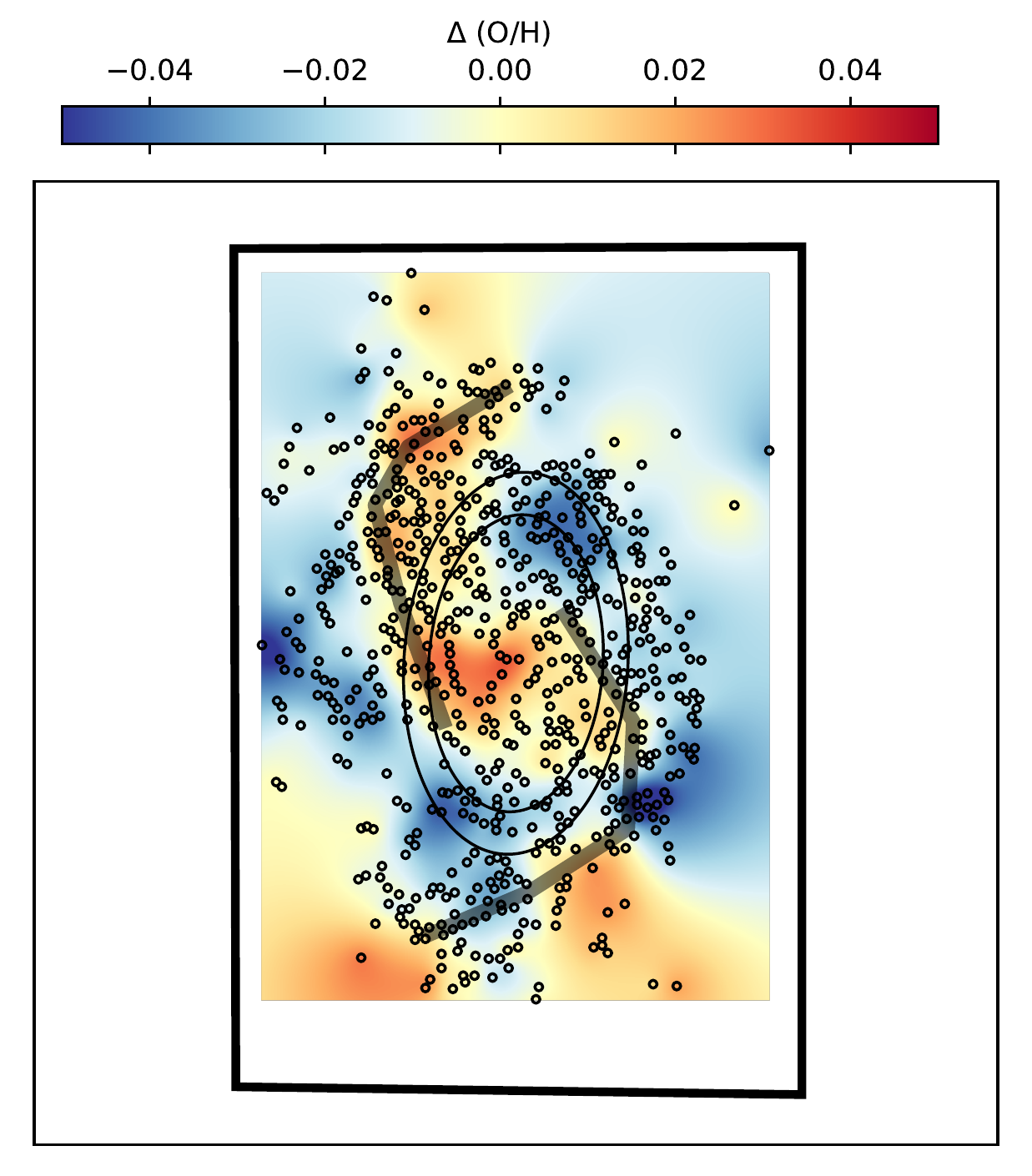}
\includegraphics[width=6in]{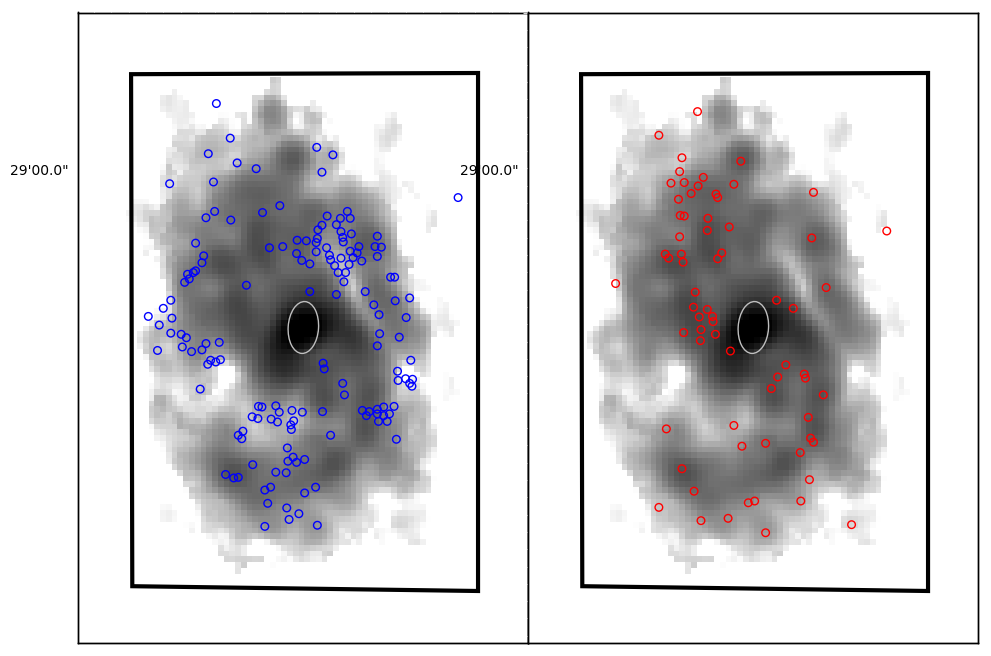}
\caption{Visualizing the azimuthal variations within NGC 1087, as described in Fig.~\ref{fig:unwrap628}.  NGC 1087 shows clear systematic enhancement broadly spread across the northern spiral arm.  Such a trend is much less obvious on the southern arm, though hinted at in the ring at fixed deprojected radius. 
\label{fig:unwrap1087}}
\end{figure*}

\begin{figure*}
\centering
\includegraphics[height=3.5in]{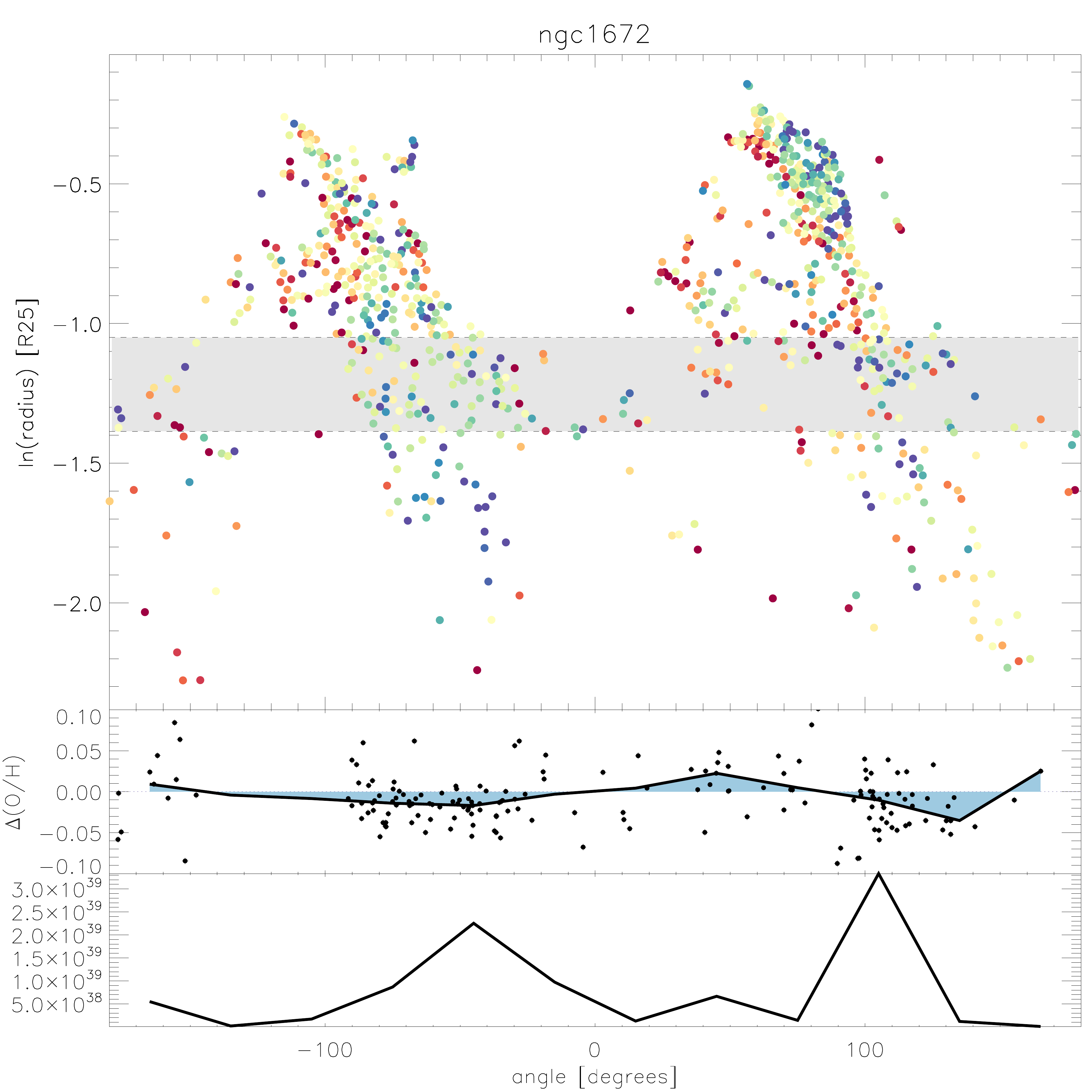}
\includegraphics[height=3.5in]{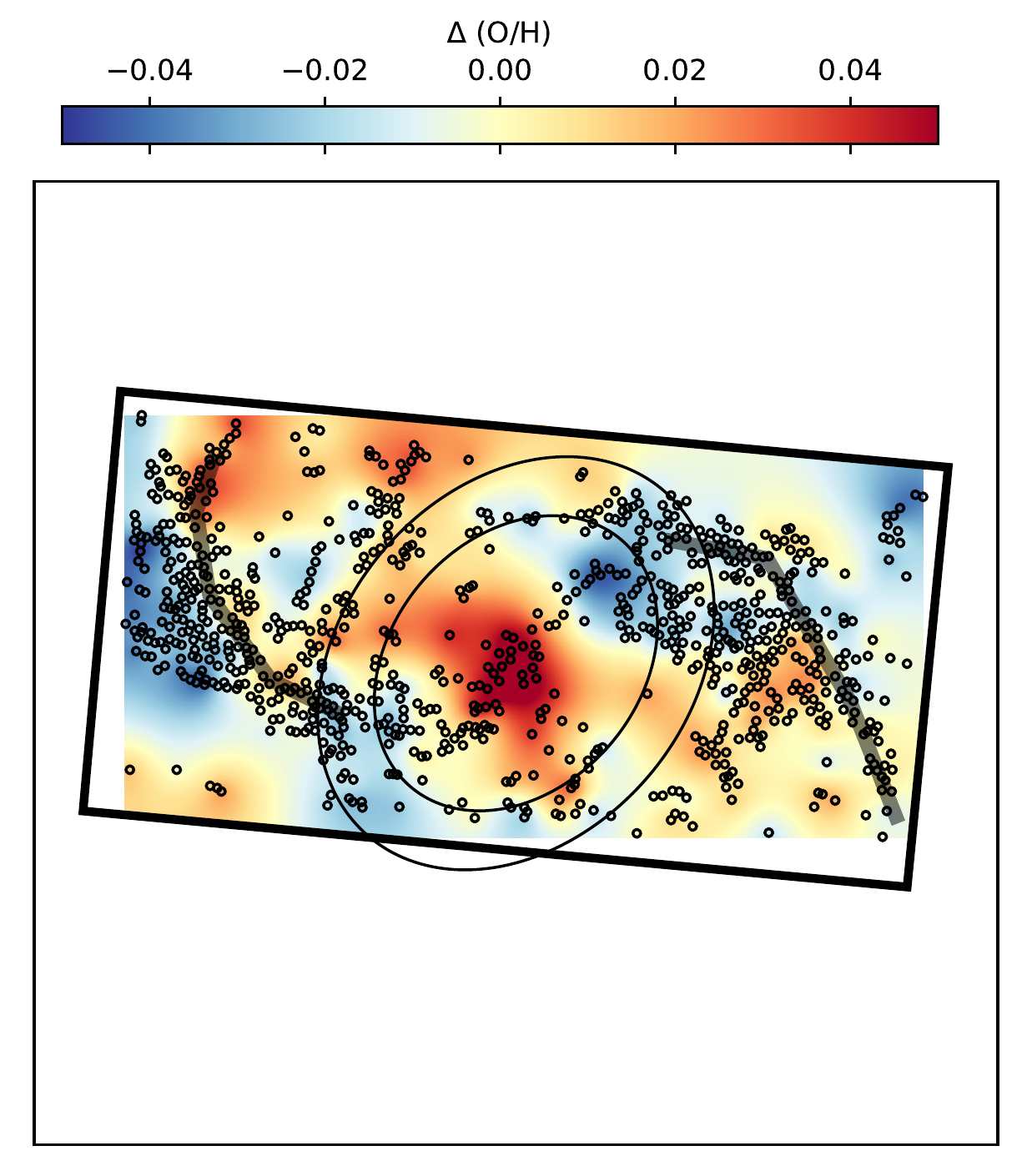}
\includegraphics[width=6in]{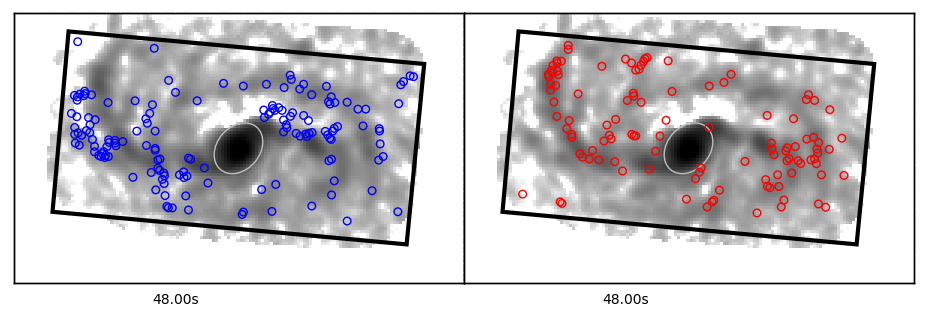}
\caption{Visualizing the azimuthal variations within NGC 1672, as described in Fig.~\ref{fig:unwrap628}.  NGC 1672 shows clear systematic enhancement along a narrow ridge in the eastern spiral arm.  No similar trend is seen on the western arm, though the spiral pattern is also less pronounced.
\label{fig:unwrap1672}}
\end{figure*}

\begin{figure*}
\centering
\includegraphics[height=3.5in]{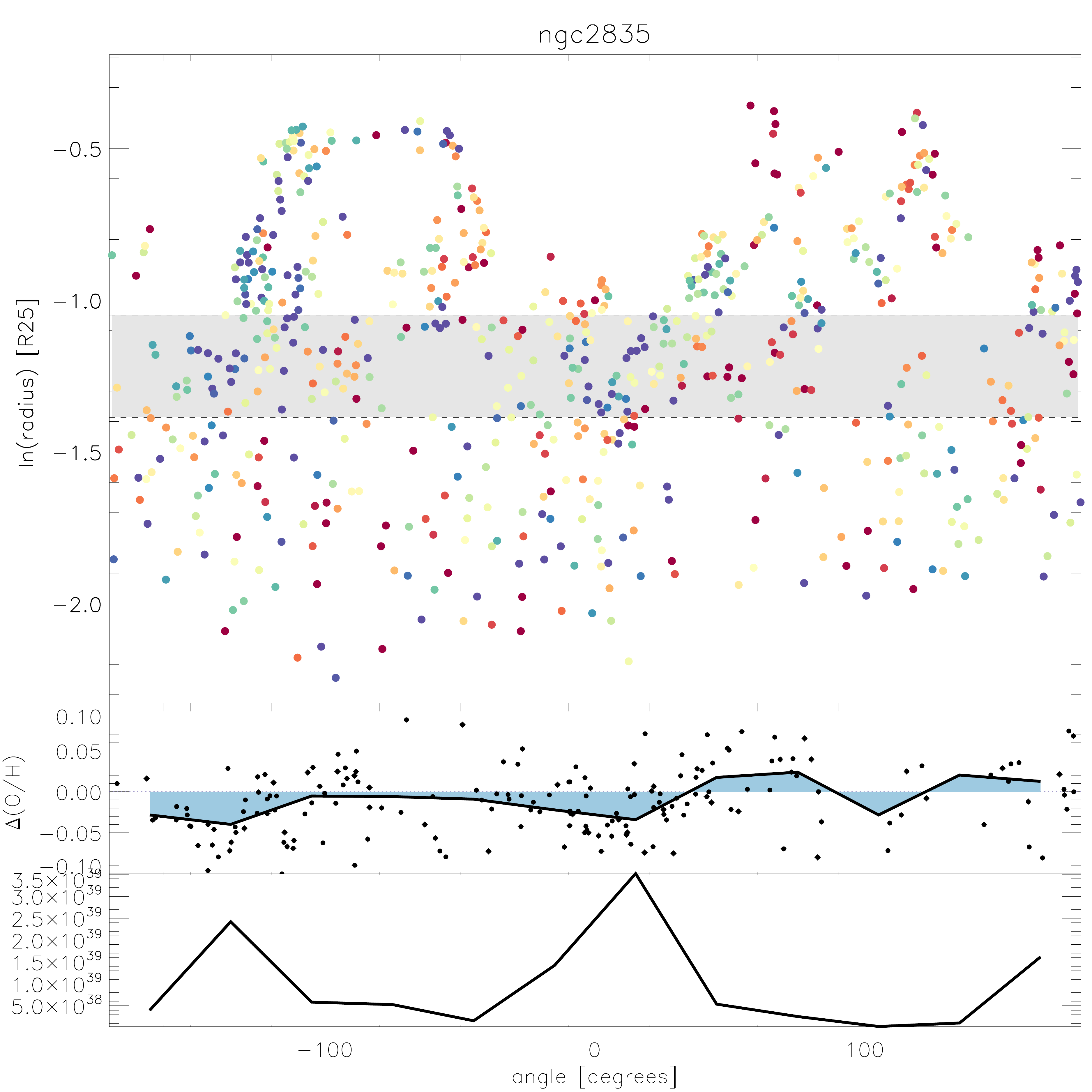}
\includegraphics[height=3.5in]{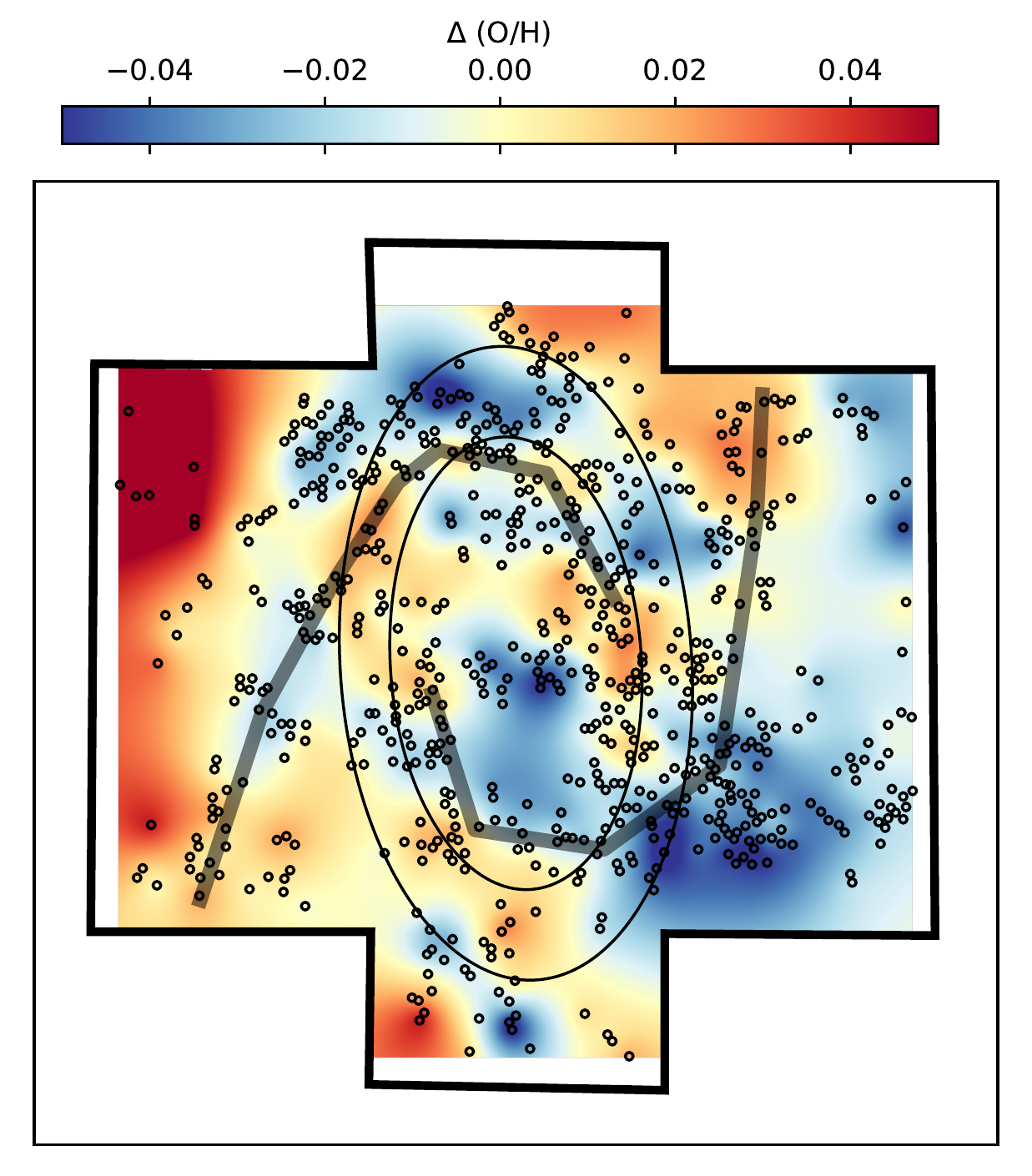}
\includegraphics[width=6in]{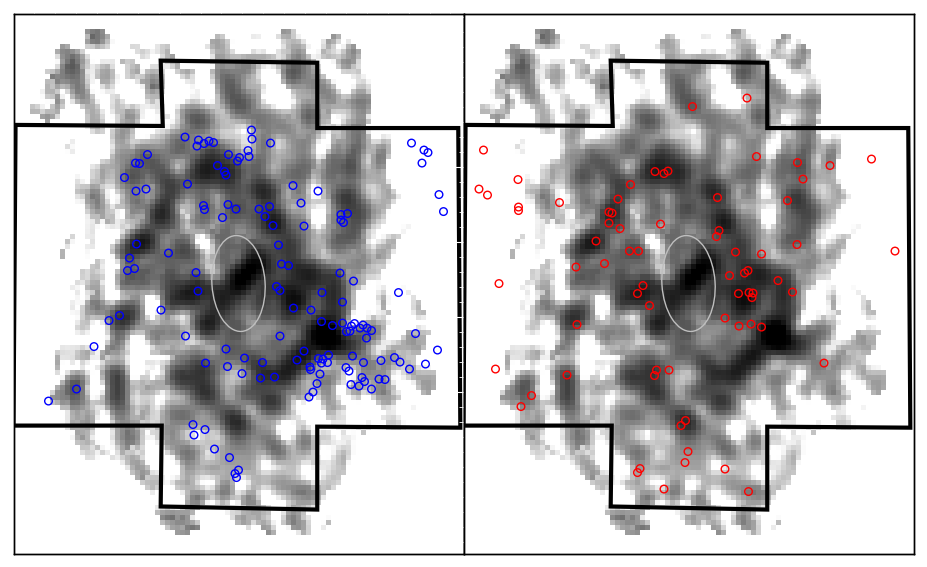}
\caption{Visualizing the azimuthal variations within NGC 2835, as described in Fig.~\ref{fig:unwrap628}.  NGC 2835 shows azimuthal variations along the ring at fixed deprojected radius with enhanced abundances that are anti-correlated with the spiral structure.  Some general preference of the enhanced regions to sit along the spiral pattern is however seen in the bottom two panels.  
\label{fig:unwrap2835}}
\end{figure*}

\begin{figure*}
\centering
\includegraphics[height=3.5in]{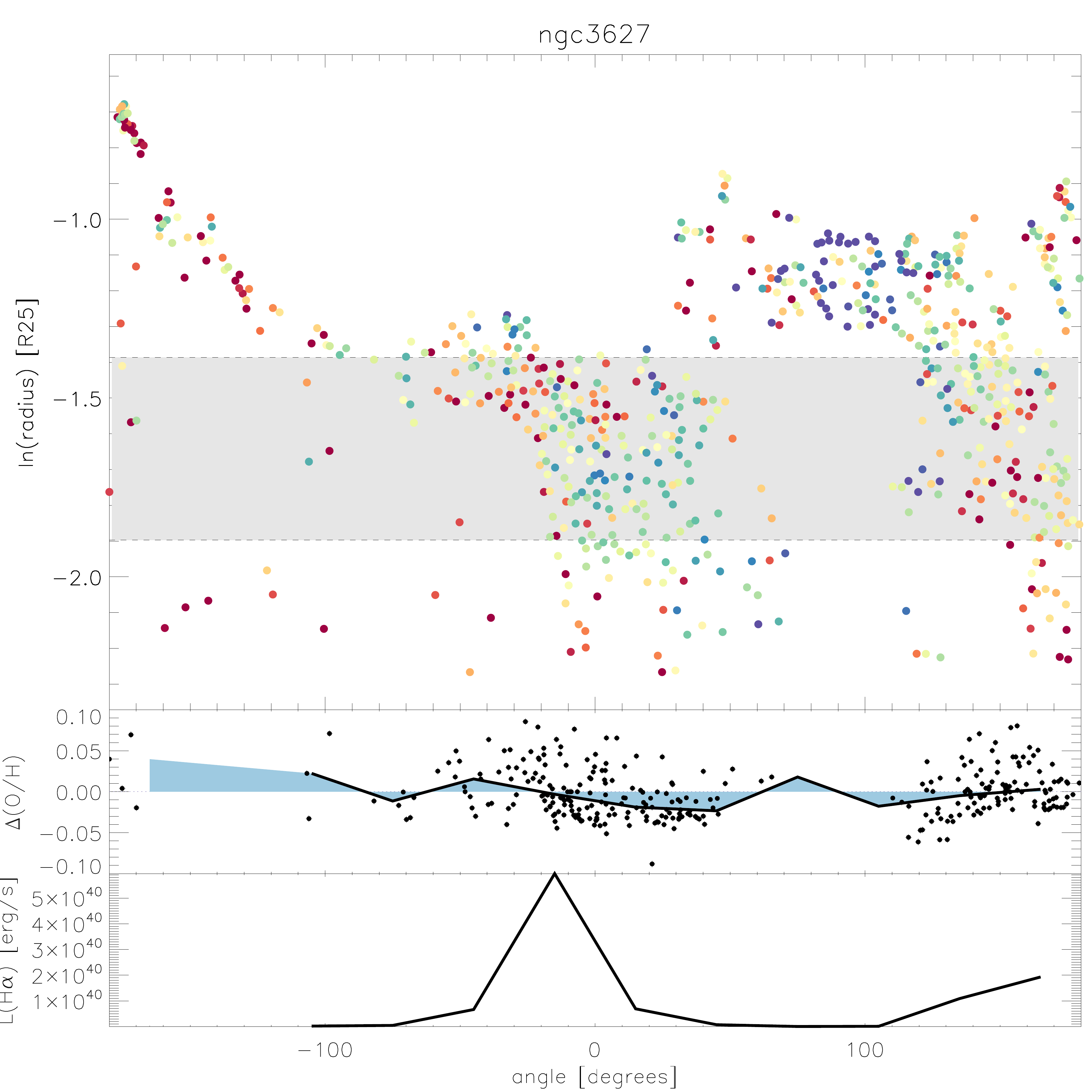}
\includegraphics[height=3.5in]{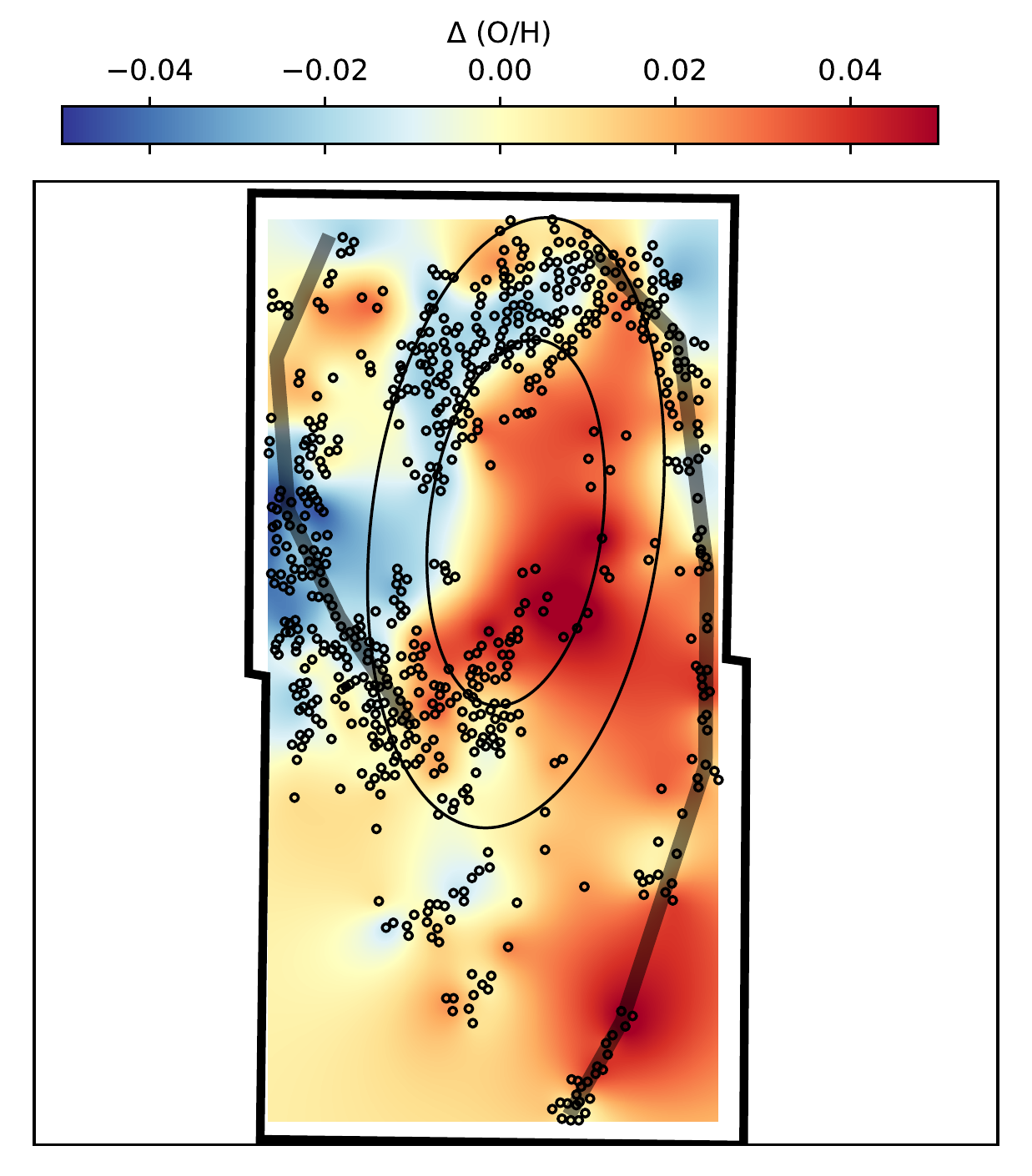}
\includegraphics[width=6in]{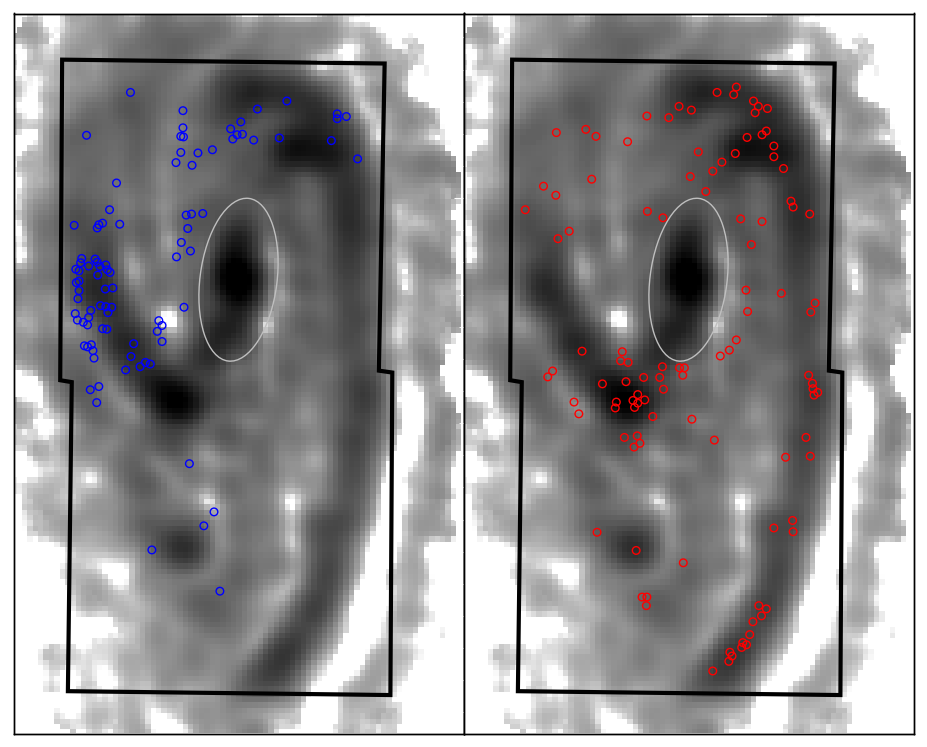}
\caption{Visualizing the azimuthal variations within NGC 3627, as described in Fig.~\ref{fig:unwrap628}.  NGC 3627 does not show any clear azimuthal variation, and our field of view is morphologically dominated by the stellar bar and strongly star-forming bar-ends.
\label{fig:unwrap3627}}
\end{figure*}

\begin{figure*}
\centering
\includegraphics[height=3.5in]{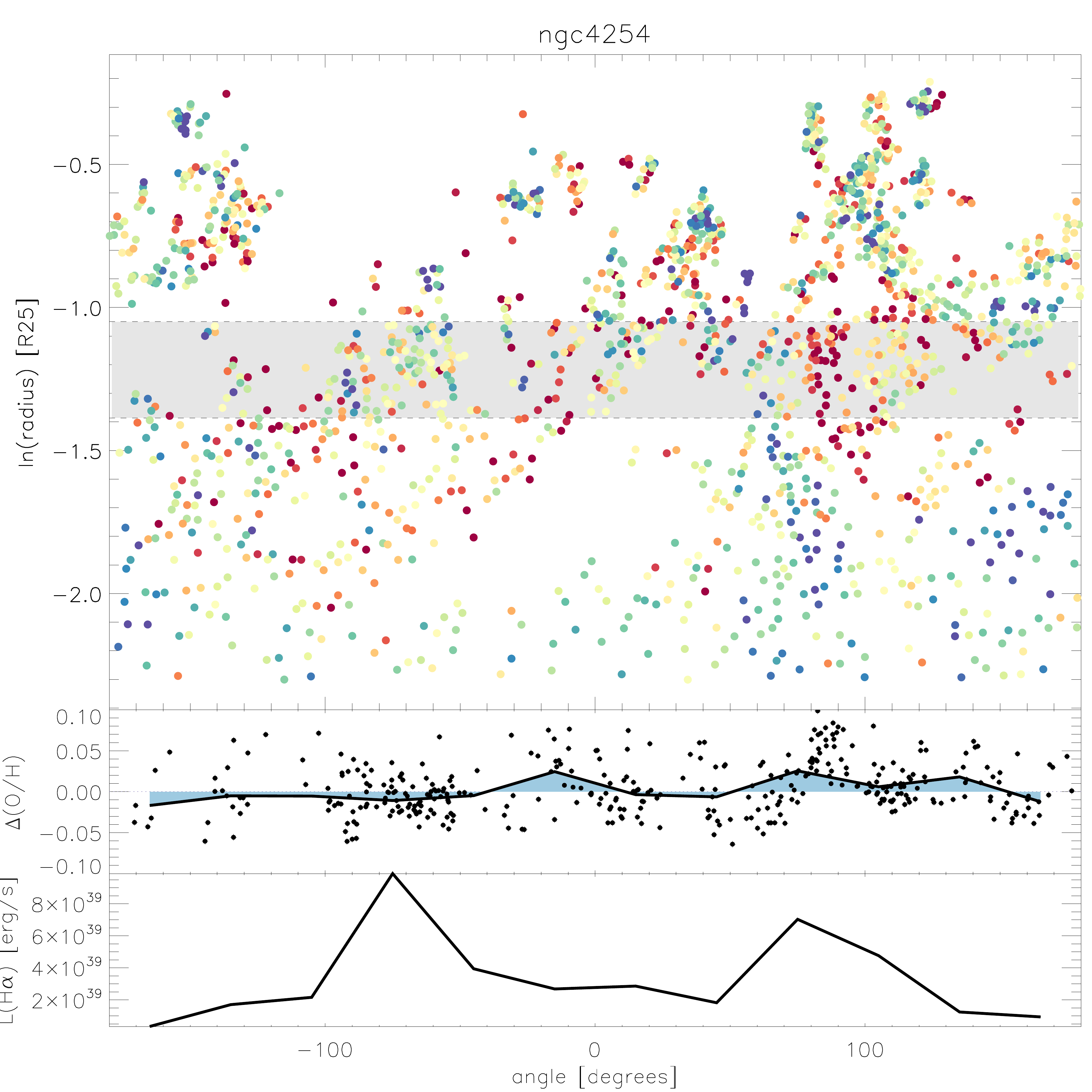}
\includegraphics[height=3.5in]{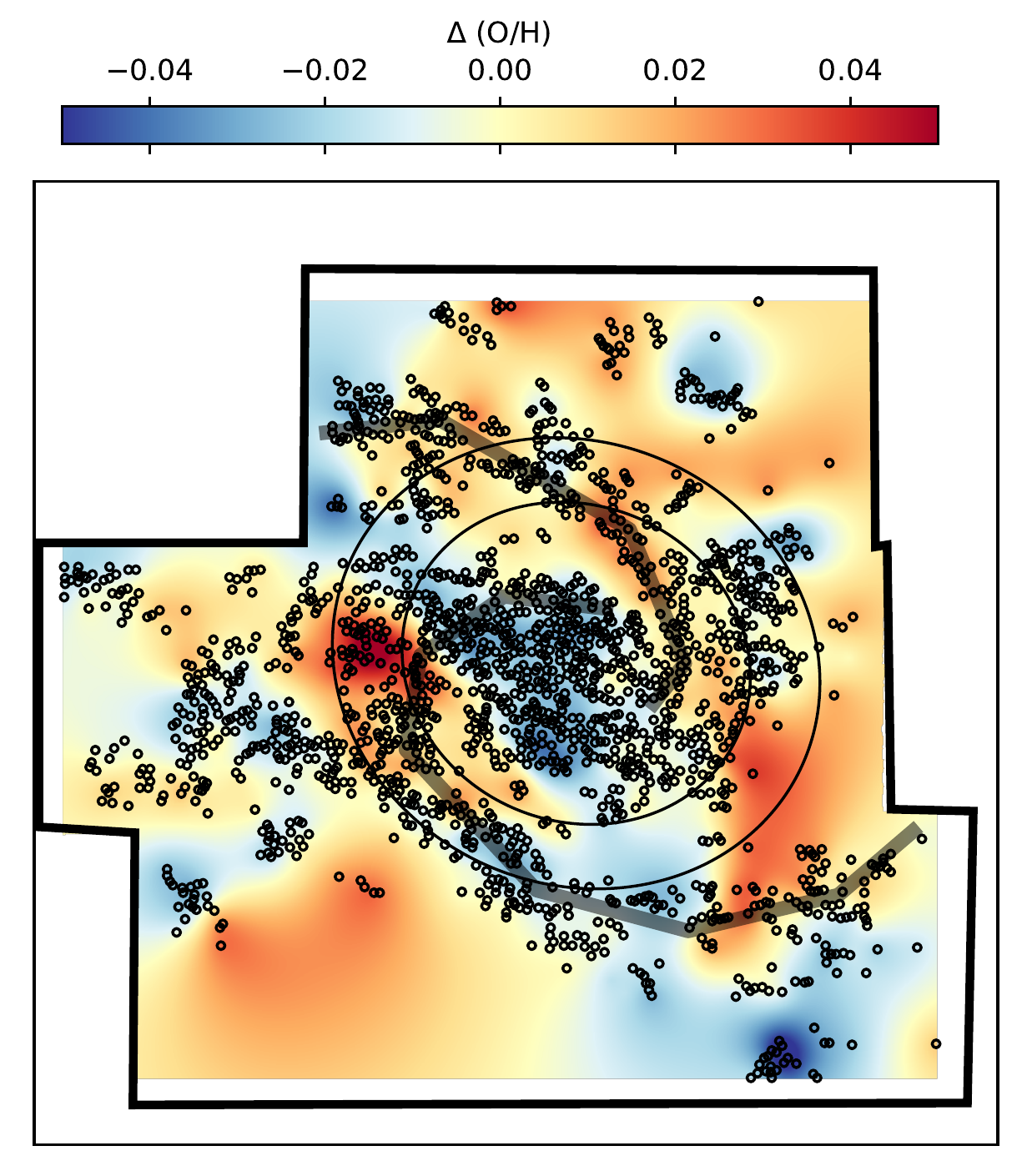}
\includegraphics[width=6in]{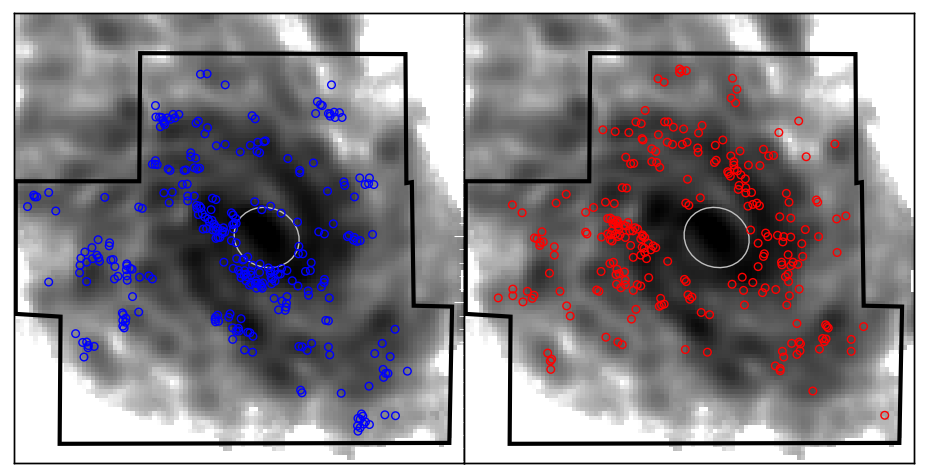}
\caption{Visualizing the azimuthal variations within NGC 4254, as described in Fig.~\ref{fig:unwrap628}.  NGC 4254 exhibits a very flocculent spiral pattern, and does not show any clear azimuthal variation.
\label{fig:unwrap4254}}
\end{figure*}

\begin{figure*}
\centering
\includegraphics[height=3.5in]{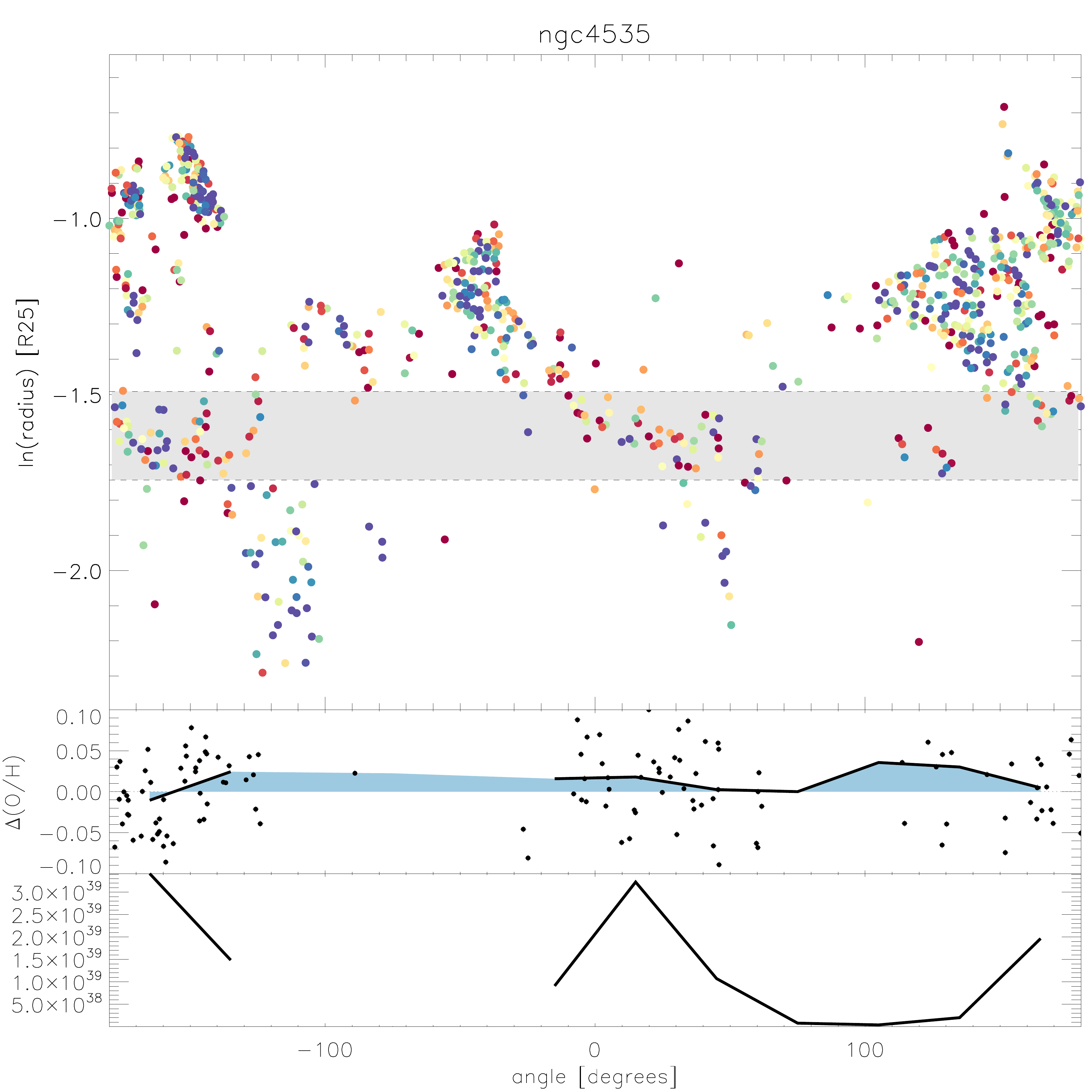}
\includegraphics[height=3.5in]{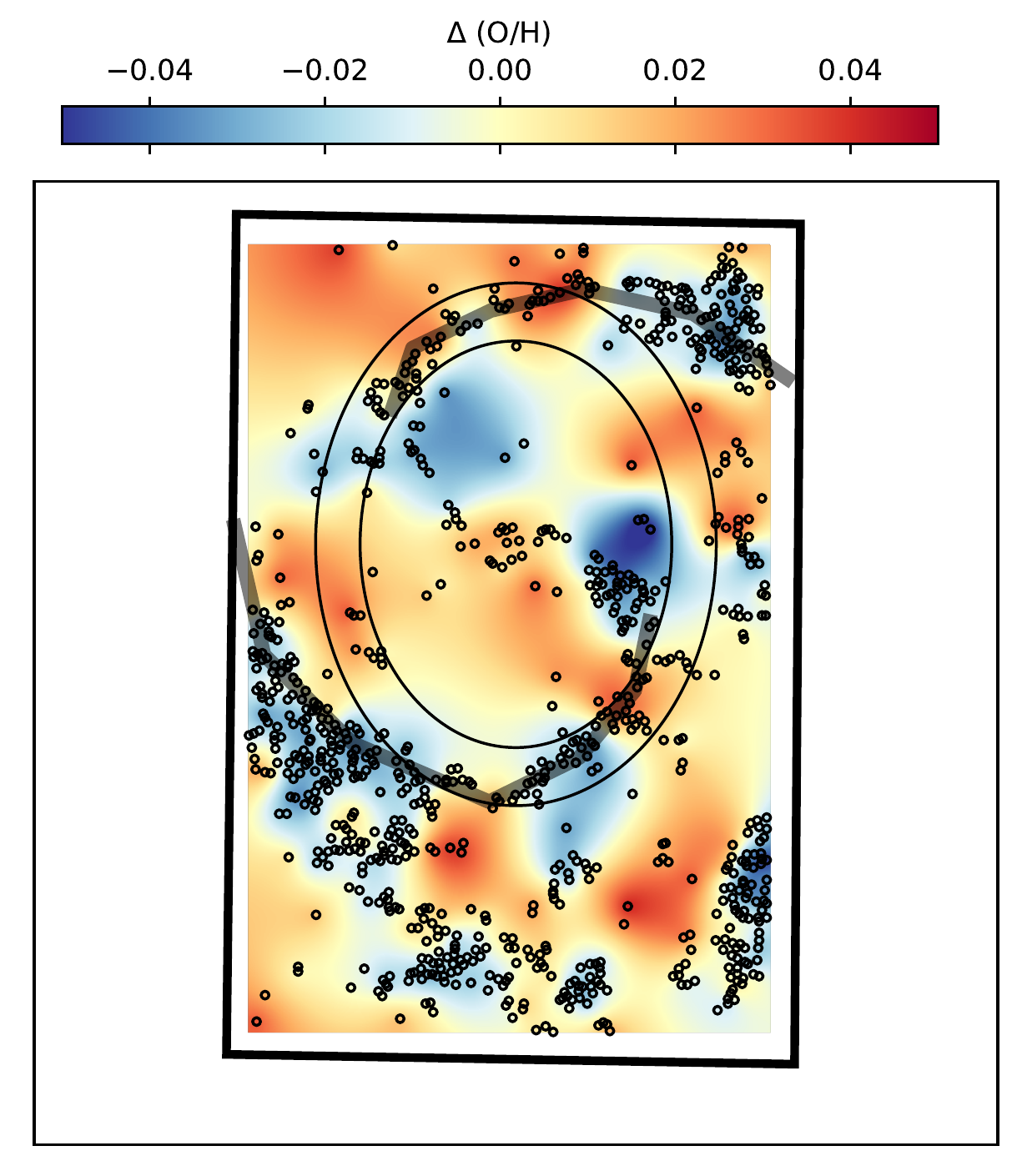}
\includegraphics[width=6in]{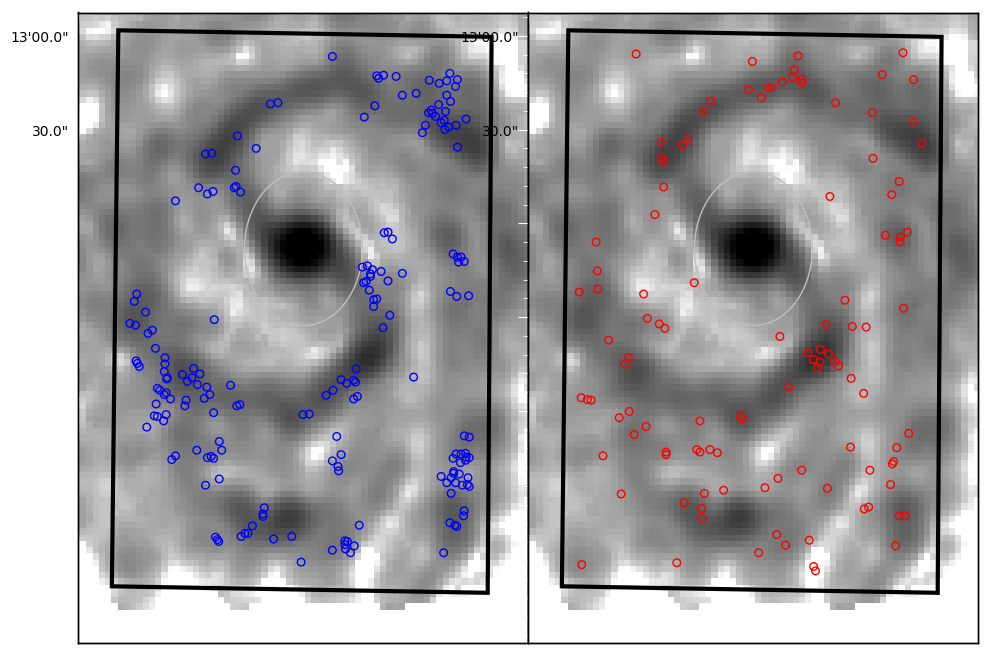}
\caption{Visualizing the azimuthal variations within NGC 4535, as described in Fig.~\ref{fig:unwrap628}.  NGC 4535 has a well defined spiral pattern and does not show any clear azimuthal variation.  It does have a ridge of enhanced regions along the northern spiral arm.  
\label{fig:unwrap4535}}
\end{figure*}

\begin{figure*}
\centering
\includegraphics[height=3.5in]{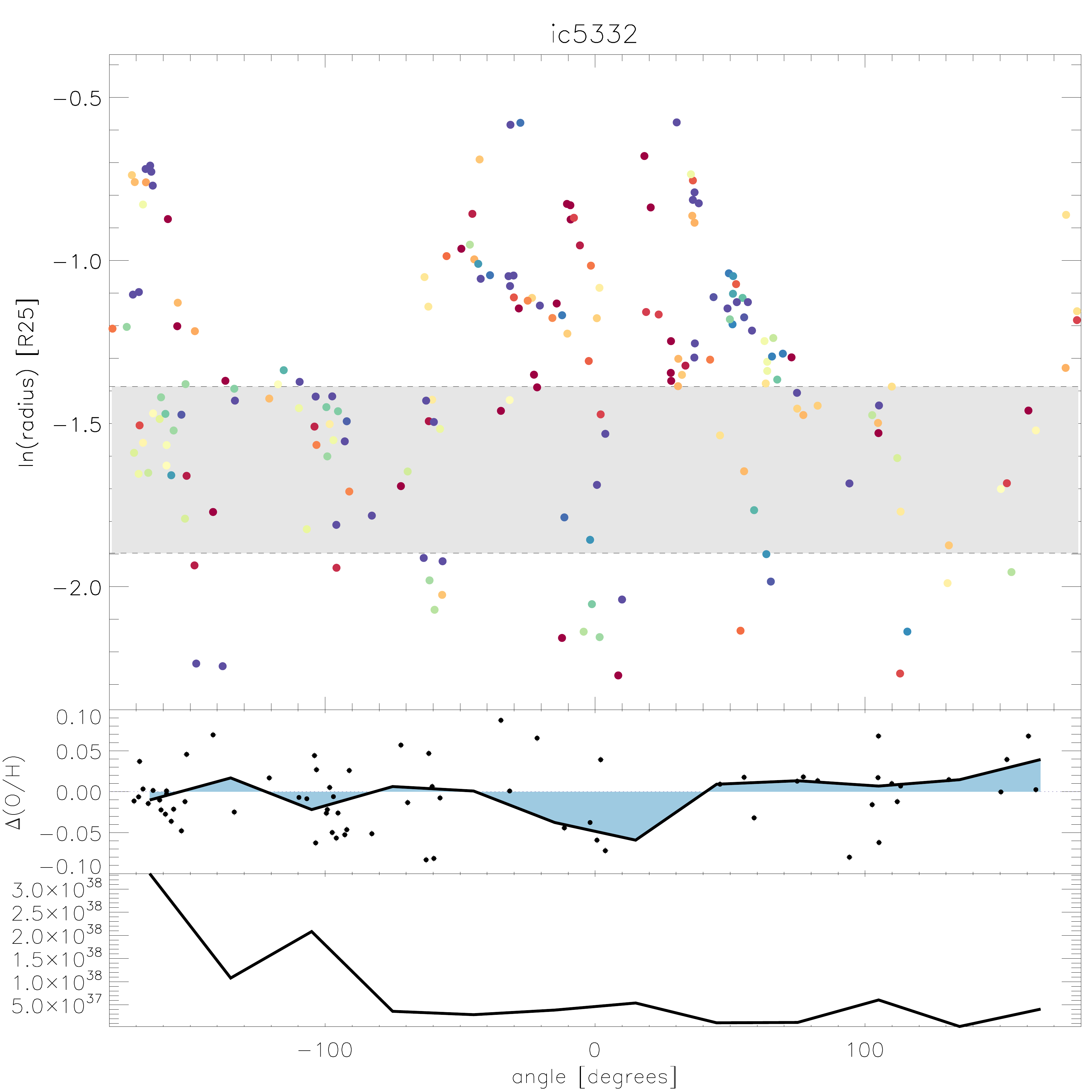}
\includegraphics[height=3.5in]{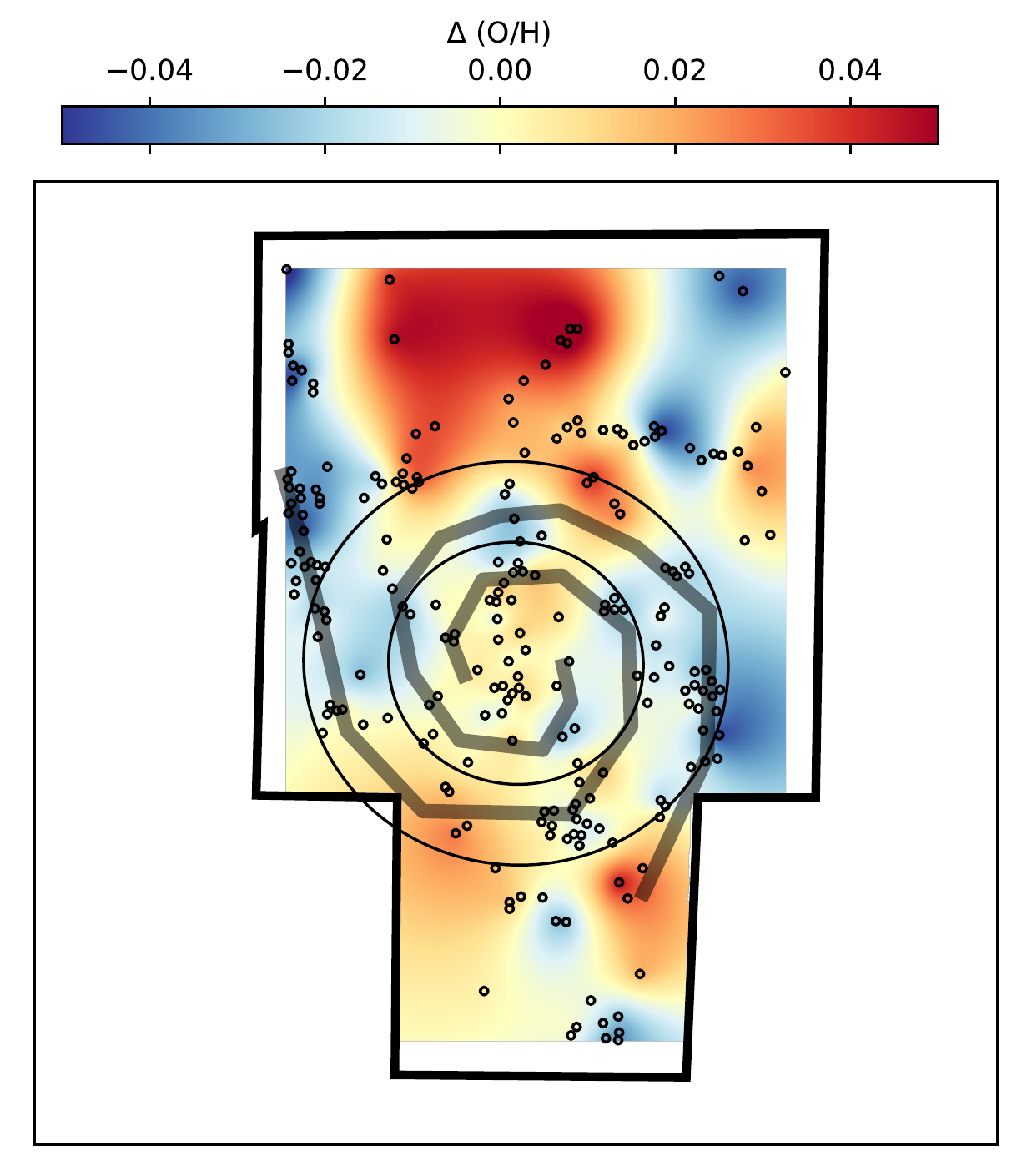}
\includegraphics[width=6in]{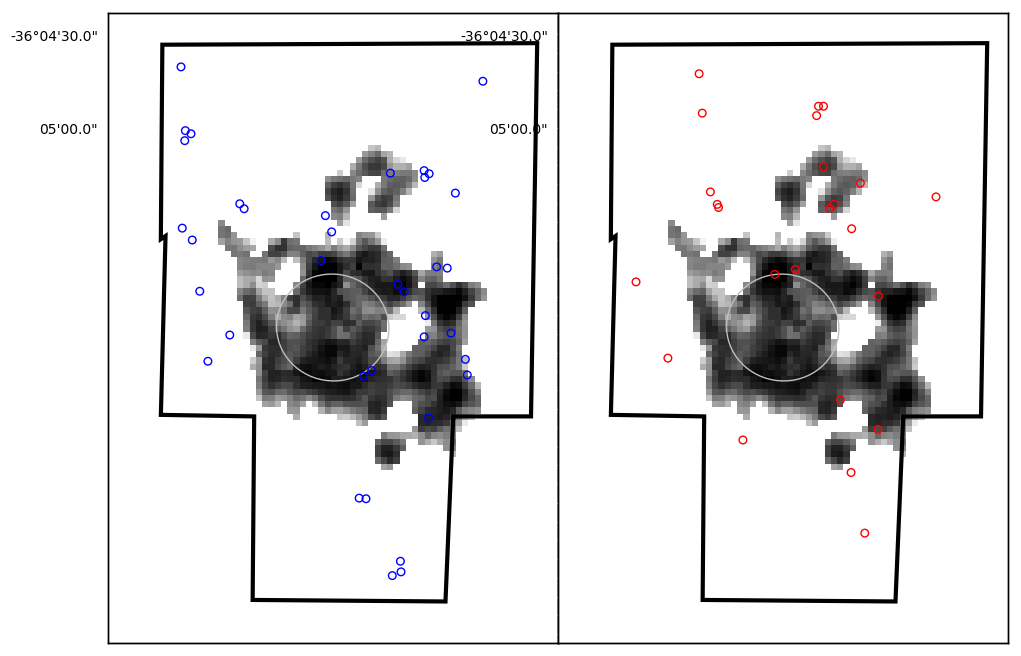}
\caption{Visualizing the azimuthal variations within IC 5332, as described in Fig.~\ref{fig:unwrap628}.  This flocculent galaxy has very little CO emission, though our ALMA coverage is well matched to the MUSE field of view. IC 5332 has a tightly wound spiral pattern and low star formation rate, and does not show any clear azimuthal variation. 
\label{fig:unwrap5332}}
\end{figure*}

\end{document}